\documentclass[12pt]{article}
\usepackage{psfig}

\def\Jou#1#2#3#4{{#1} {\bf #2}, #3 (#4)}

\def\NPA{{Nucl. Phys.} A}
\def\NPB{{Nucl. Phys.} B}
\def\PLB{{Phys. Lett.}  B}

\def\PRL{Phys. Rev. Lett.}

\def\PRD{{Phys. Rev.} D}

\def\ZPA{{Z. Phys.} A}
\def\ZPC{{Z. Phys.} C}
\def\EPJC{{Eur. Phys. J.} C}

\newfam\BMath
\font\BMathL=cmmib10 
\font\BMathl=cmmib7
\font\BMathm=cmmib5
\textfont\BMath=\BMathL \scriptfont\BMath=\BMathl
\scriptscriptfont\BMath=\BMathm

\def\B{{\fam\BMath b}}

\def\K{{\fam\BMath k}}

\def\a{\alpha}
\def\b{\beta}
\def\c{\chi}
\def\d{\delta}
\def\D{\Delta}
\def\e{\epsilon}
\def\f{\phi}
\def\g{\gamma}
\def\j{\Psi}
\def\J{\psi}
\def\l{\lambda}
\def\L{\Lambda}
\def\p{\pi}

\def\m{\mu}
\def\n{\nu}

\def\r{\rho}
\def\s{\sigma}

\def\ca{{\cal A}}
\def\cc{{\cal C}}
\def\cb{{\cal B}}

\def\cg{{\cal G}}

\def\cm{{\cal M}}

\def\cq{{\cal Q}}

\def\dd{\mbox{d}}

\def\exp{\mbox{\rm exp}}

\def\ra{\rightarrow}

\def\lra{\longrightarrow}
\def\llra{\longleftrightarrow}

\def\srm#1{\mbox{\sevenrm #1}}
\def\ssrm#1{\mbox{\fiverm #1}}

\def\dxb{[\dd x]}
\def\dyb{[\dd y]}
\def\dkb{[\dd^2 \Kp]}
\def\lqcd{\Lambda_{\rm {QCD} }}

\def\intbs#1{\frac{\dd^2 \B_#1}{(4\pi)}}

\def\intbds#1{\frac{\dd^2 \B'_#1}{(4\pi)}}
\def\Kp{{\K_\perp}}

\def\Kpn#1{\K_{\perp \,#1}}
\def\mc{(2 m_c)}
\def\ot{\otimes}

\def\be{\begin{equation}}
\def\ee{\end{equation}}
\def\bea{\begin{eqnarray}}
\def\eea{\end{eqnarray}}
\def\ba{\begin{array}}
\def\ea{\end{array}}
\def\eref#1{Eq.~(\ref{#1})}

\def\fref#1{Fig.~\ref{#1}}
\def\bfi{\begin{figure}}
\def\efi{\end{figure}}
\def\bpi#1{\begin{picture}#1}
\def\epi{\end{picture}}

\def\half{\frac{1}{2}} 

\newcommand{\ncom}{\newcommand}
\ncom{\lan}{\langle}
\ncom{\ran}{\rangle}
\ncom\fx{\!\!\!\!}
\ncom\fxt{\!\!}
\newcommand{\sla}{\hspace*{-0.20cm}/}
\newcommand{\Sla}{\hspace*{-0.32cm}/}

\font\lbigbf=cmbx10 scaled 1500
\font\sevenrm=cmr7
\font\fiverm=cmr5

\begin{document}

\begin{flushright}
\footnotesize \sffamily NUC-MINN-98/12-T
\end{flushright}

\begin{centering}
{\lbigbf Colour Octet Contribution in Exclusive P-Wave Charmonium Decay 
into Octet and Decuplet Baryons}

\vspace{0.25cm}

{S.M.H. Wong}

\em School of Physics and Astronomy, University of Minnesota, Minneapolis, \\
MN 55455, USA\footnote{present address} 

\null

Fachbereich Physik, Universit\"at Wuppertal, D-42097 Wuppertal, Germany  \\

\null

Institute of Accelerating Systems and Applications (IASA), P.O. Box 17214, \\ 
GR-10024 Athens, Greece

and

Nuclear and Particle Physics Section, University of Athens, Panepistimiopolis, \\ 
GR-15771 Athens, Greece

\end{centering}

\vspace{2cm}
\begin{abstract}

In the last years, the need of the colour octet state in inclusive P-wave 
charmonium decay has been firmly established. However, the implications
of this in the corresponding exclusive reactions have not been fully
recognized. We argue for the necessity of the colour octet in P- and
higher-wave quarkonium decay. Using a set of phenomenologically
constructed baryon wavefunctions, we consider the $\c_J$ decay into
octet and decuplet baryon-antibaryon pair. By doing so, we subject the
wavefunctions to a test of applicability. We show that the colour singlet 
component alone is insufficient to account for experimental measurements
and only by including the colour octet contribution can the partial 
theoretical decay widths be brought into range of the data.
By the present and earlier applications of the set of wavefunctions, they
show themselves to be reasonable model wavefunctions at around the 
scale $Q^2 \sim$ 10-20 GeV$^2$. 

\end{abstract}

\vfill
\eject

\section{Introduction}
\label{sec:intro}

In studying strong interactions exclusive processes, a more 
reliable method is one that bases on perturbative QCD, which 
calls to mind the likes of deep inelastic scattering.
However unlike in these inclusive reactions where the
final states of hadrons can be summed over so no explicit
knowledge of any of the hadrons is required, in exclusive processes
one is usually interested specifically only in a few final hadrons
at a time, as such the knowledge of the wavefunctions of these
hadrons is essential. 
The more accurate picture of a hadron is the Fock state expansion
where it is seen as a sum of states with increasing number of
constituents starting from the state with only the valence
quarks and/or antiquarks. The probability of the hadron being
in any of these states is given by the modulus squared of the
wavefunction associated with each of these states.
\bea |B \ran &=& \J_{\srm{valence}} |q_1^v q_2^v q_3^v \ran 
             +\J_g |q_1^v q_2^v q_3^v \; g \ran
             +\J_q |q_1^v q_2^v q_3^v \; q \bar q \ran
             + \dots                               \nonumber \\
         & & + \J_{g_1 \dots g_M q_1 \dots q_N} 
               |q_1^v q_2^v q_3^v \; g_1 \dots g_M \; 
                q_1\bar q_1 \dots q_N \bar q_N \ran
             + \dots                               \nonumber \\
     |M \ran &=& \J_{\srm{valence}} |q_1^v \bar q_2^v \ran
             +\J_g |q_1^v \bar q_2^v \; g \ran
             +\J_q |q_1^v \bar q_2^v \; q \bar q \ran
             + \dots                               \nonumber \\ 
         & & + \J_{g_1 \dots g_M q_1 \dots q_N} 
               |q_1^v \bar q_2^v \; g_1 \dots g_M \;
                q_1\bar q_1 \dots q_N \bar q_N \ran
             + \dots                               \nonumber 
\eea
Because the number of constituents is unlimited, one would 
require, in general, far too much information before an exclusive
process involving even just a few hadrons can be studied.
Fortunately as shown by Brodsky and Lepage in ref. \cite{bl} in the 
so-called standard hard scattering approach (SHSA), when a high 
momentum transfer or high $Q^2$ is involved in an exclusive process,
not only is there factorization and so the process can be 
partially calculated perturbatively, but also only the lowest 
valence Fock state wavefunctions of the hadrons are needed. The higher 
Fock states are suppressed by the large $Q^2$. Thus the large momentum 
transfer acts as a filter to let past only the lowest state of the 
hadrons. It has also been proved by Duncan and Mueller \cite{dm}
that these are also true in quarkonium decay with large timelike momentum 
transfer. 

Most recent developments in inclusive quarkonium physics have seen, 
on the other hand, the need of the next higher Fock state in P-wave 
quarkonia \cite{bod&bra&lep1,bod&bra&lep2}, 
the so-called colour octet, where the heavy $Q\bar Q$ pair is in 
a colour octet rather than the usual colour singlet as in the 
valence state. This higher state is made up of the $Q\bar Q$
plus a gluon. This is special to heavy quarkonium and
can be understood in terms of a suppression on 
the level of the wavefunction due to angular momentum as we will
explain in later sections. The advance in inclusive process
involving quarkonia has yet to fully bring about the same level
of understanding in the corresponding exclusive process.
Part of our goal in this paper is therefore to show that
colour octet is not only important in inclusive but also in
exclusive process. We will achieve this by studying the
P-wave $\c_J$ decay into baryon-antibaryon. 

As mentioned above, knowledge of hadronic wavefunctions is
very important. Even in large momentum transfer processes,
valence wavefunctions are still necessary. Since bound state wave 
equations in QCD are too hard to solve, other ways to obtain the 
solutions or the hadronic wavefunctions must be used. One way is 
via QCD moment sum rules \cite{cz1,cz,coz}. However, it has 
been shown in \cite{bjkbs} that perturbative calculation of the 
proton magnetic form factor using a range of existing nucleon distribution 
amplitudes so-obtained gives results that are, on the average, at least 
a factor of two below experimental measurements in the range of $Q^2$ up 
to 50 GeV$^2$. It is therefore clear that perturbative contribution is not 
dominant in this rather low range of $Q^2$. The same conclusion has also been 
pointed out in \cite{rad}. Nucleon wavefunctions constructed from such 
distribution amplitudes are, therefore not applicable in reactions with far 
from asymptotic values of $Q^2$, not to mention the fact that there might be 
ambiguities in the distribution amplitudes determined this way. 

In order to obtain a nucleon wavefunction that is 
suitable for application at as low as 10 GeV$^2$.  
The usually neglected but nevertheless ever present
soft overlap between the final and initial nucleon 
wavefunction contribution to the magnetic form
factor is taken to make up for the difference between
the perturbative contribution and the experimental
measurements. It was shown in \cite{bolz&kroll1} that 
a nucleon model wavefunction could indeed be constructed 
this way, using constraints from valence quark distribution, 
J/$\psi$ decay width into nucleon-antinucleon pair 
and existing data on nucleon electromagnetic form factor.  
Our motivation in this regard is to show 
that the above construction of a model nucleon wavefunction
and it generalization to other octet and decuplet baryons 
\cite{bolz&kroll2} provide a reasonable description of 
the baryonic valence Fock state wavefunctions at around
10 GeV$^2$. To do that we must apply these wavefunctions to exclusive 
processes. Our investigation will allow us to study this as well
as the colour octet contribution.

Since in $\c_J$ decay into baryon-antibaryon, only
nucleon-antinucleon in the final state has been measured,
our calculations will show first that colour singlet contribution
alone is not sufficient and by including that of the octet, the 
partial decay width can be brought into reasonable agreement
with experiment. Then with the generalization of the nucleon wavefunction
to the whole flavour octet and decuplet multiplet, we can provide 
predictions for the widths of $\c_J$ decay into other baryon pairs.

The paper is organized as follows. First we review briefly in
Sec. \ref{sec:L_supp} the angular momentum suppression in the
P-wave wavefunction and in Sec. \ref{sec:s_o} give theoretical argument 
for the inclusion of the colour octet component in exclusive decay process.
In Sec. \ref{sec:bw}, the set of phenomenologically constructed baryon
wavefunctions will be presented and explained before we use the 
improved modified hard scattering scheme (MHSA) of Botts, Li and Sterman
\cite{bs,ls} to obtain the singlet contribution in Sec. \ref{sec:decay}
and \ref{sec:cs}. The results will be compared to experiment.
In Sec. \ref{sec:co}, our method in obtaining a wavefunction of the
higher colour octet state of the $\c_J$  and the Feynman graphs of the
decay process will be given. The details in constructing the octet contribution
to the hard perturbative part $T_H$ will be explained in Sec. \ref{sec:gg}.
The actual calculation and the results are in Sec. \ref{sec:s&o_SHSA}
and \ref{sec:widths}, respectively. In the appendices, all essential details 
of the calculation, the propagators and the numerators of the graphs used 
in Sec. \ref{sec:s&o_SHSA} are given.

\section{Angular Momentum Suppression Of Charmonium Wavefunctions}
\label{sec:L_supp}

In this section, we show that there is a suppression on the
charmonium wavefunction due simply to the angular momentum 
of the heavy quark-antiquark system. This has important consequences
as we will see in the next section, where we briefly review the more
detailed argument given in \cite{wong3}. Hadronic decay of charmonium is 
through the annihilation of the charm with the anticharm, which is a short
distance process because of the heavy charmonium mass. The
annihilation size $L$ is roughly given by $L \sim 1/M$, the
inverse of the heavy mass $M$. The charmonium wavefunction, which
must enter the decay probability amplitude, is therefore needed
predominantly at $L \sim 1/M \sim 0$ for a heavy charmonium decay.
That is $\psi_S (L) \sim \psi_S (0)$, the wavefunction at the spatial 
origin. This is the case for S-wave charmonium such as $\eta_c$ or 
$J/\psi$. For a P-wave charmonium, the wavefunction in fact vanishes
at the origin, so instead one has to expand the wavefunction around the
origin. What enters the probability amplitude is actually
$\psi_P (L) \sim L\,\psi_P'(L) \sim L\,\psi_P'(0)$.
Going to momentum space, the probability amplitudes for 
a S- and P-wave decay include the wavefunction of the form
\bea \mbox{S-wave: \hskip 1.0cm} \J_S (L) \mbox{\hskip 1.850cm}
     \fx & \longrightarrow & \fx \tilde  \J_S (k)                  \nonumber \\
     \mbox{P-wave: \hskip 1.0cm} \J_P (L) \sim L \J'_P(0)
     \fx & \longrightarrow & \fx \frac{k}{M} \tilde \J_P (k)  \; , \nonumber
\eea
respectively, where $k$ is the internal momentum of the charmonium, 
the average value $\lan k \ran$ is of the order of a few hundreds MeV. So
it becomes clear that P-wave decay is suppressed by $1/M$ in comparison
to S-wave at the level of the wavefunction. This suppression is
one of the reasons why colour octet is necessary for a P-wave 
inclusive decay. We will see in the next section that it also provides
the reason for its inclusion in exclusive decays.

\section{Comparison Of The Large Scale Dependence Of Colour Singlet 
and colour octet}
\label{sec:s_o}

The simplest scheme to calculate charmonium partial decay width that is
built on the solid basis of perturbative QCD is the hard scattering
approach of Brodsky and Lepage \cite{bl}. In this scheme, decay 
probability amplitude can be factorized into hard and soft part and 
is given by a convolution of hadronic distribution amplitudes $\f$'s and 
the hard perturbative part $T_H$. For the decay of $\c_J$ into 
baryon-antibaryon, or more specifically nucleon-antinucleon, this is
\be \cm \sim f_{\c_J} \f_{\c_J}(x) \otimes f_N \f_N(x) 
    \otimes f_{\bar N} \f_{\bar N}(x) \otimes T_H(x)  \; .
\label{eq:shs}
\ee
Since the distribution amplitudes are nothing but the hadronic 
wavefunctions with their internal transverse momenta being
integrated over, there is still the decay constant $f$ accompanying 
each distribution amplitude. While the amplitude themselves are 
dimensionless, the decay constants $f$'s carry different mass dimensions 
depending on the original hadronic wavefunctions. From the fact that the 
partial decay width is given by
\be  \Gamma_{partial} \sim |\cm|^2 /M
\label{eq:width}
\ee
and there is only one mass scale $M$ in the process, namely the 
heavy charmonium mass, we can use power counting on \eref{eq:shs} to 
compare, once the mass dimensions of the decay constants are known, 
the colour singlet and octet contribution to the decay width and 
hence their relative importance to a particular hadronic decay 
process. For the determination of the mass dimension of $f$, we refer
to \cite{wong3}.
 
Examining \eref{eq:shs}, the decay constant of the valence Fock
state of nucleon $f_N$ and the colour octet charmonium decay constant
$f^{(8)}_{\c_J}$ are both 3-particle wavefunction. Therefore 
they must be of mass dimension two. The colour singlet decay 
constant $f^{(1)}_{\c_J}$, on the other hand, is a 2-particle 
wavefunction so it should be of dimension one. However, the fact
that $\c_J$ are P-wave charmonia increases this to dimension two.
The only remaining quantity in \eref{eq:shs} that has a dimension 
is $T_H$, which contains a hidden power of $M$. This power must make up 
the right dimension for $\cm$, which must be dimension one in view of 
\eref{eq:width}. So we can now collect all the dimensional quantities
in the colour singlet and octet probability amplitude and 
determine their dependence on the large scale $M$. We get
\bea \cm^{(1)} \fx & \sim & \fx 
     M \frac{f^{(1)}_{\c_J}}{M^2} \Big ( \frac{f_N}{M^2} \Big )^2  
     \sim \frac{1}{M^5}   \\
     \cm^{(8)} \fx & \sim & \fx 
     M \frac{f^{(8)}_{\c_J}}{M^2} \Big ( \frac{f_N}{M^2} \Big )^2  
     \sim \frac{1}{M^5}   \; .
\eea
It is now clear that both the singlet and octet contribution for the
P-wave $\c_J$ are weighed by the inverse fifth power of the charmonium
mass. The colour octet, although a higher Fock state of the charmonium, 
is not suppressed by the large scale of the decay process
in relation to the valence singlet state. 

If there were no suppression at the level of the P-wave wavefunction as 
explained in the previous section, for example in the case of 
$J/\psi$ decay, the dependence of the singlet contribution
on $M$ would have been $1/M^4$. Therefore the colour octet contribution
can be neglected in the decay of a S-wave but not that of a P-wave charmonium . 
For more details of the argument above, one can consult \cite{wong3}.

\section{Baryon Wavefunctions}
\label{sec:bw}

\subsection{Octet Baryons}
\label{sec:obw}

As mentioned in Sec. \ref{sec:intro}, most model wavefunctions
constructed from QCD moment sum rules \cite{cz1,cz,coz,gs,ks,bst}
lead to rather unsatisfactory distribution amplitudes at moderate $Q^2$ 
\cite{bjkbs,rad} and they certainly do not describe experimental data.
A different model wavefunction suitable for application at such low
momentum transfers was constructed in \cite{bolz&kroll1}.
This construction is based on the following form of
the nucleon wavefunction,
\bea
%
%
  \label{pstate}
  |\,p,+ \,\ran\, &
=
  \frac{\varepsilon _{a_{1}a_{2}a_{3}}}{\sqrt{3!}} 
  \int
  [{\rm d}x]
  [{\rm d}^{2}{\Kp}] &
  \Bigl\{ 
         \j^N_{123}\:|\,u_+^{a_1} u_-^{a_2} d_+^{a_3}\,\ran
       + \j^N_{213}\:|\,u_-^{a_1} u_+^{a_2} d_+^{a_3}\,\ran
  \phantom{ \Bigr\} \;\;\; }
\nonumber \\
   & & \mbox{\hskip 2.0cm} 
       - \Bigl(\j^N _{132}\, + \,
         \j^N_{231}\Bigr)\:|\,u_+^{a_1} u_+^{a_2} d_-^{a_3}\,\ran
  \Bigr\} \;\;\; 
%
\eea
which is the most general of the nucleon wavefunction 
with zero orbital angular momentum \cite{Dziem}. The nucleon,
being an isospin doublet, is also the lowest energy state 
of the baryons and is therefore reasonable for one to assume 
that it has zero orbital angular momentum. These conditions
permit the presence of only one scalar function $\j$, 
which is a function of the light-cone momentum fractions 
$x_i$ and the internal transverse momentum $\Kpn{i}$ of the 
nucleon. In the notations of \cite{bolz&kroll1}, it is given 
in terms of the nucleon decay constant, or equivalently
the wavefunction at the origin $f_N$, the distribution 
amplitude $\f_{123}$ and the function containing the
transverse momentum dependence $\Omega_N$, by
\be 
  \j_{123}(x,\Kp) =  \j (x_1,x_2,x_3;\Kpn 1,\Kpn 2,\Kpn 3) =
  \frac{1}{8\sqrt{3!}} \,  f_N (\mu_F)
  \phi^N_{123} (x,\mu_{F})\,  \Omega_N (x,\Kp) \:. 
\label{Psiansatz}
\ee
and 
\be
  \dxb = \prod_{i=1}^3 {\rm d}x_i\,\delta(1-\sum^3_{i=1} x_i) \qquad
  \dkb = \frac{1}{(16\pi^3)^{2}}\,\prod_{i=1}^3 
     {\rm d}^2 \Kpn{i} \, \delta^{(2)}(\sum^3_{i=1} \Kpn{i}) 
     \:.
\ee
are the usual constrained integration measures over momentum
fractions and the internal transverse momenta. 
The function $\Omega_N$ is conveniently taken to be 
of Gaussion form 
\be 
  \Omega_N(x,\Kp) =
  (16\pi ^{2})^{2}  \frac{a_N^{4}}{x_{1}x_{2}x_{3}}
  \exp
     \left [
            -a_N^{2} \sum_{i=1}^{3} {\Kpn{i}}^{2}/x_{i}
     \right ]\:.
\label{BLHMOmega}
\ee
where $a_N$ is a transverse size parameter which was
fitted together with the decay constant $f_N$ 
to experimental data by the procedure described in 
\cite{bolz&kroll1} to be $a_N = 0.75$ GeV$^{-1}$ and 
$f_N (\m_0) = 6.64 \times 10^{-3}$ GeV$^2$ at the reference
scale $\m_0 = 1.0$ GeV. 

The distribution amplitude $\f_{123}$ as well as
the decay constant, all have to be evolved to the 
factorization scale $\m_F$ of the process in question.
In terms of the eigenstates of the evolution equation,
the effect of the evolution is to change the coefficients
of each eigenstate by a certain power of the log of the
relevant scale. The nucleon distribution amplitude
expressed in the Appell polynomial eigenbasis is
\bea
  \f^N_{123}(x,\mu_F) &=& \f_{\rm AS}(x) \left[1 +
  \sum_{n=1}^{\infty} B^N_n(\mu_F)\,\tilde \f^n_{123}(x,\mu_F) \right]
  \\
  &=& \f_{\rm{AS}}(x) 
      \left [1 + \frac{3}{4} \tilde \f^1_{123}(x)
               + \frac{1}{4} \tilde \f^2_{123}(x) \right ]
      \;\;\; = \; 60 x_1 x_2 x_3 [1+3 x_1]  \; .
\label{DAentw}
\eea
Under a change of scale both the coefficients $B^N_n$ of the
expansion and the decay constant are scaled by the following \
factors of logarithm of the relevant scale $\m_F$
\be
  f_N(\mu_F)  =  f_N(\mu_0)\,\left(
    \frac{\ln(\mu_0/\lqcd)}{\ln(\mu_F/\lqcd)} \right)^{2/3\beta_0}
     \hspace{-1mm}, 
  B^N_n(\mu_F)  =  B^N_n(\mu_0)\,\left(
    \frac{\ln(\mu_0/\lqcd)}{\ln(\mu_F/\lqcd)} \right)^{\tilde
    \gamma_n/\beta_0}  
\hspace{-8mm} 
\label{BnFOevol}
\ee
where $\tilde \g_n$ are the reduced anomalous dimensions and $\b_0$ 
is the first coefficient of the $\b$ function. The last line in 
\eref{DAentw} is the expression for $\f^N_{123}$ at the scale 
$\m_F=\m_0$.

As shown in \cite{bolz&kroll2}, the form of the 
nucleon wave function need not be restricted to the 
SU(2) isospin doublet. On the contrary, one can extend
it to the complete SU(3) flavour octet. The simplest 
way to do this, with SU(3) flavour symmetry breaking
by the heavier strange quark mass taken into account,
is to assume the complete baryon octet shares the same
octet transverse size parameter, $a_{B_8} = 0.75$ GeV 
and the octet decay constant, 
$f_{B_8} = 6.64 \times 10^{-3}$ GeV$^2$. Then the
flavour symmetry breaking effects are all put into the
octet distribution amplitudes $\f_{123}^{B_{8}}$ and they
manifest themselves as uneven distribution of the 
light-cone momentum fractions amongst the valence quarks. 
As shown in \cite{bolz&kroll2}, introducing an additional 
exponential dependence on the strange constituent quark 
mass $m_s$, of the form
\be \exp \left (-{{a_{B_8}^2 m_s^2} \over x_j} \right )
\ee
in the distribution amplitude for each strange quark with 
label $j$ suffices for the purpose. Using several different
values of $m_s$, representative sets of the expansion 
coefficients $B^{B_8}$, of the octet distribution amplitudes,
$\f^{B_8}_{123}$ can be obtained. The set 3 in \cite{bolz&kroll2},
obtained with $m_s = 350$ MeV, shows the most promise and 
will be used in the following investigations.  

The octet baryon wavefunctions with positive helicities
can then be expressed as follows. 
\bea
  |\,B_8\;,+ \,\ran\, & =
  \frac{\varepsilon _{a_{1}a_{2}a_{3}}}{\sqrt{3!}} 
  \int \dxb \dkb &
  \Bigl\{ 
         \j^{B_8} _{123}\:|\,f_{1+}^{a_1} f_{1-}^{a_2} f_{2+}^{a_3}\,\ran
       + \j^{B_8} _{213}\:|\,f_{1-}^{a_1} f_{1+}^{a_2} f_{2+}^{a_3}\,\ran
  \phantom{ \Bigr\} \;\;\; }
\nonumber \\
       & & {\hskip 2.0cm}
       - \Bigl(\j^{B_8} _{132}\, + \,
         \j^{B_8} _{231}\Bigr)\:|\,f_{1+}^{a_1} f_{1+}^{a_2} f_{2-}^{a_3}\,
         \ran
  \Bigr\} \;\;\;  ,
\label{octstate} \\
  |\,\;\L\;,+ \,\ran\, & =
  \frac{\varepsilon _{a_{1}a_{2}a_{3}}}{\sqrt{2}} 
  \int \dxb \dkb  & 
  \Bigl\{ 
         \j^{\L} _{123}\:|\,u_+^{a_1}\; d_-^{a_2}\; s_+^{a_3}\,\ran
       - \j^{\L} _{213}\:|\,u_-^{a_1}\; d_+^{a_2}\; s_+^{a_3}\,\ran
  \phantom{ \Bigr\} \;\;\; }
\nonumber \\
       & &  {\hskip 2.0cm}
       + \Bigl(\j^{\L} _{132}\, - \,
         \j^{\L} _{231}\Bigr)\:|\,u_+^{a_1}\; d_+^{a_2}\; s_-^{a_3}\,
         \ran
  \Bigr\} \;\;\;  ,
\label{Lamstate}
\eea
where $f^{a_i}_j$ stands for the quark flavour of quark $j$ with
colour $a_i$ and $\j^{B_8}_{123}$ is the corresponding 
scalar functions in \eref{Psiansatz} of the octet baryons.
The $\L$ is a slightly different member of the flavour octet multiplet.
Being an isospin singlet, it has to vanish under the action of
SU(2)$_{\rm isospin}$ and therefore has different signs between
the different wavefunction components.

\subsection{Decuplet Baryons}
\label{sec:dbw}

As we mentioned in the introduction, our interest in this
investigation is the $\c_J$ decay into baryon-antibaryon
pair. Unlike the octet baryon-antibaryon which can
coupled only to spin S=1, a decuplet baryon-antibaryon
pair can coupled to S=2 as well. This leads to the
interesting potential possibility of $\c_2$ with $S_z=2$
to decay into a coupled decuplet baryon pair with the same
total third component of the spin. However within 
perturbative QCD, this is not possible due to 
quark and gluon couple via a vector coupling. This results
in the well known helicity conservation or helicity
sum rule, which forces the outgoing baryon-antibaryon to
have zero total helicity \cite{bl2}. Or in other words, they must 
be in a total spin one state. For the same reason,
$\c_0$ decay into baryon-antibaryon is forbidden.
Therefore for the decuplet baryons, all we need are the
helicity $+1$ decuplet baryon wavefunctions. 

Starting from the $\D^{++}$, the simplest distribution
amplitude which is symmetric between the three u-quarks
is the asymptotic distribution amplitude $\f_{AS}(x)$.
Using this as the starting point, one can likewise
generalized to the whole decuplet baryon multiplet 
and introduce SU(3) flavour symmetry breaking in the 
same manner as in Sec. \ref{sec:obw} by using an 
exponential $m_s$ dependence. This again
yields several sets of representative expansion
coefficients $B^{B_{10}}_n$, for the decuplet
baryon distribution amplitudes $\f^{B_{10}}_{123}(x)$. 
They are listed in Table \ref{tab:decuplet_B}. The decuplet 
wavefunctions can be expressed in a similar fashion as before 
\bea
  |\Delta^{++},+ \,\ran\, & =
  \frac{\varepsilon _{a_{1}a_{2}a_{3}}}{\sqrt{2}} 
  \int \dxb \dkb  & \,
    \j^{\Delta} _{123}\:|\,u_+^{a_1}\; u_-^{a_2}\; u_+^{a_3}\,\ran 
   \phantom{  
   + \j^{\Delta} _{213}\:|\,u_-^{a_1}\; u_+^{a_2}\; d_+^{a_3}\,\ran
            \Bigr\} \;\;\;}
  \label{Del++state} \\
%
%
  |\;B_{10}\;,+ \,\ran\, & =
  \frac{\varepsilon _{a_{1}a_{2}a_{3}}}{\sqrt{3!}} 
  \int  \dxb \dkb &
  \Bigl\{ 
         \j^{B_{10}} _{123}\:|\,f_{1+}^{a_1} f_{1-}^{a_2} f_{2+}^{a_3}\,
         \ran
       + \j^{B_{10}} _{213}\:|\,f_{1-}^{a_1} f_{1+}^{a_2} f_{2+}^{a_3}\,
         \ran
  \phantom{ \Bigr\} \;\;\; }
\nonumber \\
       & & \mbox{\hskip 3.40cm} +
         \j^{B_{10}} _{132}\:|\,f_{1+}^{a_1} f_{1+}^{a_2} f_{2-}^{a_3}\,
         \ran
  \Bigr\} \; .\;\;  
\label{Del+state}
\eea
and now the scalar functions are 
\be
  \j^{B_{10}} _{123}(x,\Kp) =  \frac{f_{B_{10}}(\mu_F)}{24\sqrt{2}}\,
  \f^{B_{10}}_{123}(x,\mu _{F})\,  \Omega_{B_{10}} (x,\Kp)\:. 
\label{PsiDansatz}
\ee
As shown in \cite{bolz&kroll2}, using the assumption that
the nucleon and the delta have the same valence Fock state
probabilities, the decuplet decay constant 
and the transverse size parameter can take a range of values.
We take as representative values 
$f_{B_{10}}(\m_0) = 0.0143$ GeV$^2$ and $a_{B_{10}} = 0.80$ GeV$^{-1}$.

\begin{table}
\begin{center}
\begin{tabular}{|c||c|c|c|c|c|}
\hline
           &  $B_1$   &  $B_2$   & $B_3$    &  $B_4$   &  $B_5$    \\
\hline\hline
$\Delta$   &  0.000 &  0.000 &  0.000 &  0.000 &  0.000  \\ \hline
$\Sigma^*$ & -0.547 &  0.182 & -0.216 & -1.081 &  0.062  \\ \hline
$\Xi^*$    &  0.540 & -0.180 & -0.382 &  1.742 & -0.413  \\ \hline
\end{tabular}
\end{center}
\caption{The expansion coefficients of the distribution amplitudes
$\f^{B_{10}}_{123}$ of the octet baryons considered in the $\c_J$ decay.
The parameters associated with this set of coefficents are 
$f_{B_{10}}(\m_0) = 0.0143$ GeV$^2$ and $a_{B_{10}} = 0.80$ GeV$^{-1}$.}
\label{tab:decuplet_B}
\end{table}

\section{$\c_J$ Decay In The Modified Hard Scattering Approach}
\label{sec:decay}

Since the $\c_0$ decay into baryon-antibaryon is forbidden 
by angular momentum conservation and the not-yet-confirmed 
$^1 P_1$ state $h_c$ cannot preserve C-parity and parity 
simultaneously in this decay mode \cite{murg&melis1} 
in a perturbative approach, only $\c_1$ and $\c_2$ may have finite 
partial decay widths \footnote{In practice, $\c_0$ has
a surprisingly large upper bound on the partial width of 
the decay channel in question \cite{pdg} and the $h_c$
may also have non-zero partial decay width. This should
be attributed to non-perturbative soft physics or 
higher twist effects since mass corrections 
alone should not yield such large width
assuming that the experimental width is near the upper
limit.} into baryon-antibaryon pair.
The helicity amplitudes in covariant form in terms of the
baryon-antibaryon spinors $u_B(p,\l)$ and $v_B(p,\l)$ 
suitable for our consideration are, for $\c_1$
\be 
   \cm^{1}_{\l_1 \l_2 \l} = 
   \bar u_B(p_1,\l_1) \, \cb_1 \, \g^\n \, v_B(p_2,\l_2) 
   \, \e_\n(\l) \; ,
\label{eq:hamp1}
\ee
and for $\c_2$
\be
   \cm^{2}_{\l_1 \l_2 \l} = 
   \bar u_B(p_1,\l_1) \, \cb_2 \, \g^\n \, v_B(p_2,\l_2) 
   \, \e_{\m \n}(\l) \, {{(p_2^\m-p_1^\m)} \over {M_{\c_2}} } \; ,
\label{eq:hamp2}
\ee
where $\e_\m$ and $\e_{\m\n}$ are the polarization vector
and tensor of $\c_1$ and $\c_2$, respectively and 
$\cb_J$ are their corresponding decay form factors.
Note that \eref{eq:hamp1} and \eref{eq:hamp2} are the only covariant
form permitted for the corresponding helicity amplitudes. Using
only $\g^\m$, $(p_2-p_1)^\m$ and $g^{\m\n}$ to form a vector and a
symmetric tensor, they are the only form that can be constructed
which still respect helicity conservation.
The decay widths into baryon-antibaryon are therefore
\be
   \Gamma (\c_1\rightarrow B\bar B) 
   = { {\r_{\rm p.s.}(M_B/M_{\c_1})} \over {16 \p M_{\c_1}} } \; 
     \frac{1}{3} \; \sum_{\l's} \left |\cm^1_{\l_1 \l_2 \l} \right |^2
   = { {\r_{\rm p.s.}(M_B/M_{\c_1}) \, m_c^2} \over {3 \p M_{\c_1}} } \; 
     \; \left | \cb^B_1 \right |^2  \; ,
\label{eq:gam_1}
\ee
and
\be
   \Gamma (\c_2\rightarrow B \bar B) 
   = { {\r_{\rm p.s.}(M_B/M_{\c_2})} \over {16 \p M_{\c_2}} } \;
     \frac{1}{5} \; \sum_{\l's} \left |\cm^2_{\l_1 \l_2 \l} \right |^2 
   = { {\r_{\rm p.s.}(M_B/M_{\c_2}) \, m_c^2} \over {10 \p M_{\c_2}} } \; 
     \; \left | \cb^B_2 \right |^2  \; .
\label{eq:gam_2}
\ee
Since in the standard or modified hard scattering scheme, baryons
are treated as massless in comparison to the large scale $M$ of
the process, phase space must be corrected. This is taken care of by 
the phase space factor in the above equations given by 
$\r_{\rm p.s.}(z)= \sqrt{1-4 z^2}$.

Within the SHSA, one has factorization 
by which the soft infrared physics is contained in the light-cone 
wavefunctions and a perturbatively calculable hard scattering amplitude 
$T_H$. In the present problem, the hard scale is set by twice the
charm quark mass $2m_c$, rather than by the charmonium mass
$M_{\c_J}$, because as we mentioned before, the $c\bar c$ 
pair annihilates at a much smaller size than that of the charmonium.
This explains the appearance of $m_c^2$ in \eref{eq:gam_1} and
\eref{eq:gam_2}. The decay amplitudes or equivalently,  
the decay form factors $\cb^J$ are expressed as a convolution 
of the hadron wavefunctions and the hard scattering amplitude 
$T_H$. Based on this approach, there are already a number of
work on charmonium decay into nucleon-antinucleon 
\cite{andrik,dam&tso&berg,coz2,murg&melis2}. However, we 
consider these as incomplete for the following reasons.
First, they used a number of nucleon wavefunctions, 
which did not describe the correct physics at the scale of 
the order of $M_{\c_J}$. As shown in \cite{bjkbs}, none of these 
wavefunctions are able to describe data of the nucleon magnetic 
form factor. Second, their treatments of $\a_s$ are 
ambiguous given that the decay widths depend on $\a_s^6$,
any small changes in the value of $\a_s$ used will change the
width considerably. It is therefore not difficult to
obtain a width that match the experimental decay widths.
All one has to do is to choose the right value of $\a_s$. 
This is very arbitrary in our opinion. A better way is to
determine the scale at which $\a_s$ should be evaluated
by using the virtualities of the internal exchanged gluons.
However, these virtualities in the SHSA depend on the light-cone 
momentum fractions. One will encounter problems as the end 
point regions is approached when some of these gluon virtualities 
drop down to $\lqcd^2$. $\a_s$ will become large and the 
perturbative part of the SHSA breaks down. Other treatment of
this problem such as arbitrarily freezing $\a_s$ at some values as 
the virtualities become small or using an equally arbitrary gluon mass 
is not well justified. Third, as we have already discussed 
due to the development first shown in \cite{bod&bra&lep1,bod&bra&lep2}
that the contributions from the next higher Fock state of the 
P-wave charmonium, where the $c\bar c$ pair is in a colour octet, 
are comparable to that of the lowest colour singlet contribution 
because of suppression by angular momentum. To the best of our knowledge,
colour octet contributions have not been taken into account
in most exclusive reaction involving P- and higher wave charmonium.
In the case of the decay into light pseudoscalar mesons, this
has been worked out recently within the SHSA in \cite{bks} and 
within the MHSA in \cite{bks2}.

To deal with the above deficiencies of the previous calculations, 
the phenomenologically constructed nucleon wavefunction and 
its generalization to the whole of the octet and decuplet 
baryon multiplets \cite{bolz&kroll1,bolz&kroll2} should 
provide better baryon wavefunctions at around 10 GeV$^2$. 
The problem with the coupling can be dealt with successfully 
with the MHSA although it complicates the calculation with
the additional, but necessary transverse momentum dependence 
for a self-consistent description. The above mentioned
end point problem in which one runs into the infrared 
non-perturbative region within a perturbative scheme is
cured by the introduction of radiative corrections in the
form of Sudakov factor which spans the energy range between the 
lower factorization scale $\m_F$ and the higher hard 
scale of the process in question. The Sudakov factor
with transverse size dependence was first calculated in
\cite{bs,ls} and was shown to cure the problem in the
case of the pseudoscalar meson scattering.
In the case of baryons, this will provide a cure 
of the above problem only if one supplements the 
Sudakov suppression factor with the MAX prescription for 
deciding the infrared cutoff scale \cite{bjkbs}. 
This amounts to choosing the largest transverse separation
scale as the infrared cutoff, which is physical in the
sense that very long wavelength gluons cannot resolve
a ``small'' hadron which is colour singlet as far as the
gluon is concerned.

\section{Colour Singlet Contribution}
\label{sec:cs}

P-wave charmonium decays dominantly through annihilation
into gluons, which is a short distance process set by
the scale of the charmonium mass $M_{\c_J}$. The non-perturbative 
information of the bounded system must come from and be parametrized 
by the wavefunction of the $\c_J$ at small spatial separation, usually 
taken to be at the origin. For P-wave charmonium, the vanishing of the
wavefunction at the origin forces the substitution of the wavefunction
there by the first derivative of the radial wavefunction
\be  R'_P(0) = {{4 i \sqrt{\p m_c}} \over {3\sqrt{3}} }
     \int {{\dd^3 \K \, \K^2} \over {(2\p)^3 2 M_{\c_{J=1,2}}} } 
     \tilde \j^{(1)}_{J=1,2} (k) 
             = i \sqrt{\frac{16 \p m_c}{3}} \; f^{(1)}_{\c_{J=1,2}}  \; ,
\ee
where $|R'_P(0)| = 0.22$ GeV$^{5/2}$ and $\j^{(1)}_{J=1,2}(k)$ 
are functions of the internal relative momentum of the $c\bar c$ 
system and are the reduced wavefunctions of the $\c_J$, that means a 
power of $k$ has been extracted and put into the covariant spin part
of the $\c_J$ wavefunctions. This is the form of the wavefunction
commonly used for $\c_J$. The related but more general form
of the colour singlet wavefunctions of the $\c_J$ for $J=1,2$ are 
\bea
    |\c_1^{(1)}, p \ran &=& {\d_{ab} \over \sqrt{3}}
    \int {{\dd^3 \K} \over {(2\p)^3 2 M_{\c_1}}} 
    \tilde \j^{(1)}_1 (k) \, S^{(1)}_1(p,k) \; |c\bar c; k,p\ran \; ,  \\
    |\c_2^{(1)}, p \ran &=& {\d_{ab} \over \sqrt{3}}
    \int {{\dd^3 \K} \over {(2\p)^3 2 M_{\c_2}}} 
    \tilde \j^{(1)}_2 (k) \, S^{(1)}_2(p,k) \; |c\bar c; k,p\ran \; ,
\eea
and the covariant spin wave function of $\c_1$ and $\c_2$ 
expanded up to $O(k^2)$ are
\bea 
    S^{(1)}_1 (p,k) &=& {{-i} \over {2 M_{\c_1}}}
          \left [p \sla + M_{\c_1} 
            -  \frac{2}{M_{\c_1}} p \sla K \Sla \right ] 
             \; \e_{\m \n \a \b} \; p^\m \e^\n K^\a \g^\b \\
    S^{(1)}_2 (p,k) &=& \frac{1}{\sqrt{2}}
          \left [(p \sla + M_{\c_2} ) \g_\m 
            +  \frac{2}{M_{\c_2}} 
              [(p \sla + M_{\c_2} ) K_\m - p \sla K \Sla \g_\m]
          \right ] \e^{\m \n} K_\n  \; ,   
\eea
where $K\cdot p= 0$. The $\d_{ab}$ in the above equations is to 
ensure the $c\bar c$ is indeed in a colour singlet state.

Since $\c_J$ are even under charge conjugation, they can 
annihilate into two or three gluons at leading order
O($\a_s^3$) \cite{nov&etal}. The possible types of diagrams 
for $c\bar c$ annihilation into three light quark-antiquark 
pairs are shown in \fref{f:sing_fig}. As discussed in
\cite{nov&etal}, a colour singlet quark-antiquark system
has C-parity $(-1)^{L+S}$, so P-wave spin-1 charmonia are all
even C-parity states. Since strong interactions respect
C-parity conservation, the intermediate two or three gluons
must also be even under charge conjugation. Two gluons in
a colour singlet state is automatically in an even C-parity 
state so \fref{f:sing_fig}(a) and (b) are possible at the two 
gluon stage. However in colour space, an examination of the 
colour structure of the baryon wavefunctions \eref{octstate} 
and \eref{Del+state} show that exchanges between any two quark lines 
are symmetric but in \fref{f:sing_fig} (b) there is an antisymmetric 
three-gluon coupling so it is eliminated. For the decay via three 
gluons in \fref{f:sing_fig} (c), it is possible for
three gluons to be even under C-parity through
a $f^{abc}$ coupling, but again symmetry in colour space
between the three light quark lines forces the three gluons
to couple via $d^{abc}$, which violates C-parity conservation.
We are therefore left only with graphs of type (a).
\bfi
\centerline{
\psfig{figure=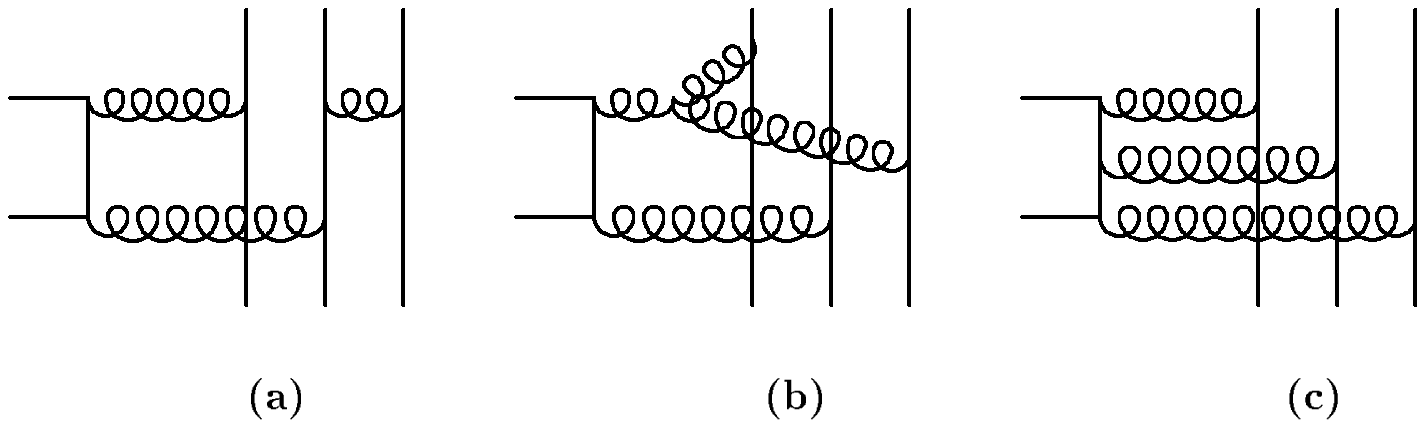,width=5.0in}
}
\caption{Basic graphs that could contribute to $\c_J$ 
colour singlet decay. But actually, only graphs of type 
(a) can contribute.}
\label{f:sing_fig}
\efi
These can further be divided into four groups. Each group
can be obtained from those shown in \fref{f:sing} by permutations 
of the three light quark lines. Since it is usual to treat the 
heavy $c\bar c$ as a non-relativistic system, the heavy quarks 
share the energy and momentum equally or in other words, 
the distribution amplitude of the charmonium is taken to be a 
delta function which peaks at one-half. With this assumption, 
the hard scattering amplitude $T_H$ can be worked out. 
For P-wave decays, one has to keep the relative momentum $K$ of the 
$c\bar c$ system and expand the hard part around $K=0$ since 
only terms quadratic in $K$ survive the $\K$ integration. 

Taking \fref{f:sing} (a) as an example, with $x_i, \; y_i$ the 
momentum fractions and $\Kpn{i}, \; \Kpn{i}'$ the internal 
transverse momenta of the valence quarks in the baryon and 
antibaryon respectively, the virtualities of the internal lines 
are
\be
\begin{array}{rlll}
      \cg_1 &= x_1 y_1 \; (4 m_c^2) -(\Kpn{1}+\Kpn{1}')^2 
            &= \tilde \cg_1 -(\Kpn{1}+\Kpn{1}')^2     \\
      \cg_2 &= (1-x_1) (1-y_1) \; (4 m_c^2) -(\Kpn{1}+\Kpn{1}')^2 
            &= \tilde \cg_2 - (\Kpn{1}+\Kpn{1}')^2    \\
      \cg_3 &= x_3 y_3 \; (4 m_c^2) -(\Kpn{3}+\Kpn{3}')^2    
            &= \tilde \cg_3 - (\Kpn{3}+\Kpn{3}')^2   \\
  \cq_{\;\,}&= (1-x_1) y_3 \; (4 m_c^2) - (\Kpn{1}-\Kpn{3}')^2 
            &= \tilde \cq -  (\Kpn{1}-\Kpn{3}')^2  \\
      \cq_c &= 2 [x_1 (1-y_1) + y_1 (1-x_1)] \; m_c^2 
                + (\Kpn{1}+\Kpn{1}')^2    
            &= \tilde \cq_c + (\Kpn{1}+\Kpn{1}')^2 
\end{array}
\label{eq:virt}
\ee
\bfi
\centerline{
\psfig{figure=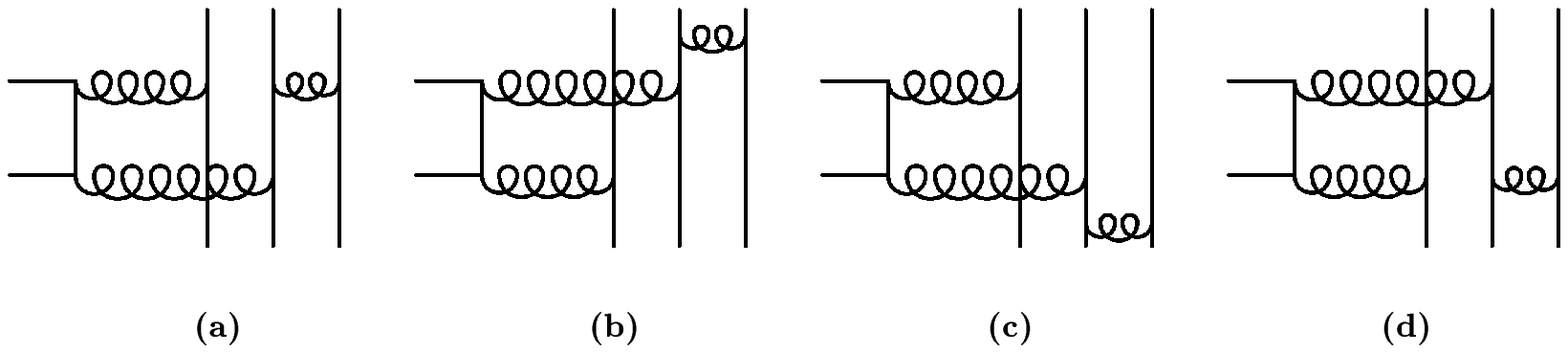,width=5.0in}
}
\caption{Graphs of type \fref{f:sing_fig} (a) can be
divided further into four groups.}
\label{f:sing}
\efi
In fact, \fref{f:sing} (a) and (b) give identical contributions
and the similar contributions from \fref{f:sing} (c) and (d) 
can be obtained by interchanging $x$ and $y$. One gets, after
sorting and integrating out $K$, for the hard scattering amplitude
\bea T^J_H (x,y,\Kp,\Kp') & = & 
    {{2^{10} \sqrt{2} \; (4\p)^3 \; m_c^5 \; \a_s(t_1) \a_s(t_2) \a_s(t_3)} 
    \over {9\sqrt{3} \; (\cg_1+i\e)\,(\cg_2+i\e)\,(\cg_3+i\e)\,(\cq+i\e)
                     \, (\cq_c-i\e) }} 
    \nonumber \\ 
    & \times &  y_3
    \left  [  (1-2 x_1)^{2-J} 
            + {{(-1)^{J-1} \; 2 m_c^2 \; x_1 (x_1-y_1)} \over {(\cq_c-i\e)}}
    \right ] \nonumber \\
    & + & (x \llra y )  \; .
\label{eq:s_hard}
\eea
To improve perturbation theory via renormalization group, the square 
of the renormalization scale $\m_R$ for each running coupling is set to 
one of the virtualites $t_i, i=1,2,3$, which are chosen by \cite{bs,ls}
\be
    t_1 = \mbox{max}(\tilde \cg_1,\tilde \cq_c,1/b_1^2,1/{b'}_1^2) \;\;\;\;\;
    t_2 = \mbox{max}(\tilde \cg_2,\tilde \cq,  1/b_2^2,1/{b'}_2^2) \;\;\;\;\;
    t_3 = \mbox{max}(\tilde \cg_3,1/b_3^2,1/{b'}_3^2)
\ee 
with $n_f=4$ and $\lqcd =220$ MeV.

These must be convoluted with the wavefunctions to obtain the
colour singlet decay form factors. In transverse separation space
and in terms of the expressions,
\bea 
    \cc^J_1 (x,y) &=& y_3 \; (1-2 x_1)^{2-J}    \\
    \cc^J_2 (x,y) &=& (-1)^{J-1} \; y_3 \; x_1 (x_1-y_1) 
\eea
the hard scattering part $\hat T^J_H$ is 
\bea  \hat T^J_H (x,y,\B,\B') \fx &=& \fx
      {{2^{9} \sqrt{2} \; m_c} \over {9\sqrt{3}} } \; \;
      \a_s(t_1) \a_s(t_2) \a_s(t_3) 
      \; \; \d^2 (\B_{1}-\B_{1}'-\B_{3}+\B_{3}') 
      \nonumber \\ \fx & & \fx \times
      \frac{i\p}{2} H^{(1)}_0 (\sqrt{x_3 y_3} (2 m_c) |\B_3|) \; \;      
      \frac{i\p}{2} H^{(1)}_0 (\sqrt{(1-x_1) y_3} (2 m_c) |\B_1-\B_1'|)
      \nonumber \\ \fx & & \fx \times
      \Bigg \{ 
      i\p \bigg [ {{H^{(1)}_0 (\sqrt{x_1 y_1} (2 m_c) |\B_1'|)}
                  \over {(x_1+y+1)(1-x_1-y_1)}}
               \Big ( \cc^J_1(x,y) 
                     + {{\cc^J_2(x,y)} \over {x_1+y_1}}
               \Big )       \nonumber \\ \fx & & \; \; \; \; \;
                -{{H^{(1)}_0 (\sqrt{(1-x_1)(1-y_1)} (2 m_c) |\B_1'|)}
                  \over {(1-x_1-y_1)(2-x_1-y_1)}}
               \Big ( \cc^J_1(x,y) 
                     + {{\cc^J_2(x,y)} \over {2-x_1-y_1}}
               \Big )       
          \bigg ]           \nonumber \\ \fx & & 
              -\frac{4}{\p} 
                 {{K^{(1)}_0 (\sqrt{x_1 (1-y_1) + y_1 (1-x_1)} m_c |\B_1'|)}
                  \over {(x_1+y_1)(2-x_1-y_1)}}  \nonumber \\ 
              \fx & & \;\;\;\; 
               \Big ( \cc^J_1 (x,y) 
                     + {{2 \; \cc^J_2 (x,y)} 
                        \over {(x_1+y_1)(2-x_1-y_1)}}
               \Big )       \nonumber \\ \fx & & \;
      -{{2 m_c |\B_1'| \; \cc^J_2(x,y) \; 
         K^{(1)}_1(\sqrt{x_1 (1-y_1) + y_1 (1-x_1)} m_c |\B_1'|)}  
        \over {\sqrt{x_1 (1-y_1) + y_1 (1-x_1)} \; (2-x_1-y_1) (x_1+y_1)}}
      \Bigg \}    \nonumber \\ \fx & & 
       + \Big ( (x, \; \B) \llra (y, \; \B') \Big ) \; \; \; .
\eea
With this, the decay form factor can be expressed as 
\bea \cb_J^{B \; (1)} 
    &=& -i {{\sqrt{3} |R_p'(0)| \; \s_J} \over {8\sqrt{\p} m_c^{3/2}} }
    \int \dxb \dyb \; \intbs{1} \intbs{3} \intbds{1} \intbds{3}
    \; \hat T^J_H (x,y,\B,\B')    \nonumber \\
    & & \mbox{\hskip 2.5cm}
    \exp [-S(x,y,\B,\B',2m_c)] \; \; \; 
    \| \hat\j^B(x,\B) \hat\j^B(y,\B') \| \; 
\label{eq:s_df}
\eea
where $\s_J = 1/\sqrt{2}, 1$ for $J=1,2$ respectively, and 
Sudakov correction factor evaluated at the scale of $2m_c$
is included in the convolution. As mentioned earlier, the presence 
of this radiative correction in the intermediate scale range, 
together with the MAX prescription for the infrared cut-off in the 
Sudakov factor, $\tilde b = \mbox{max}(b_1,b_2,b_3,b_1',b_2',b_3')$,
renders the whole approach self-consistent. Actually, since there 
are two hadrons in the final state, one can have a separate infrared 
scale for each hadron, for example
$\tilde b  = \mbox{max}(b_1,b_2,b_3)$ and 
$\tilde b' = \mbox{max}(b_1',b_2',b_3')$. However, numerically this 
would make no difference so we merge the two into one scale. 
The factorization scale $\m_F$ in the wavefunction is then set to
$\m_F = 1/\tilde b$ as usual.

The spin structure of the contribution from the diagrams of 
\fref{f:sing} requires the $J_z$ of the charmonium to be equal to 
$S_z$ of the quark line that is not attached to other light quark lines
via gluon, i.e. the first quark line from the left of the three
vertical lines in any of the figures in \fref{f:sing}. This
permits two possibilities for the helicities, that is (+,+,--) and
(+,--,+) of the u and d-quark lines in \fref{f:sing}
from left to right. The hard scattering amplitudes 
are, however, identical for the two arrangements of
helicities. This gives the following sums of products of the 
Fourier transform of the scalar functions $\j^B_{123}$
of the baryon-antibaryon, given in \eref{Psiansatz} and 
\eref{PsiDansatz}, represented in \eref{eq:s_df}
by $\| \hat\j^B(x,\B) \hat\j^B(y,\B') \|$.
\bea
    \| \hat\j^{B_8}(x,\B) \hat\j^{B_8}(y,\B') \| \;\;
  \fx &=& \fx 2 \;
  \bigg \{\; 
          \hat\j^{B_8}_{123}(x,\B) \hat\j^{B_8}_{123}(y,\B')  
         +\hat\j^{B_8}_{321}(x,\B) \hat\j^{B_8}_{321}(y,\B')  
          \nonumber \\
  \fx & &  +
      \Big(\hat\j^{B_8}_{123}(x,\B ) 
         + \hat\j^{B_8}_{321}(x,\B ) \Big)\! 
      \Big(\hat\j^{B_8}_{123}(y,\B')  
         + \hat\j^{B_8}_{321}(y,\B') \Big)
          \nonumber \\ 
  \fx & &  +
   (2 \llra 3 )
  \bigg \}   \;\; \\
    \| \hat\j^{\L}(x,\B) \hat\j^{\L}(y,\B') \| \; \; \; \;
  \fx &=& \fx 6 \;
  \bigg \{\; 
          \hat\j^{\L}_{123}(x,\B) \hat\j^{\L}_{123}(y,\B')  
         +\hat\j^{\L}_{321}(x,\B) \hat\j^{\L}_{321}(y,\B')  
          \nonumber \\
  \fx & &  +
      \Big(\hat\j^{\L}_{123}(x,\B ) 
         - \hat\j^{\L}_{321}(x,\B ) \Big)\! 
      \Big(\hat\j^{\L}_{123}(y,\B')  
         - \hat\j^{\L}_{321}(y,\B') \Big)
          \nonumber \\ 
  \fx & &  +
    (2 \llra 3)  
  \bigg \}  \;\; \\
    \| \hat\j^{B_{10}}(x,\B) \hat\j^{B_{10}}(y,\B') \|
  \fx &=& \fx 3 \;
  \bigg \{\; 
          \hat\j^{B_{10}}_{123}(x,\B) \hat\j^{B_{10}}_{123}(y,\B')  
         +\hat\j^{B_{10}}_{321}(x,\B) \hat\j^{B_{10}}_{321}(y,\B')  
          \nonumber \\
  \fx & & 
         +\hat\j^{B_{10}}_{132}(x,\B) \hat\j^{B_{10}}_{132}(y,\B')  
         +\hat\j^{B_{10}}_{231}(x,\B) \hat\j^{B_{10}}_{231}(y,\B')           
  \bigg \}  \;\;
\eea

The results for the colour singlet contributions in $\c_1$
and $\c_2$ decay into nucleon-antinucleon are shown 
in Table \ref{tab:sing_nucl}
\footnote{Our present numbers supersede those previously reported 
in \cite{wong1}.}  together with the experimental 
measurements. It is clear that the singlet contribution
is insufficient even with the smaller results of the BES collaboration
\cite{bes} to account for the experimental measurements. For uncertainties
in the theoretical estimate of the colour singlet contribution, the 
consistency of this calculation within MHSA, the choice of proton 
wavefunction and the value of the proton decay constant used etc., 
we refer to \cite{wong3}. We will discuss more about the differences between 
the experimental results in later section. The very important colour octet 
contribution will be investigated in the following section.

\begin{table}
\begin{center}
\begin{tabular}{||c||c|c|c||}
\hline
  $J$ & $\Gamma^{(1)}(\c_J\ra p\bar p)$ (eV)         
      & PDG (eV) \cite{pdg} &  BES (eV) \cite{bes}  \\
\hline \hline
 1    & \ 2.53  & \  75.68  & \ 37.84  \\ \hline
 2    &  16.58  &   200.00  &  118.00  \\ \hline
\end{tabular}
\end{center}
\caption{Clearly, the colour singlet contributions are insufficient in
explaining the experimental data of $\c_J$ decay into $p\bar p$.}
\label{tab:sing_nucl}
\end{table}

\section{Colour Octet Contribution}
\label{sec:co}

In colour octet $c\bar c$ decays into baryon-antibaryon, 
there is a constituent gluon in the initial state so 
the C-parity arguments given in Sec. \ref{sec:cs} no longer
hold. The diagrams in \fref{f:sing_fig} then form the bases
of three different contributing groups. The other possibilities 
are from the graphs where the $c\bar c$ pair annihilates
into a single gluon which would not be possible, if the pair
were in a colour singlet state. These additional groups 
are shown in \fref{f:oct}.

\bfi
\centerline{
\psfig{figure=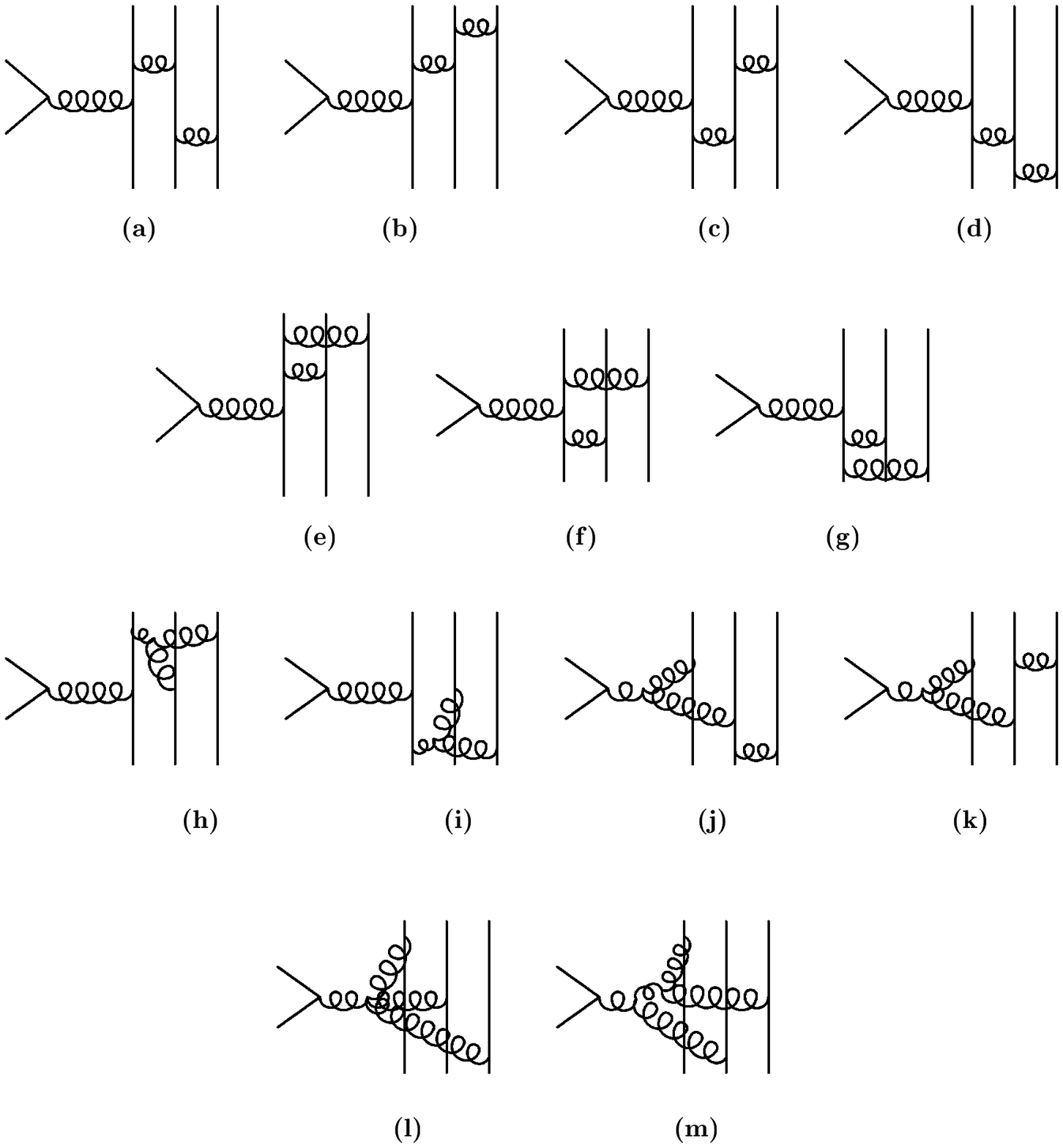,width=6.0in}
}
\caption{In addition to the graphs of type \fref{f:sing_fig},
these form the bases of further contributions in the colour
octet decay channel.}
\label{f:oct}
\efi

In order to form colour singlet baryon-antibaryon in the final states, 
this net colour from the constituent gluon must be neutralized. 
One could allow it to enter directly into one of the final baryons as
a constituent of a higher Fock state of the latters. 
However, within the hard scattering approach, at least one hard 
gluon must be exchanged between all the constituent partons, 
therefore such contribution involving the next higher Fock state of 
the baryons will be suppressed by the hard scale of the process and 
also by the smaller probability of the next Fock state. In any case, 
bringing in a higher state unnecessarily will introduce additional
unknown wavefunctions. It is therefore best to avoid it. The introduction 
of the colour octet in the charmonium system is, on the contrary, well 
founded and is indeed necessary as shown in 
\cite{bod&bra&lep1,bod&bra&lep2} and as we argued earlier. 
The alternative for colour neutralization is done by
attaching the constituent gluon onto all possible 
places in the diagrams of \fref{f:sing_fig}, \fref{f:sing} 
and \fref{f:oct}. This will generate about 9--11 diagrams
for each group and the light quark lines will also have to 
be permutated and so altogether there are over two hundred diagrams. 
Fortunately, the tedious algebra can be handled completely by
the computer program FORM, and one can arrange the constituent 
gluon to be attached automatically to all possible places in each 
basic graph and worked out the colour factors as well as the perturbative 
hard scattering part entirely by using this program. In fact, none of the 
algebra needs to be worked out by hand.

The colour octet wavefunction of the charmonium is 
\be
  |\chi_{\c_J}^{(8)},p \ran\, = \, \frac{t^a_{c\bar c}}{2} \,
                      f_{\c_J}^{(8)}\,\int [\dd z] \;
                      \f_{\c_J}^{(8)}(z_1,z_2,z_3) \, S_{J\n}^{(8)}(p) 
                      \; |c\bar c g ; p \ran \ .
\label{coc}
\ee
As explained in \cite{bolz&kroll2}, the distribution amplitude
of the octet charmonium $\f_{\c_J}^{(8)}$ is conveniently taken to be 
delta functions that peak at the light-cone momentum $z_3 =z =0.15$ for 
the constituent gluon and $z_1 =z_2 =(1-z)/2$ for the heavy quarks
while the octet decay constant $f_{\c_J}^{(8)}$ are obtained from fits
in \cite{bks,bks2}. The covariant spin wavefunctions for the octet states 
are given in \cite{bks2}
\bea
  S_{1\r}^{(8)} (p)\fx &=& \fx \frac{-i}{2 M_{\c_1}} (p \sla+M_{\c_1})
                        \; \e_{\m\n\r\s} p^\m \e^\n \g^\s 
                \nonumber \\
  S_{2\r}^{(8)} (p)\fx &=& \fx \frac{1}{\sqrt{2}}\,( p\sla\,+\,
                        M_{\c_2}) \; \e_{\s\r} \gamma^{\s} \ . 
\label{csw8}
\eea
As we will take the spin of the charmonium to be pointing upward without 
loss of generality, we found it more straight forward to write these as
\bea
  S^{(8)(+)}_{1\r} (p)\fx &=& \fx \frac{1}{2} (p \sla+M_{\c_1})
                    ( \e \sla^{(+)}_{c\bar c} \, \e^{(0)}_\r 
                     -\e \sla^{(0)}_{c\bar c} \e^{(+)}_\r )           \\
  S^{(8)(+)}_{2\r} (p)\fx &=& \fx \frac{1}{2} (p \sla+M_{\c_2})
                    ( \e \sla^{(+)}_{c\bar c} \, \e^{(0)}_\r 
                     +\e \sla^{(0)}_{c\bar c} \e^{(+)}_\r )   
\label{csw8_our}
\eea
instead. The $\e_\r$ is the polarization vector of the constituent gluon
and $\e^\m_{c\bar c}$ is the spin S=1 vector of the colour octet $c\bar c$ 
system. Then the numerator of any contributing graphs to $\c_J$ decay can 
all be expressed in the form of 
\be  N^J = \ca + (-1)^J \ca'   
\ee
and the difference in the numerator between the $\c_1$ and $\c_2$ system 
comes entirely from the sign. In the appendix where we list all the 
contributions from the graphs of each group, the numerators are all in 
the above format. 

To deal with the colour octet contributions, bearing in mind that the 
advantages of the dynamical setting of the renormalization scales and 
the built-in Sudakov suppression of the end-point problematic regions of
the distribution amplitudes, one could use again the modified hard 
scattering scheme \cite{bs,ls} as we did in the singlet contribution in the 
previous section. However, these advantages are obtained at the expense of 
complicating the expressions and the calculations of the perturbative
hard part $T_H(x,\Kp)$ by including the internal transverse momenta 
in the propagators. These transverse momenta will have to be integrated 
out subsequently by convoluting with the hadronic wavefunctions. 
Therefore the number of integration variables can be quite high
especially when we are dealing with baryon and antibaryon which contain
three constituent quarks and antiquarks even at the valence level 
and also there are many diagrams to consider. 
To keep things simple, it is advisable to return to the standard
scheme \cite{bl} so as to deal only with distribution amplitudes
and $T_H(x)$ without the internal transverse momenta and not the more 
complicated wavefunctions and $T_H(x,\Kp)$. This is what we will do 
below. We will discuss getting the graphs and the expressions for the 
individual colour octet contribution to the amplitude in the next 
section within the standard scheme.

\section{Getting The Graphs And Calculating $T_H$ Of The \\
Colour Octet Contributions}
\label{sec:gg}

In the appendices, we give the graphs and expressions of our 
calculation for the colour octet contribution. The diagrams 
can be divided into ten basic groups. Each group is based on 
one basic graph in which the constituent gluon from the colour 
octet component of the charmonium has not yet been drawn or inserted. 
Because of the restriction of C-parity, in the absence of the constituent 
gluon, these groups individually may or may not exist in the colour 
singlet contribution to the decay. In the latter case, they survive solely 
because of the presence of the constituent gluon. These graphs 
are those already presented in \fref{f:sing_fig} and \fref{f:oct}. 
The group associated with each basic graph is generated by attaching the 
constituent gluon to all possible places on the basic graph except on the 
initial $c$-quark or $\bar c$-quark when they have just emerged from
the charmonium. 

The numerator of the hard part $T_H$ of each diagram from each 
group will be given in the appendices, while the denominator which is 
essentially a product of the propagators in each diagram, will not be 
written out explicitly individually but can be derived from those of the 
basic graphs (these will be presented with each group) by following some 
simple rules that followed from the momentum flow through the graph.
Given that the momenta of the $\c_J$, and those of the 
outgoing baryon and antibaryon are $P_{\c_J}$, $P_B$ and 
$P_{\bar B}$ respectively, the outgoing momenta of the 
constituents of the baryon and antibaryon will be assigned
the momenta $p_{q_i}= x_i P_B$, $p_{\bar q_i} = y_i P_{\bar B}$ 
with $i=1,2,3\; $. Here $x_i$ and $y_i$ are the momentum
fraction of the $i$ light quark line of the outgoing 
baryon-antibaryon subjected to $\sum_i x_i = \sum_i y_i = 1$.
In the basic graphs, the constituent gluon is not included
so the momentum of $\c_J$ shared by the charm-anticharm is
$p_c = z_1 P_{\c_J}$ and $p_{\bar c}=z_2 P_{\c_J}$ with
$z_1+z_2 =1$. Obviously energy-momentum conservation requires 
$P_{\c_J} = P_B +P_{\bar B}$. The product of the gluon propagators in
each of the basic graphs is expressed with these momentum 
configurations with $2 P_B \cdot P_{\bar B} = \mc^2 \gg P_B^2, P_{\bar B}^2 \sim 0$.
On the insertion of the constituent gluon with momentum fraction
$z$, the condition $z_1+z_2 =1$ is now replaced by $z+z_1+z_2 =1$.
Therefore the heavy quark pair will have only $(1-z) P_{\c_J}$ 
instead of the full $P_{\c_J}$ to annihilate into gluons. 
The consequence is that in the new graphs the momenta flowing along the 
path or lines connecting the initial heavy quark 
line of the charmonium to the insertion point of the constituent 
gluon must be reduced by $z P_{\c_J}$. The momenta in any
intermediate heavy quark line(s) which only exist in those graphs where
the $c\bar c$ annihilates into two or more gluons, may also need to be 
so-adjusted but that depends on the insertion point. This momentum
shift in the basic graphs brings about a similar shift in the momenta 
carried by the propagators along the same connecting path. 
In general, this shift can be done by replacing one of the three
($x_i, y_i$) pairs in the affected propagators including the
intermediate heavy quark one by ($x_i-z, y_i-z$). 

The momentum fractions and the quark and gluon lines are labelled
as follows for simpler description and they apply to all basic graphs. 
The top (bottom) heavy quark line on the left is the initial charm (anticharm) 
quark line. The three vertical light quark lines running from the bottom 
to the top are light quark line 1 to 3 from left to right. Light quarks
(antiquark) of the baryon (antibaryon) emerge from the top (bottom). 
The fraction $x_i$ and $y_i$ label that of the quark and antiquark
of the $i$ light quark line. The gluon line coming directly or 
indirectly from the charmonium and entering the $i$ light quark line
is gluon line $i$, G$i$. Whereas the labelling of the vertical light 
quark line is for the entire line, each of these gluon lines connects 
only two vertices and ends at these vertices. There will be only 
one gluon line entering each light quark line even though each light 
quark line may be attached to more than one gluon line. In this case, 
only one such line enters the quark line and the other is leaving it. 
The gluon line is then labelled with G$i$ if it is entering light quark 
line $i$ and not leaving it as we mentioned above. \fref{f:label_ex}
illustrates our labelling scheme with some examples.
\bfi
\centerline{
\psfig{figure=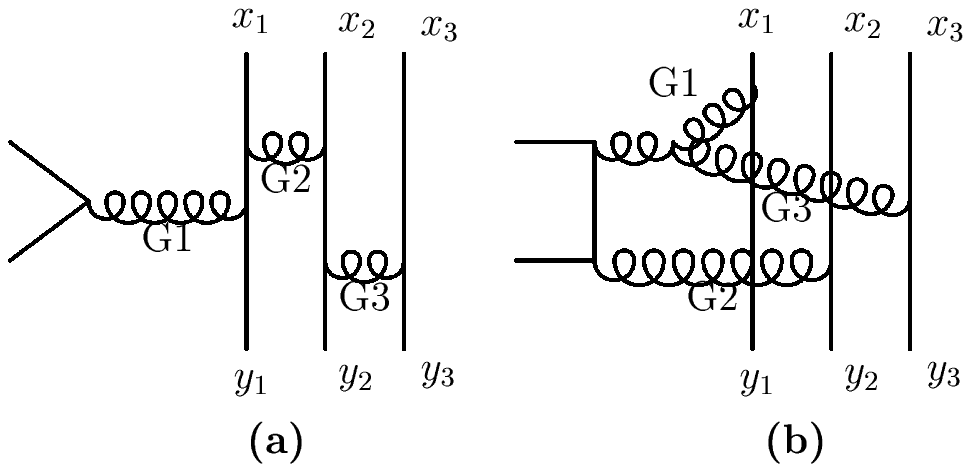,width=3.80in}
}
\caption{Our labelling scheme as applied to (a) Group 2 and (b) Group 4.}
\label{f:label_ex}
\efi
There are however gluon lines that enter a 3- or 4-gluon
vertex instead of a light quark line, see Group 4 and 6 for example. 
These are not labelled. The labelling is for the purpose of identifying the 
graphs created from the basic one by the insertion of the constituent gluon.
The insertion point (I.P.) of the constituent gluon will uniquely
identify each so-created graph and these graphs will form a group 
based on the basic one. Taking the $\c_1$ and $\c_2$ to be spinning up. 
The allowed helicities of the three light quark lines $(\l_1,\l_2,\l_3)$ 
are $(+,+,-)$, $(+,-,+)$ and $(-,+,+)$. These three helicities 
configurations of each graph do not necessarily have the same or permutation 
related numerator as in the colour singlet case. The presence of the 
constituent gluon makes this not possible in general. The corresponding
numerators are listed in the tables in the appendices.  

As mentioned above, each graph created from a basic graph by adding
the constituent gluon is identified by the insertion point
where the gluon from the colour octet state of the charmonium joins the
basic graph. The group of a basic graph is created by attaching the
colour octet gluon to all possible places on the basic graph with the
exception of the initial heavy quark or antiquark line. Also the 
creation of a graph with an intermediate state with one gluon and nothing 
else is forbidden by C-parity. More explicitly it means that when the basic 
graph contains the $c\bar c$ annihilation into one single gluon, the graph 
with the constituent gluon attached to this gluon is not possible. Note 
that this kind of basic graphs by themselves cannot exist without the 
constituent gluon because of C-parity and so they do not contribute to the 
colour singlet contribution. 

The complete graphs created from the basic ones will be 
labelled as follows. If the constituent gluon attaches to one of the 
light quark lines, it can be with one of the three quarks (antiquarks) 
that forms the baryon (antibaryon). These graphs are labelled as
U$i$ or L$i$ indicating that the constituent gluon joins the upper (top) 
or lower (bottom) part of the quark line $i$. If the quark line has one or 
more intermediate off-shell middle segment, which is or are separated from 
the upper and lower part by quark-gluon-antiquark vertex or vertices, to 
which the octet gluon attaches itself, these are labelled M$i$ for one 
intermediate off-shell quark line on quark line $i$, for example the quark 
line 2 of Group 1 (sees \fref{f:grp1}), and UM$i$ or LM$i$ denoting upper 
or lower part of the middle off-shell line if two intermediate middle 
segments exist on the light quark line $i$, for example the quark line 1 
of Group 8 (see \fref{f:grp8}). Evidently, the constituent gluon can 
be attached to the gluon lines as well. These graphs are labelled with 
G$i$ (not to be confused with the gluon line labelling) if the new graph 
is created by joining the gluon to the gluon line $i$ by a 3-gluon vertex. 
If the gluon is attached instead to an unlabelled gluon line, the graph will 
be called GR as the constituent gluon attaches to the remaining unlabelled
gluon line. Only one such graph per group is possible at maximum
and these always vanish due to colour. We will not discuss these graphs
further. Whenever a 3-gluon vertex exists in a basic graph, there is 
the possibility of attaching the constituent gluon to this vertex
to create a complete graph with a 4-gluon vertex. These will be
labelled as 4G for one such possibility or 4G1 and 4G2 when two
such possibilities exist as in Group 9 (see \fref{f:grp9}). There remains 
attaching the gluon to the heavy intermediate heavy quark line(s) when this
exists. We denote this simply by graph Q for only one heavy quark line
or UQ and LQ in the presence of two heavy quark lines, one situated 
above the other on the basic graph, meaning the gluon is attached
to the upper or lower heavy quark line. This latter is possible
only for Group 5.  

The rules for getting the product of propagators for each new graph
from that of the basic graphs are as follows. In determining
which of the pair ($x_i, y_i$) in the propagators to replace 
by ($x_i-z, y_i-z$), it is the $i$ pair for graphs labelled U$i$, 
L$i$, G$i$, M$i$, UM$i$ or LM$i$.  
The replacement should only be made to propagators along the line
connecting the heavy quark line to the insertion point whether
they contain ($x_i, y_i$) explicitly or not. Modification must be
made to gluon or light quark propagators through the relation
$x_i=1-x_j-x_k$ and $y_i=1-y_j-y_k$ where $j\neq k\neq i$ and also
to any charm propagator only if it contains the pair ($x_i, y_i$)
explicitly. So applying this to graph U1 of Group 5 illustrated
in \fref{f:eg1}(a), only the propagators associated with the gluon 
line G1 and the upper heavy quark line in \fref{f:grp5} need this 
replacement. The product becomes
\bea & \frac{1}{\{ (z_1+z-x_1)(z_1+z-y_1) -1/4 \} \mc^2} 
       \frac{1}{\{ (z_2-x_3)(z_2-y_3) -1/4 \} \mc^2}      
       \frac{1}{(x_1-z)(y_1-z) \mc^2 +\r^2} \frac{1}{x_2 y_2 \mc^2}
       \frac{1}{x_3 y_3 \mc^2}         \; . \nonumber \\
     & 
\eea
If it is graph L2 of Group 5 in \fref{f:eg1}(b), only the propagator 
of gluon line G2 needs to be changed and we have
\bea & \frac{1}{\{ (z_1-x_1)(z_1-y_1) -1/4 \} \mc^2} 
       \frac{1}{\{ (z_2-x_3)(z_2-y_3) -1/4 \} \mc^2}
       \frac{1}{x_1 y_1 \mc^2} \frac{1}{(x_2-z)(y_2-z) \mc^2 +\r^2}
       \frac{1}{x_3 y_3 \mc^2}         \; . \nonumber \\
     & 
\eea 
As a third example, applying this to \fref{f:eg1}(c) that is graph
L3 of Group 1 gives
\bea & \frac{1}{(1-z)^2 \mc^2} \frac{1}{(x_3-z)(1-y_1-z) \mc^2 +\r^2}
       \frac{1}{x_1 y_1 \mc^2} \frac{1}{(1-x_1-z)(1-y_1-z) \mc^2 +\r^2}
       \frac{1}{(x_3-z)(y_3-z) \mc^2 + \r^2}  \; . \nonumber \\
     &
\label{eq:grp1_ex}
\eea
It must be mentioned that in the appendices where details 
of each group are given, the propagators are written down without
the usual $i\e$. This is to save on typing but it should be understood
implicitly that a $i\e$ should be present in each propagator. This also
applies to the expressions here. Also, we have inserted a term $\r^2$ in
the denominators of those propagators which carry two poles which cannot 
be handled by the $i\e$ prescription. The $\r^2$ is understood to be
the mean-squared internal momentum of the baryons $\r^2 = \lan \Kp^2\ran$.
It is used here to prevent two possible poles in any one propagator to occur 
simultaneously. This problem has been treated in the same manner in 
\cite{bks} whenever the $i\e$ prescription failed. The actual values
of $\r^2$ depend on the baryon wavefunctions. We used 
$\r_{(8)}$ = 415.0 MeV and $\r_{(10)}$ = 389.0 MeV for the octet 
and decuplet baryons, respectively. These values are obtained
from the respective wavefunctions. 

\bfi
\centerline{
\psfig{figure=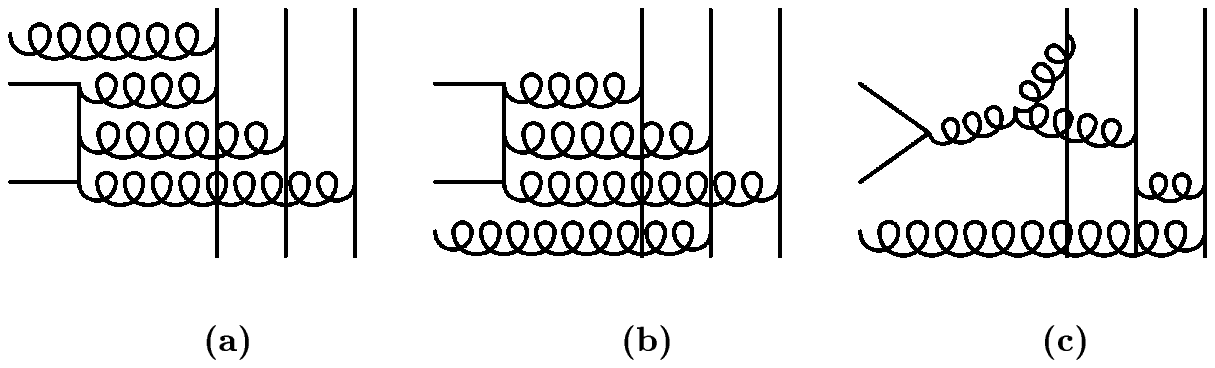,width=4.0in}
}
\caption{Examples of complete colour octet graphs. (a) graph U1 and (b)
graph L2 of Group 5, and (c) graph L3 of Group 1.}
\label{f:eg1}
\efi

\bfi
\centerline{
\psfig{figure=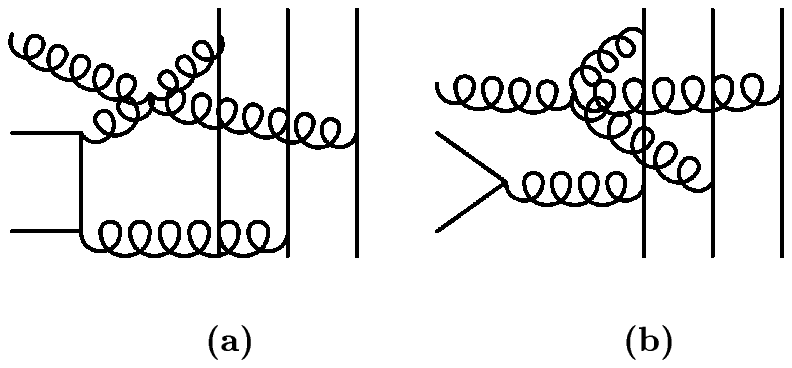,width=3.0in}
}
\caption{Examples of complete colour octet graphs with 4-gluon vertex.
(a) graph 4G of Group 4 and (b) graph 4G of Group 10.}
\label{f:eg2}
\efi

For 4G graphs, if the unlabelled gluon line attaches to the 3-gluon
vertex connecting gluon line G$i$ and G$j$, then either replace
($x_i, y_i$) with ($x_i-z, y_i-z$) or ($x_j, y_j$) with ($x_j-z, y_j-z$) 
in the gluon propagator and do the same in the heavy quark propagator but
in this case only when either pair of momentum fractions appears 
explicitly is sufficient. Applying this to Group 4, we get
\bea & \frac{1}{\{ (z_2-x_2)(z_2-y_2) - 1/4 \} \mc^2} 
       \frac{1}{(1-x_2-z)(1-y_2-z) \mc^2 +\r^2 } 
       \frac{1}{x_1 y_1 \mc^2} \frac{1}{x_2 y_2 \mc^2}
       \frac{1}{x_3 y_3 \mc^2}  \; , \nonumber \\
     &         
\eea
and to Group 10, the product of propagator become
\bea & \frac{1}{(1-z)^2 \mc^2} \frac{1}{(1-z)(1-y_1-z) \mc^2}        
       \frac{1}{(1-x_1-z)(1-y_1-z) \mc^2+\r^2} \frac{1}{x_2 y_2 \mc^2} 
       \frac{1}{x_3 y_3 \mc^2}  \; . \nonumber \\
     & 
\label{eq:grp10_ex}
\eea
\eref{eq:grp1_ex} and \eref{eq:grp10_ex} are fine examples of
the necessity of having to make the shift even if either pair of 
momentum fractions does not appear explicitly in the light quark and gluon
propagators. In these example, the replacement has to be made even for
$x_1+x_2+x_3 = y_1+y_2+y_3 = 1$ to $1-z$ because the momentum fractions
are completely hidden. For the remaining Q, UQ and LQ graphs, it is only 
necessary to replace $z_1$ by $z_1+z$ and $z_2$ must be left alone.

One must not forget that the insertion of the constituent gluon will
introduce an additional propagator (X-Prop.) to the basic 
graph. This is, however, not true for the 4G graphs but these have
a factor $\mc^2$ less than the other graphs and so a factor of $1/\mc^2$
must be multiplied to them so that a global prefactor can be written down
for all graphs. The extra propagator or the extra factor $1/\mc^2$ is given 
in the last table of numerators of each group under the column X-Prop.
Similarly, the structure of group 6 is such that it is a factor
$\mc^2$ less than the other group from the start because these graphs
have one less propagator. In order to keep the same overall factor for all 
groups, $1/\mc^2$ has been multiplied to the product of the basic propagators 
of this group. This extra factor of course does not correspond to any 
propagator but is merely a compensation factor. It must be mentioned
that not all graphs generated from the basic graphs exist. Some
vanish for one reason or another. For these, the additional propagators
are not given in the tables but have the entry n. g. (not given)
instead.

There exist several more groups that we have not drawn out explicitly 
in the appendices. Their basic graphs may be obtained from those of 
Group 1, 2 and 2', 3, 4, 7 and 10 by making one simple change to them. 
For example, attaching the line G2 to quark line 1 at a point below 
instead of above the line G1 on Group 2 and 2', or moving line G3
in Group 1 and 3 to above the $qg\bar q$-vertex on quark line 2, or 
flipping any of the basic graphs of Group 3, 4, 7 and 10 upside down. 
However, these extra groups may be included by giving a factor of 
two on each of their associated group because these are related to each 
other by a simple change of variables. Finally, we must mentioned that 
the graphs in each group must be subjected to further permutations of the 
three light quark lines to give further graphs. But these also can be taken 
care of by a numerical factor so each possible insertion point will generate
only one graph and there is no need to divide the graphs any further
or add any more groups.

\section{Colour Singlet And Octet Contribution In The Standard Scheme}
\label{sec:s&o_SHSA}

\subsection{Colour Singlet Contribution}
\label{sec:cs_SHSA}

The procedure to obtain the colour singlet contribution in the
standard scheme is quite similar to the one we used in Sec. \ref{sec:cs}.
Remembering that in the standard scheme, the internal transverse
momenta are taken to be negligible in comparison with the virtualities
in all propagators in the perturbative hard part,
so $T_H(x)$ is free from any $\Kp$'s and the latter can be integrated over
each wavefunction to give the corresponding distribution amplitude. 
So setting the $\Kp$'s in $\cg_i$, $\cq$ and $\cq_c$ in \eref{eq:virt} 
to zero, $T^J_H$ is now independent of $\Kp$ and $\Kp'$
\bea T^J_H (x,y) & = & 
    {2 \sqrt{2} \; (4\p)^3 \; (\a_s(m_c))^3
    \over {9\sqrt{3} m_c^5 \; x_1 x_3 y_1 y_3 \, (1-x_1)^2 (1-y_1)  
    \, [x_1 (1-y_1) +y_1 (1-x_1)] }}
    \nonumber \\ 
    & \times &  
    \left  [  (1-2 x_1)^{2-J} 
            + { {(-1)^{J-1} \; x_1 (x_1-y_1)} \over 
                {x_1 (1-y_1) +y_1 (1-x_1) }      }
    \right ] \nonumber \\
    & + & (x \llra y )  \; .
\label{eq:s_hard_SHSA}
\eea
The renormalization scale $\m_R$ in the above equation has been set at the 
constant scale $m_c$ in $\a_s$ because each gluon takes approximately 
$M_{\c_J}/2 \approx m_c$ from the charmonium and so the virtuality
is roughly $m_c^2$. Now Fourier transforming the transverse momentum 
independent $T^J_H$ to transverse position-space yields several delta 
functions which force all $\B$'s and $\B'$'s to the origin. As a consequence,
following from its definition and derivation, the Sudakov factor has to be set 
to unity. The decay form factor defined in Sec. \ref{sec:decay} of $\c_J$ 
into $B-\bar B$ becomes 
\bea \cb_J^{B \; (1)} 
    &=& -i {{\sqrt{3} |R_p'(0)| \; \s_J} \over {8\sqrt{\p} m_c^{3/2} (4\p)^4} }
    \int \dxb \dyb  \; T^{J\; (1)}_H (x,y) \; 
    \| \hat\j^B(x,0) \hat\j^B(y,0) \| 
\label{eq:s_df_SHSA}
\eea
where $\s_J ={1/\sqrt{2}, 1}$ for $J =1,2$ as before.
This has to be combined with that of the colour octet contribution to be
discussed below in accordance with our theoretical arguments given
in Sec. \ref{sec:s_o} to give the true partial decay width.

\subsection{Colour Octet Contribution}
\label{sec:co_SHSA}

The total colour octet contribution to $\cb_J^{B\; (8)}$ has to be
the sum over all contributing graphs from each group and over 
all possible helicity configurations of the light quarks and 
antiquarks in the outgoing baryon-antibaryon given in the appendices
\bea \cb_J^{B \; (8)} 
    &=& \sum_{\l_1,\l_2,\l_3 =\pm} 
        {{f^{(8)}_{\c_J} \; \s_J} \over{2 m_c\; (4\p)^4}}
        \int \dxb \dyb \; \{T^{J\; (8)}_H (x,y) \}_{\l_1,\l_2,\l_3} \;
        \| \hat\j^B(x,0) \hat\j^B(y,0) \|_{\l_1,\l_2,\l_3}  \; .
        \nonumber \\
\eea
The helicity dependent hard perturbative parts are 
\bea \{T^{J\; (8)}_H (x,y) \}_{\l_1,\l_2,\l_3} 
    &=& i\; (4\p\a_s(m_c))^3 \sqrt{4\p\a^{soft}_s} \; \mc^7 \nonumber \\
    & & \times \sum_{{g \in Groups} \atop {m \in Members}} 
        S_g P_{g\,m}(x,y) \{N^J_{g\,m}(x,y)\}_{\l_1,\l_2,\l_3}
        \nonumber \\
\eea
where $P_{g\,m}(x,y)$ is the product of propagators of the member
$m$ =U1, U2, U3, L1, L2, L3, $\dots$ etc. of the group $g$ =1, 2, 2', 
$\dots$, 10. They can be obtained from the product of
the basic propagators of each group as discussed in Sec. \ref{sec:gg}. 
The $S_g$ is a symmetry factor for the group $g$ to take care of similar 
potential groups that are related to $g$ by simple change of variables.
For the groups listed in the appendices, $S_g$ =\{2,2,2,4,2,1,1,2,1,1,2\}
for the group $g$ =1, 2, 2', 3, $\dots$, 10, respectively.
The coupling $\a^{soft}_s$, taken to be equal to $\p$, is that of attaching 
the constituent gluon to the basic graphs. It needs special treatment
because of its different nature in comparison to the rest of the $\a_s$'s
\cite{bks2}. The $\{N_{g\,m}(x,y)\}_{\l_1,\l_2,\l_3}$ in the above equation
contains the helicity dependence and denotes the numerator of the 
member $m$ of the group $g$. These numerators are given in the appendices. 
Note that the calculation is really only gauge-invariant to order $z^2$
in the numerator (see ref. \cite{bks2}) because the $c\bar c$ are treated
as on-shell, therefore all $z^3$ or higher terms in the numerator have
to be dropped. The remaining helicity dependent product of the scalar 
functions $\j^B_{123}(x,0)$ of the baryon wavefunctions are given below.

\bea   \| \hat\j^{B_8}(x,0) \hat\j^{B_8}(y,0) \|_{+,+,-}  \;\;
  \fx &=& \fx 2 \;
  \bigg \{\; 
          \hat\j^{B_8}_{132}(x,0) \hat\j^{B_8}_{132}(y,0)  
         +\hat\j^{B_8}_{231}(x,0) \hat\j^{B_8}_{231}(y,0)  
          \nonumber \\
  \fx & &  +
      \Big(\hat\j^{B_8}_{132}(x,0) 
         + \hat\j^{B_8}_{231}(x,0) \Big)\! 
      \Big(\hat\j^{B_8}_{132}(y,0)  
         + \hat\j^{B_8}_{231}(y,0) \Big)
  \bigg \}   \;\; \nonumber \\  \\
       \| \hat\j^{B_8}(x,0) \hat\j^{B_8}(y,0) \|_{+,-,+}  \;\;
  \fx &=& \fx 2 \;
  \bigg \{\; 
          \hat\j^{B_8}_{123}(x,0) \hat\j^{B_8}_{123}(y,0)  
         +\hat\j^{B_8}_{321}(x,0) \hat\j^{B_8}_{321}(y,0)  
          \nonumber \\
  \fx & &  +
      \Big(\hat\j^{B_8}_{123}(x,0) 
         + \hat\j^{B_8}_{321}(x,0) \Big)\! 
      \Big(\hat\j^{B_8}_{123}(y,0)  
         + \hat\j^{B_8}_{321}(y,0) \Big)
  \bigg \}   \;\; \nonumber \\  \\
       \| \hat\j^{B_8}(x,0) \hat\j^{B_8}(y,0) \|_{-,+,+}  \;\;
  \fx &=& \fx 2 \;
  \bigg \{\; 
          \hat\j^{B_8}_{213}(x,0) \hat\j^{B_8}_{213}(y,0)  
         +\hat\j^{B_8}_{312}(x,0) \hat\j^{B_8}_{312}(y,0)  
          \nonumber \\
  \fx & &  +
      \Big(\hat\j^{B_8}_{213}(x,0) 
         + \hat\j^{B_8}_{312}(x,0) \Big)\! 
      \Big(\hat\j^{B_8}_{213}(y,0)  
         + \hat\j^{B_8}_{312}(y,0) \Big)
  \bigg \}   \;\; \nonumber \\ 
\eea
\bea   \| \hat\j^{\L}(x,0) \hat\j^{\L}(y,0) \|_{+,+,-} \; \; \; \;
  \fx &=& \fx 6 \;
  \bigg \{\; 
          \hat\j^{\L}_{132}(x,0) \hat\j^{\L}_{132}(y,0)  
         +\hat\j^{\L}_{231}(x,0) \hat\j^{\L}_{231}(y,0)  
          \nonumber \\
  \fx & &  +
      \Big(\hat\j^{\L}_{132}(x,0 ) 
         - \hat\j^{\L}_{231}(x,0 ) \Big)\! 
      \Big(\hat\j^{\L}_{132}(y,0)  
         - \hat\j^{\L}_{231}(y,0)  \Big)
  \bigg \}  \;\; \nonumber \\
\eea
The other two helicity arrangements of this quantity for the $\Lambda$ 
follow the pattern above for the other octet baryons. For the decuplet 
baryons, they are
\bea   \| \hat\j^{B_{10}}(x,0) \hat\j^{B_{10}}(y,0) \|_{+,+,-}
  \fx &=& \fx 3 \;
  \bigg \{\; 
          \hat\j^{B_{10}}_{132}(x,0) \hat\j^{B_{10}}_{132}(y,0)  
         +\hat\j^{B_{10}}_{231}(x,0) \hat\j^{B_{10}}_{231}(y,0)  
  \bigg \}  \;\;  \nonumber \\
    \| \hat\j^{B_{10}}(x,0) \hat\j^{B_{10}}(y,0) \|_{+,-,+}
  \fx &=& \fx 3 \;
  \bigg \{\; 
          \hat\j^{B_{10}}_{123}(x,0) \hat\j^{B_{10}}_{123}(y,0)  
         +\hat\j^{B_{10}}_{321}(x,0) \hat\j^{B_{10}}_{321}(y,0)  
  \bigg \}  \;\;  \nonumber \\
       \| \hat\j^{B_{10}}(x,0) \hat\j^{B_{10}}(y,0) \|_{-,+,+}
  \fx &=& \fx 3 \;
  \bigg \{\; 
          \hat\j^{B_{10}}_{213}(x,0) \hat\j^{B_{10}}_{213}(y,0)  
         +\hat\j^{B_{10}}_{312}(x,0) \hat\j^{B_{10}}_{312}(y,0)  
  \bigg \}  \;\;. \nonumber \\
\eea
With these notations, the colour octet contribution looks rather
simple. One only has to sum up all graphs from each and every group
and then perform the integrations.

\null
\section{The Widths Of $\c_J$ Decay Into Baryons-Antibaryons}
\label{sec:widths}

With our method discussed in the previous sections, colour octet
contribution which we argued to be necessary in additional to the
singlet contribution to getting the correct P-wave $\c_J$ partial 
decay widths can be included. Before giving the numerical results,
it must be mentioned that of all the baryons we considered, 
only the decay into proton-antiproton is measured so the majority
of our results are in fact predictions. Moreover,
the most recently reported measured values \cite{bes} differ by as 
much as a factor of two from those in the Particle Data Tables
\cite{pdg}. This is also true in the case of the decay into pseudoscalar
mesons as noted already in \cite{bks2}. Therefore we are content 
with results that lie somewhere in between these measurements. 
Until the situation improves, the colour octet decay constants fitted 
in \cite{bks,bks2} cannot be more accurately determined hence also
the current results. In any case, part of our goal is to show that
explicit calculations do indeed support the theoretical arguments,
that is the colour octet contribution in P-wave $\c_J$ decays cannot
be neglected both in inclusive as well as in exclusive process. 

\begin{table}[t]
\begin{center}
\begin{tabular}{||c||r|r||c||r|r||}
\hline
 Octet    & \multicolumn{2}{|c||}{$\Gamma^{(1)+(8)}$ (eV)} &
 Decuplet & \multicolumn{2}{|c||}{$\Gamma^{(1)+(8)}$ (eV)} \\ \cline{2-3}\cline{5-6}
 Baryons  & J=1 & J=2 & Baryons  & J=1 & J=2               \\ 
\hline \hline
 $\c_J \rightarrow N       \bar N      $ & (56.27) & 154.19     &
 $\c_J \rightarrow \Delta\, \bar \Delta   $ &  33.49  & 124.62 \\
 $\c_J \rightarrow \Sigma  \bar \Sigma $ & 28.42   &  97.69     &
 $\c_J \rightarrow \Sigma^* \bar \Sigma^* $ &  18.46  &  71.09 \\
 $\c_J \rightarrow \Xi     \bar \Xi    $ & 21.49   &  72.62     &
 $\c_J \rightarrow \Xi^*    \bar \Xi^*    $ &   9.42  &  41.16 \\ 
 $\c_J \rightarrow \Lambda \bar \Lambda$ & 33.64   &  69.19     &
                                         &         &            \\ \hline
\end{tabular}
\end{center}
\caption{The partial decay widths for $\c_J$ decay into octet 
and decuplet baryon-antibaryon pairs. The width of $\c_1 \lra N\bar N$ 
in parenthesis is to indicate that this value is a fit unlike all partial
widths of $\c_2$ which are predictions. Based on this fit, the rest of 
$\c_1$ widths are also predictions.}
\label{tab:results}
\begin{center}
\begin{tabular}{||c||r|r|r||}
\hline
     & \multicolumn{3}{|c||}{Branching ratios ($\times 10^{-5}$)}  \\ \cline{2-4}
     & Br$^{(1)+(8)}$   &   PDG   &   BES   \\ 
\hline \hline
 $\c_1 \rightarrow p       \bar p      $ & (6.39) &   8.60 $\pm$ 1.2 
                                         &  4.30 $\pm$ 2.3 $\pm$ 2.9 \\
 $\c_2 \rightarrow p       \bar p      $ &  7.71  &  10.00 $\pm$ 1.0 
                                         &  5.90 $\pm$ 3.1 $\pm$ 3.3 \\ \hline
\end{tabular}
\end{center}
\caption{Comparing our results with the measured widths from the PDG 
\cite{pdg} data and from the BES collaboration \cite{bes}.
This branching ratio of $\c_1$ is a fit.}
\label{tab:results_cf}
\end{table}

Our results are shown in Table \ref{tab:results} for octet and
decuplet baryons. The numerical parameters used to obtain these 
results which were given throughout this paper have been collected 
together again in appendix \ref{a:np} so that any interested readers 
do not have to search through the paper for their values.
Only the kinematically plausible decuplet 
baryons with the lowest masses are considered. In getting this
results, we used the colour octet decay constant 
$f^{(8)}_{\c_2} = 0.9 \times 10^{-3}$ GeV$^2$ fitted in \cite{bks}
for the decay of $\c_2 \rightarrow \p \p$ within the standard 
hard scattering approach. We found that in order to obtain 
reasonable agreement of the $\c_1 \rightarrow p\bar p$ 
decay, it must have a smaller colour octet decay constant. 
Therefore we use $f^{(8)}_{\c_1} = 0.225 \times 10^{-3}$ GeV$^2$.
Note that this does not contradict the result in \cite{bks},
since $\c_1$ cannot decay into $\p\p$ because of parity.
and the decay constant $f^{(8)}_{\c_1}$ is therefore unconstrained
in \cite{bks}. The smaller colour octet decay constant for the $\c_1$ 
than those of the $J=0,2$ partners is not too surprising given 
that the odd-spin P-wave charmonium is a somewhat different heavy 
quark-antiquark system. Table \ref{tab:results} shows that
the decay widths decrease roughly with increasing value of the baryon 
masses or with the reduction in the available phase space.

In Table \ref{tab:results_cf}, the branching ratios of the decays 
into $p\bar p$ is shown against the experimental measurements. 
As mentioned above, the disagreement between PDG and BES is sizable. 
This is true not only in the ratios into $p\bar p$ but also in the
case of a pair of pseudoscalars in the final state. Since our
results are based on fit to the latters, until the experimental 
situation improves, we have to settle for our current results. 

From Sec. \ref{sec:cs} and Table \ref{tab:results_cf}, it is
clear that only by including colour octet can the decay widths
of the P-wave charmonia be brought in range of the experimental
measurements. It would be better to show this in the modified
hard scattering scheme which has the advantage of the
dynamical setting of renormalization scales by the process 
itself. But due to the complexity of the calculations, we
have to revert back to the standard scheme. Remembering
this dynamical setting of scales is only possible if
the Landau pole in $\a_s$ is suppressed by the Sudakov
factor or better yet if it is not present at all. This
suggests the use of one of the analytic models for $\a_s$
which are free from the problem of the Landau pole 
\cite{grun,dkt,ss,web}. The one suggested in \cite{ss}
is particularly appealing because of its relative simplicity. 
By combining $\a_s^{\srm{analytic}}$ with the standard hard 
scattering approach, a simpler scheme than the modified
one can be constructed, but nevertheless still preserving
the best features of that scheme. Under the new scheme,
the amplitude would be given by
\be \cm \sim f_{\c_J} \f_{\c_J}(x) \ot f_N \f_N(x) 
     \ot f_{\bar N} \f_{\bar N}(x)                     
     \ot T_H(x,\a^{\ssrm{analytic}}_s(x))   \; .
\label{eq:smhs}
\ee
We have also tried using this semi-modified scheme and again
both colour singlet and octet must be included. The details 
will be given elsewhere \cite{wong2}. A somewhat similar
use of this analytic $\a_s$ model in exclusive process to study 
pion form factors can be found in \cite{kss}.

Thus having satisfied ourselves with the genuine necessity and the
correctness of the inclusion of the colour octet state in P-wave 
charmonium decays, we can now generalize the arguments to even higher 
wave quarkonia. Remembering that the P-wave $Q\bar Q$ wavefunction 
has a $1/M$ suppression due to angular momentum, so one can
deduce in a straight forward manner that for a D-wave, the wavefunction
would be doubly suppressed by $1/M^2$ in relation to a S-wave.
Then to calculate the exclusive decay of a D-wave quarkonium,
one would need to include not only the colour singlet valence
state and the next higher colour octet state now with the
heavy fermions in a P-wave, but the next-next higher
state must also be included for a consistent calculation. 

Finally as mentioned in the introduction, nucleon wavefunction
constructed from QCD sum rules has large disagreement
with the magnetic form factor measurements below 50 GeV$^2$.
Therefore while lacking an alternative method to derive wavefunction
in a fundamental way, we will have to satisfy with 
phenomenological constructions. We have shown together with
\cite{bolz&kroll2} that the so-constructed wavefunction and its 
generalization to the flavour octet and decuplet baryons provide a set 
of reasonable model wavefunctions. They would no doubt provide a useful
basis for future study of other exclusive processes involving 
baryons at moderate $Q^2$.

\section*{Acknowledgments}

The author would like to thank most of all P. Kroll for introducing him 
to the very interesting subjects of the hard scattering scheme and the colour 
octet picture of quarkonium. The author benefited greatly from his initial 
guidance of this research and from many useful discussions. The author also 
thanks the computational physics group of K. Schilling in Fachbereich 
Theoretische Physik, Universit\"at Wuppertal for using their valuable computer 
resources and for the Nuclear and Particle Physics Section and IASA 
institute at the University of Athens for kind hospitality where part of this 
work was done. All algebras, including those of the Dirac and colour, 
have been performed with the computer program FORM v1.0. written by 
J.A.M. Vermaseren. This work was supported by the European Training 
and Mobility of Researchers Programme under contract 
no. ERB-FMRX-CT96-0008.

\vfill
\eject

\appendix

\section{Appendix: Basic Graphs And Contributions For \\
colour octet}
\label{a:basic}

\subsection{Group 1}
\label{a:grp1}

\bfi[t]
\setlength{\unitlength}{1.0mm}
\bpi{(100,40)}

\put( 0,40){\line(1,0){160}}
\put( 0, 0){\line(1,0){160}}
\put( 0, 0){\line(0,1){40}}
\put(50, 0){\line(0,1){40}}
\put(160, 0){\line(0,1){40}}

\put(5,5){
\psfig{figure=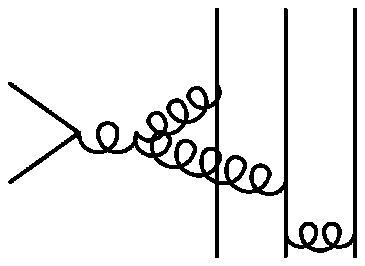,width=4.0cm}}

\put(60,30){Basic Propagators:}
\put(60,20){$  \frac{1}{\mc^2} \frac{1}{x_3 (1-y_1) \mc^2}
               \frac{1}{x_1 y_1 \mc^2} \frac{1}{(1-x_1)(1-y_1) \mc^2}
               \frac{1}{x_3 y_3 \mc^2}  $}
                      
\epi
\caption{Basic graph of Group 1}
\label{f:grp1}
\efi

\null
\begin{table}[h]
\centerline{
\ba[t]{|c|c|c|} \hline\hline
 \mbox{I.P.} & \multicolumn{2}{c|}{\mbox{Num. for graphs of Group 1 with}} \\
             & \multicolumn{2}{|c|}{(\l_1,\l_2,\l_3) = (+,+,-)} \\ \cline{2-3}
             & \mbox{\hspace{2.4cm} J=1 \hspace{2.4cm}} 
             & \mbox{J=2}                                       \\ \hline
 \mbox{U3} & \emptyset & \emptyset                              \\ \hline
 \mbox{L3} & \multicolumn{2}{c|}{
               \sqrt{2} (x_3-z)(y_3-z)(1+y_1-z) }               \\ \hline
 \mbox{U2} & \multicolumn{2}{c|}{
              -\frac{4 \sqrt{2}}{3} x_3
               \{ (x_2-z)(1+y_1-z) -(-1)^J 2 z (x_1-y_1) \} }   \\ \hline
 \mbox{M2} & \multicolumn{2}{c|}{
              -\frac{4 \sqrt{2}}{3} x_3 (x_3-z)(1+y_1-z) }      \\ \hline 
 \mbox{L2} & \multicolumn{2}{c|}{
               \frac{\sqrt{2}}{3} (x_3-z)(y_2-z)(1+y_1-z) }     \\ \hline
 \mbox{U1} & \multicolumn{2}{c|}{
               \frac{4 \sqrt{2}}{3} x_3
               \{ (x_1-z)(1-y_1-2 z) +(-1)^J z (x_1-y_1 +3(1-z)) \} }
                                                                \\ \hline
 \mbox{L1} & \multicolumn{2}{c|}{
              -\frac{4 \sqrt{2}}{3} x_3
               \{ (1+y_1-2 z) (y_1-z) -(-1)^J z (1-2 y_1 +z) \}}\\ \hline
 \mbox{G3} & \multicolumn{2}{c|}{
              -\sqrt{2} (x_3-y_3)(x_3-z)(1+y_1-z) }             \\ \hline
 \mbox{G2} & \multicolumn{2}{c|}{\ba[t]{c@{\,}l} 
               \frac{4 \sqrt{2}}{3} x_3 
               \{ & 2 (1-z^2) -y_1 (3+y_1-6 z) +x_1 (3-y_1-2 z) \\
                  &-(-1)^J 2 (x_1-y_1)(1-y_1-2 z) \} \ea  }     \\ \hline
 \mbox{G1} & \multicolumn{2}{c|}{\ba[t]{c@{\,}l} 
               \frac{4 \sqrt{2}}{3} x_3 
               \{ & 2 (x_1-y_1)(1+y_1-2 z)                      \\
                  &+(-1)^J ( 2z (2-z) +x_1 (1 -y_1+ 2 z) +y_1 (3 -y_1 -6 z)) \} \ea }
                                                                \\ \hline
 \mbox{4G} & \emptyset & -\frac{16 \sqrt{2}}{3} x_3             \\
 \hline\hline
\ea
}
\caption{(a) Numerators of the graphs of group 1 with helicity (+,+,--).}
\label{tab:grp1a}
\end{table}

\addtocounter{table}{-1}

\null
\begin{table}
\centerline{
\ba[t]{|c|c|c|} \hline\hline
 \mbox{I.P.} & \multicolumn{2}{c|}{\mbox{Num. for graphs of Group 1 with}} \\
             & \multicolumn{2}{|c|}{(\l_1,\l_2,\l_3) = (+,-,+)} \\ \cline{2-3}
             & \mbox{\hspace{2.4cm} J=1 \hspace{2.4cm}} 
             & \mbox{J=2}                                       \\ \hline
 \mbox{U3} & \emptyset & \emptyset                              \\ \hline
 \mbox{L3} & \multicolumn{2}{c|}{ (+,+,-) + (-,+,+) }           \\ \hline
 \mbox{U2} & \multicolumn{2}{c|}{
              -\frac{4 \sqrt{2}}{3} x_3 (x_2-z)(1+y_1-z) }      \\ \hline
 \mbox{M2} & \multicolumn{2}{c|}{
              -\frac{4 \sqrt{2}}{3} x_3 
               \{ (x_3-z)(1+y_1-z) +(-1)^J 2 (x_1-y_1)(1-y_1-z) \} }
                                                                \\ \hline
 \mbox{L2} & \multicolumn{2}{c|}{
               \frac{\sqrt{2}}{3} (x_3-z)(y_2-z)(1+y_1-z) }     \\ \hline
 \mbox{U1} & \multicolumn{2}{c|}{
               \frac{4 \sqrt{2}}{3} x_3
               \{ (x_1-z)(1-y_1-2 z) +(-1)^J z (x_1-y_1 +3(1-z)) \} }
                                                                \\ \hline
 \mbox{L1} & \multicolumn{2}{c|}{
              -\frac{4 \sqrt{2}}{3} x_3
               \{ (1+y_1-2 z) (y_1-z) -(-1)^J z (1-2 y_1 +z) \}}\\ \hline
 \mbox{G3} & \multicolumn{2}{c|}{ (+,+,-) + (-,+,+) }           \\ \hline
 \mbox{G2} & \multicolumn{2}{c|}{\ba[t]{c@{\,}l} 
               \frac{4 \sqrt{2}}{3} x_3 
               \{ & 2 (1-z^2) -y_1 (3+y_1-6 z) +x_1 (3-y_1-2 z) \\
                  &-(-1)^J 2 (x_1-y_1)(1-y_1-2 z) \} \ea  }     \\ \hline
 \mbox{G1} & \multicolumn{2}{c|}{\ba[t]{c@{\,}l} 
               \frac{4 \sqrt{2}}{3} x_3 
               \{ & 2 (x_1-y_1)(1+y_1-2 z)                      \\
                  &+(-1)^J ( 2z (2-z) +x_1 (1 -y_1+ 2 z) +y_1 (3 -y_1 -6 z)) \} \ea }
                                                                \\ \hline 
 \mbox{4G} & \emptyset & -\frac{16 \sqrt{2}}{3} x_3             \\
 \hline\hline
\ea
}
\caption{(b) Numerators of the graphs of group 1 with helicity (+,--,+).}
\label{tab:grp1b}
\end{table}

\addtocounter{table}{-1}

\null
\begin{table}
\centerline{
\ba[t]{|c|c|c|c|} \hline\hline
 \mbox{I.P.} & \multicolumn{2}{c|}{\mbox{Num. for graphs of Group 1 with}} 
             & \mbox{X-Prop.}                                   \\
             & \multicolumn{2}{|c|}{(\l_1,\l_2,\l_3) = (-,+,+)} 
             &                                                  \\ \cline{2-3}
             & \mbox{\hspace{1.2cm} J=1 \hspace{1.2cm}} 
             & \mbox{J=2} &                                     \\ \hline
 \mbox{U3} & \emptyset & \emptyset                              
           & \mbox{n. g.}                                       \\ \hline
 \mbox{L3} & \multicolumn{2}{c|}{
            -(-1)^J 2\sqrt{2} z(x_1-y_1)(1-y_1-z) } 
           & \frac{1}{-z(y_3-z) \mc^2 }                         \\ \hline
 \mbox{U2} & \emptyset & \emptyset                              
           & \frac{1}{-z(x_2-z) \mc^2 }                         \\ \hline
 \mbox{M2} & \emptyset & \emptyset                              
           & \frac{1}{x_3 (1-y_1) \mc^2 }                       \\ \hline
 \mbox{L2} & \multicolumn{2}{c|}{
             (-1)^J \frac{2 \sqrt{2}}{3} z (x_1-y_1)(1-y_1-z) }      
           & \frac{1}{-z(y_2-z) \mc^2 }                         \\ \hline 
 \mbox{U1} & \emptyset & \emptyset                              
           & \frac{1}{-z(x_1-z) \mc^2 }                         \\ \hline
 \mbox{L1} & \emptyset & \emptyset                              
           & \frac{1}{-z(y_1-z) \mc^2 }                         \\ \hline
 \mbox{G3} & \multicolumn{2}{c|}{
            -(-1)^J \sqrt{2} (x_1-y_1)(x_3+z)(1-y_1-z) }       
           & \frac{1}{x_3 y_3 \mc^2 }                           \\ \hline
 \mbox{G2} & \emptyset & \emptyset                              
           & \frac{1}{(1-x_1)(1-y_1) \mc^2 }                    \\ \hline
 \mbox{G1} & \emptyset & \emptyset
           & \frac{1}{x_1 y_1 \mc^2 }                           \\ \hline
 \mbox{4G} & \emptyset & \emptyset
           & \frac{1}{\mc^2 }                                   \\
 \hline\hline
\ea
}
\caption{(c) Numerators and the additional propagators of the graphs of 
group 1 with helicity (--,+,+).}
\label{tab:grp1c}
\end{table}

\vfill
\eject

\subsection{Group 2 and Group 2'}
\label{a:grp22}

\subsubsection{Group 2}
\label{a:grp2}

\bfi[t]
\setlength{\unitlength}{1.0mm}
\bpi{(100,40)}

\put( 0,40){\line(1,0){160}}
\put( 0, 0){\line(1,0){160}}
\put( 0, 0){\line(0,1){40}}
\put(50, 0){\line(0,1){40}}
\put(160, 0){\line(0,1){40}}

\put(5,5){
\psfig{figure=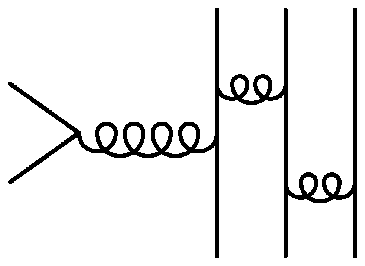,width=4.0cm}}

\put(60,30){Basic Propagators:}
\put(60,20){$  \frac{1}{\mc^2} \frac{1}{(1-y_1) \mc^2} 
               \frac{1}{(1-x_1)(1-y_1) \mc^2} \frac{1}{x_3 (1-y_1) \mc^2}
               \frac{1}{x_3 y_3 \mc^2} $}
                      
\epi
\caption{Basic graph of Group 2}
\label{f:grp2}
\efi

\null
\begin{table}[h]
\centerline{
\ba[t]{|c|c|c|c|c|} \hline\hline
 \mbox{I.P.} & \multicolumn{4}{c|}
               {\mbox{Num. for graphs of Group 2 with}\; (\l_1,\l_2,\l_3)} 
                                                                \\ \cline{2-5}
             & \multicolumn{2}{c|}{(+,+,-)}
             & \multicolumn{2}{c|}{(+,-,+)}                     \\ \cline{2-5}
             & \mbox{\hspace{0.5cm} J=1 \hspace{0.5cm}} 
             & \mbox{J=2} 
             & \mbox{\hspace{0.6cm} J=1 \hspace{0.6cm}} 
             & \mbox{J=2}                                       \\ \hline
 \mbox{U3} & \emptyset & \emptyset & \emptyset & \emptyset      \\ \hline
 \mbox{L3} & \emptyset & \emptyset 
           & \multicolumn{2}{c|}{
             -(-1)^J \frac{32 \sqrt{2}}{27} z (1-y_1-z)^2 }     \\ \hline 
 \mbox{U2} & \multicolumn{2}{c|}{
             (-1)^J \frac{40 \sqrt{2}}{27} x_3 z (1-y_1-z) }
           & \emptyset & \emptyset                              \\ \hline
 \mbox{M2} & \emptyset & \emptyset
           & \multicolumn{2}{c|}{
             -(-1)^J \frac{32 \sqrt{2}}{27} x_3 (1-y_1-z)^2 }   \\ \hline
 \mbox{L2} & \emptyset & \emptyset & \emptyset & \emptyset      \\ \hline
 \mbox{U1} & \multicolumn{4}{c|}{
             (-1)^J \frac{8 \sqrt{2}}{27} x_3 z (1-y_1-z) }     \\ \hline
 \mbox{M1} & \emptyset & \emptyset & \emptyset & \emptyset      \\ \hline
 \mbox{L1} & \emptyset & \emptyset & \emptyset & \emptyset      \\ \hline
 \mbox{G3} & \emptyset & \emptyset & \emptyset & \emptyset      \\ \hline
 \mbox{G2} & \multicolumn{4}{c|}{\ba[t]{c@{\,}l} 
             \frac{4 \sqrt{2}}{3} x_3 
             \{ & (1-z)(2(1-y_1)-z)                             \\
                & -(-1)^J (1 -y_1 -2 z)(1 -y_1 -z) \} \ea }     \\ 
 \hline\hline
\ea
}
\caption{(a) Numerators of the graphs of group 2 with helicity 
(+,+,--) and (+,--,+).}
\label{tab:grp2a}
\end{table}

\addtocounter{table}{-1}

\null
\begin{table}
\centerline{
\ba[t]{|c|c|c|c|} \hline\hline
 \mbox{I.P.} & \multicolumn{2}{c|}{\mbox{Num. for graphs of Group 2 with}} 
             & \mbox{X-Prop.}                                   \\
             & \multicolumn{2}{|c|}{(\l_1,\l_2,\l_3) = (-,+,+)} 
             &                                                  \\ \cline{2-3}
             & \mbox{\hspace{0.6cm} J=1 \hspace{0.6cm}} 
             & \mbox{J=2} &                                     \\ \hline
 \mbox{U3} & \emptyset & \emptyset                              
           & \mbox{n. g.}                                       \\ \hline
 \mbox{L3} & \multicolumn{2}{c|}{
             -(-1)^J \frac{32 \sqrt{2}}{27} z (1-y_1-z)^2 }     
           & \frac{1}{-z(y_3-z) \mc^2 }                         \\ \hline
 \mbox{U2} & \emptyset & \emptyset                              
           & \frac{1}{-z(x_2-z) \mc^2 }                         \\ \hline
 \mbox{M2} & \emptyset & \emptyset
           & \frac{1}{x_3 (1-y_1) \mc^2 }                       \\ \hline
 \mbox{L2} & \multicolumn{2}{c|}{
              (-1)^J \frac{32 \sqrt{2}}{27} z (1-y_1-z)^2 }                              
           & \frac{1}{-z(y_2-z) \mc^2 }                         \\ \hline
 \mbox{U1} & \emptyset & \emptyset                              
           & \frac{1}{-z(x_1-z) \mc^2 }                         \\ \hline
 \mbox{M1} & \emptyset & \emptyset                              
           & \mbox{n. g.}                                       \\ \hline
 \mbox{L1} & \emptyset & \emptyset                             
           & \mbox{n. g.}                                       \\ \hline
 \mbox{G3} & \emptyset & \emptyset                      
           & \mbox{n. g.}                                       \\ \hline
 \mbox{G2} & \emptyset & \emptyset                      
           & \frac{1}{(1-x_1)(1-y_1) \mc^2}                     \\ 
 \hline\hline
\ea
}
\caption{(b) Numerators and the additional propagators of the graphs of 
group 2 with helicity (--,+,+).}
\label{tab:grp2b}
\end{table}

\addtocounter{table}{-1}

\clearpage
\vfill
\eject

\subsubsection{Group 2'}
\label{a:grp2d}

\bfi[t]
\setlength{\unitlength}{1.0mm}
\bpi{(100,40)}

\put( 0,40){\line(1,0){160}}
\put( 0, 0){\line(1,0){160}}
\put( 0, 0){\line(0,1){40}}
\put(50, 0){\line(0,1){40}}
\put(160, 0){\line(0,1){40}}

\put(5,5){
\psfig{figure=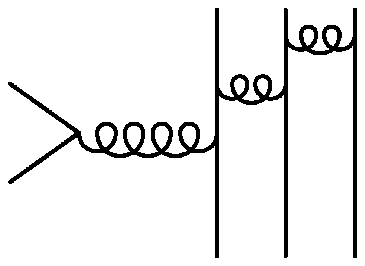,width=4.0cm}}

\put(60,30){Basic Propagators:}
\put(60,20){$  \frac{1}{\mc^2} \frac{1}{(1-y_1) \mc^2} 
               \frac{1}{(1-x_1)(1-y_1) \mc^2} \frac{1}{(1-x_1) y_3 \mc^2}
               \frac{1}{x_3 y_3 \mc^2} $}
                      
\epi
\caption{Basic graph of Group 2'}
\label{f:grp2d}
\efi

\null
\begin{table}[h]
\centerline{
\ba[t]{|c|c|c|} \hline\hline
 \mbox{I.P.} & \multicolumn{2}{c|}{\mbox{Num. for graphs of Group 2' with}} \\
             & \multicolumn{2}{|c|}{(\l_1,\l_2,\l_3) = (+,+,-)} \\ \cline{2-3}
             & \mbox{\hspace{1.6cm} J=1 \hspace{1.6cm}} 
             & \mbox{J=2}                                       \\ \hline
 \mbox{U3} & \multicolumn{2}{c|}{
               \frac{22 \sqrt{2}}{27} (1-z)(x_3-z)(y_3-z) }     \\ \hline
 \mbox{L3} & \multicolumn{2}{c|}{
              -\frac{32 \sqrt{2}}{27} (1-z)(y_3-z)^2 }          \\ \hline
 \mbox{U2} & \multicolumn{2}{c|}{
              -\frac{14 \sqrt{2}}{27} (1-z)(x_2-z)(y_3-z) }     \\ \hline
 \mbox{M2} & \multicolumn{2}{c|}{
              -\frac{40 \sqrt{2}}{27}  y_3 (1-z)(y_3-z) }       \\ \hline
 \mbox{L2} & \multicolumn{2}{c|}{
              -\frac{32 \sqrt{2}}{27} y_3 
               \{ (1-z)(y_2-z) +(-1)^J z(1-y_1-z) \} }          \\ \hline
 \mbox{U1} & \multicolumn{2}{c|}{
              -\frac{8 \sqrt{2}}{27} y_3 (1-z)(x_1-z) }         \\ \hline
 \mbox{M1} & \multicolumn{2}{c|}{
               \frac{64 \sqrt{2}}{27} y_3
               \{ (1-z) -(-1)^J (1-y_1-z) \} }                  \\ \hline
 \mbox{L1} & \multicolumn{2}{c|}{
               \frac{64 \sqrt{2}}{27} y_3 \{ y_1-z +(-1)^J z \}}\\ \hline
 \mbox{G3} & \multicolumn{2}{c|}{
               2 \sqrt{2} (1-z)(x_3-y_3)(y_3-z) }               \\ \hline
 \mbox{G2} & \multicolumn{2}{c|}{\ba[t]{c@{\,}l} 
              -\frac{4 \sqrt{2}}{3} y_3 
               \{ & (1-z)(x_1-y_1 -3 z)                         \\
                  & -(-1)^J (1-x_1-2 z)(1-y_1-z) \} \ea }       \\ 
 \hline\hline
\ea
}
\caption{(a') Numerators of the graphs of group 2' with helicity 
(+,+,--).}
\label{tab:grp2da}
\end{table}

\addtocounter{table}{-1}

\null 
\begin{table}
\centerline{
\ba[t]{|c|c|c|} \hline\hline
 \mbox{I.P.} & \multicolumn{2}{c|}{\mbox{Num. for graphs of Group 2' with}} \\
             & \multicolumn{2}{|c|}{(\l_1,\l_2,\l_3) = (+,-,+)} \\ \cline{2-3}
             & \mbox{\hspace{1.4cm} J=1 \hspace{1.4cm}} 
             & \mbox{J=2}                                       \\ \hline
 \mbox{U3} & \multicolumn{2}{c|}{ (+,+,-) + (-,+,+) }           \\ \hline
 \mbox{L3} & \multicolumn{2}{c|}{
              -\frac{32 \sqrt{2}}{27} (1-z)(y_3-z)^2 }          \\ \hline
 \mbox{U2} & \multicolumn{2}{c|}{
              -\frac{14 \sqrt{2}}{27} (1-z)(x_2-z)(y_3-z) }     \\ \hline
 \mbox{M2} & \multicolumn{2}{c|}{\ba[t]{c@{\,}l} 
              -\frac{40 \sqrt{2}}{27} y_3 
               \{ & (1-z)(y_3-z)                                \\
                  & -(-1)^J (1-x_1-z)(1-y_1-z) \} \ea }         \\ \hline 
 \mbox{L2} & \multicolumn{2}{c|}{
              -\frac{32 \sqrt{2}}{27} y_3 (1-z)(y_2-z) }        \\ \hline
 \mbox{U1} & \multicolumn{2}{c|}{
              -\frac{8 \sqrt{2}}{27} y_3 (1-z)(x_1-z) }         \\ \hline
 \mbox{M1} & \multicolumn{2}{c|}{
               \frac{64 \sqrt{2}}{27} y_3
               \{ (1-z) -(-1)^J (1-y_1-z) \} }                  \\ \hline
 \mbox{L1} & \multicolumn{2}{c|}{
               \frac{64 \sqrt{2}}{27} y_3 \{ y_1-z +(-1)^J z \}}\\ \hline
 \mbox{G3} & \multicolumn{2}{c|}{ (+,+,-) + (+,-,+) }           \\ \hline
 \mbox{G2} & \multicolumn{2}{c|}{\ba[t]{c@{\,}l}           
              -\frac{4 \sqrt{2}}{3} y_3 
               \{ & (1-z)(x_1-y_1 -3 z)                         \\
                  & -(-1)^J (1-x_1-2 z)(1-y_1-z) \} \ea }       \\
 \hline\hline
\ea
}
\caption{(b') Numerators of the graphs of group 2' with helicity (+,--,+).}
\label{tab:grp2db}
\end{table}

\addtocounter{table}{-1}

\null 
\begin{table}
\centerline{
\ba[t]{|c|c|c|c|} \hline\hline
 \mbox{I.P.} & \multicolumn{2}{c|}{\mbox{Num. for graphs of Group 2' with}} 
             & \mbox{X-Prop.}                                   \\
             & \multicolumn{2}{|c|}{(\l_1,\l_2,\l_3) = (-,+,+)} 
             &                                                  \\ \cline{2-3}
             & \mbox{\hspace{1.2cm} J=1 \hspace{1.2cm}} 
             & \mbox{J=2} &                                     \\ \hline
 \mbox{U3} & \multicolumn{2}{c|}{
        (-1)^J \frac{22 \sqrt{2}}{27} z (1-x_1-z)(1-y_1-z) }    
           & \frac{1}{-z(x_3-z) \mc^2 }                         \\ \hline 
 \mbox{L3} & \emptyset & \emptyset 
           & \frac{1}{-z(y_3-z) \mc^2 }                         \\ \hline
 \mbox{U2} & \multicolumn{2}{c|}{
        (-1)^J \frac{14 \sqrt{2}}{27} z(1-x_1-z)(1-y_1-z) } 
           & \frac{1}{-z(x_2-z) \mc^2 }                         \\ \hline
 \mbox{M2} & \emptyset & \emptyset 
           & \frac{1}{(1-x_1) y_3 \mc^2 }                       \\ \hline  
 \mbox{L2} & \emptyset & \emptyset 
           & \frac{1}{-z(y_2-z) \mc^2 }                         \\ \hline
 \mbox{U1} & \emptyset & \emptyset 
           & \frac{1}{-z(x_1-z) \mc^2 }                         \\ \hline  
 \mbox{M1} & \emptyset & \emptyset 
           & \frac{1}{(1-y_1) \mc^2 }                           \\ \hline  
 \mbox{L1} & \emptyset & \emptyset 
           & \frac{1}{-z(y_1-z) \mc^2 }                         \\ \hline  
 \mbox{G3} & \multicolumn{2}{c|}{
             (-1)^J \sqrt{2} (1-x_1-z)(1-y_1-z)(y_3+z) }        
           & \frac{1}{x_3 y_3 \mc^2 }                           \\ \hline
 \mbox{G2} & \emptyset & \emptyset 
           & \frac{1}{(1-x_1)(1-y_1) \mc^2 }                    \\
 \hline\hline
\ea
}
\caption{(c') Numerators and the additional propagators of the graphs of 
group 2' with helicity (--,+,+).}
\label{tab:grp2dc}
\end{table}

\vfill
\eject

\subsection{Group 3}
\label{a:grp3}

\bfi[t]
\setlength{\unitlength}{1.0mm}
\bpi{(100,40)}

\put( 0,40){\line(1,0){160}}
\put( 0, 0){\line(1,0){160}}
\put( 0, 0){\line(0,1){40}}
\put(50, 0){\line(0,1){40}}
\put(160, 0){\line(0,1){40}}

\put(5,5){
\psfig{figure=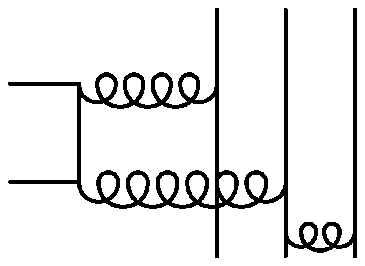,width=4.0cm}}

\put(60,30){Basic Propagators:}
\put(60,20){$  \frac{1}{\{ (z_1-x_1)(z_1-y_1) -1/4 \} \mc^2} 
               \frac{1}{x_3 (1-y_1) \mc^2}
               \frac{1}{x_1 y_1 \mc^2}  $}
\put(60,12){$  \times \frac{1}{(1-x_1)(1-y_1) \mc^2}
                      \frac{1}{x_3 y_3 \mc^2}  $}
\epi
\caption{Basic graph of Group 3}
\label{f:grp3}
\efi

\null 
\begin{table}[h]
\centerline{
\ba[t]{|c|c|c|} \hline\hline
 \mbox{I.P.} & \multicolumn{2}{c|}{\mbox{Num. for graphs of Group 3 with}} \\
             & \multicolumn{2}{|c|}{(\l_1,\l_2,\l_3) = (+,+,-)} \\ \cline{2-3}
             & \mbox{\hspace{1.9cm} J=1 \hspace{1.9cm}} 
             & \mbox{J=2}                                       \\ \hline
 \mbox{U3} & \multicolumn{2}{c|}{
              -\frac{10 \sqrt{2}}{27} (x_3-z)^2 (2 y_1+z) }     \\ \hline
 \mbox{L3} & \multicolumn{2}{c|}{
               \frac{8  \sqrt{2}}{27} (x_3-z)(y_3-z)(2 y_1+z) } \\ \hline              
 \mbox{U2} & \multicolumn{2}{c|}{
               -\frac{4 \sqrt{2}}{27} x_3 
                \{ (x_2-z)(2 y_1+z) -(-1)^J 2z (x_1-y_1) \}  }  \\ \hline
 \mbox{M2} & \multicolumn{2}{c|}{
               -\frac{14 \sqrt{2}}{27} x_3 (x_3-z) (2 y_1+z) }  \\ \hline
 \mbox{L2} & \multicolumn{2}{c|}{
               -\frac{4 \sqrt{2}}{27} (x_3-z)(y_2-z)(2 y_1+z)}  \\ \hline
 \mbox{U1} & \multicolumn{2}{c|}{
                \frac{14 \sqrt{2}}{27} x_3 
                \{ (x_1-z)(2y_1-z)+(-1)^J 2 z \} }              \\ \hline
 \mbox{L1} & \multicolumn{2}{c|}{
               -\frac{4 \sqrt{2}}{27} x_3 
                \{ (2 y_1-z)(y_1-z)-(-1)^J 2 z (1-2 y_1+z) \} } \\ \hline
 \mbox{G3} & \multicolumn{2}{c|}{
               -\frac{2 \sqrt{2}}{3} (x_3-z)(x_3-y_3)(2 y_1+z)} \\ \hline
 \mbox{G2} & \multicolumn{2}{c|}{\ba[t]{c@{\,}l} 
               -\frac{\sqrt{2}}{3} x_3
                \{ & (x_1+y_1-2 z)(2 y_1+z)-2 (2 x_1+z)         \\
                   & +(-1)^J (x_1 -y_1) (1-y_1 -2 z) \} \ea }   \\ \hline
 \mbox{G1} & \multicolumn{2}{c|}{\ba[t]{c@{\,}l} 
                \frac{2 \sqrt{2}}{3} x_3
                \{ & (x_1-y_1)(2 y_1-z)                         \\ 
                   & -(-1)^J ((x_1+z)(2 y_1-z)-(x_1+y_1+2 z)) \} \ea }
                                                                \\ \hline
 \mbox{Q}  & \multicolumn{2}{c|}{\ba[t]{c@{\,}l} 
               -\frac{\sqrt{2}}{27} x_3 
                \{ & \half (2y_1+z)(2y_1-z) - (x_1+y_1+z)       \\  
          & +(-1)^J (\half (2x_1+z)(2y_1-z) - (x_1+y_1+z)  ) \} \ea }
                                                                \\ 
 \hline\hline
\ea
}
\caption{(a) Numerators of the graphs of group 3 with helicity (+,+,--).}
\label{tab:grp3a}
\end{table}

\addtocounter{table}{-1}

\null 
\begin{table}
\centerline{
\ba[t]{|c|c|c|} \hline\hline
 \mbox{I.P.} & \multicolumn{2}{c|}{\mbox{Num. for graphs of Group 3 with}} \\
             & \multicolumn{2}{|c|}{(\l_1,\l_2,\l_3) = (+,-,+)} \\ \cline{2-3}
             & \mbox{\hspace{2.2cm} J=1 \hspace{2.2cm}} 
             & \mbox{J=2}                                       \\ \hline
 \mbox{U3} & \multicolumn{2}{c|}{
              -\frac{10 \sqrt{2}}{27} (x_3-z)^2 (2 y_1+z) }     \\ \hline
 \mbox{L3} & \multicolumn{2}{c|}{
               \frac{8  \sqrt{2}}{27} 
             \{ (x_3-z)(y_3-z)(2y_1+z)-(-1)^J 2z(x_1-y_1)(1-y_1-z) \} }
                                                                \\ \hline
 \mbox{U2} & \multicolumn{2}{c|}{
              -\frac{4  \sqrt{2}}{27} x_3(x_2-z)(2y_1+z)  }     \\ \hline
 \mbox{M2} & \multicolumn{2}{c|}{
               -\frac{14 \sqrt{2}}{27} x_3
             \{ (x_3-z)(2y_1+z)+(-1)^J 2 (x_1-y_1)(1-y_1-z) \} }\\ \hline
 \mbox{L2} & \multicolumn{2}{c|}{
               -\frac{4 \sqrt{2}}{27} (x_3-z)(y_2-z)(2 y_1+z) } \\ \hline
 \mbox{U1} & \multicolumn{2}{c|}{
                \frac{14 \sqrt{2}}{27} x_3 
                \{ (x_1-z)(2y_1-z)+(-1)^J 2 z \} }              \\ \hline
 \mbox{L1} & \multicolumn{2}{c|}{
               -\frac{4 \sqrt{2}}{27} x_3 
                \{ (2 y_1-z)(y_1-z)-(-1)^J 2 z (1-2 y_1+z) \} } \\ \hline
 \mbox{G3} & \multicolumn{2}{c|}{ (+,+,-) + (-,+,+) }           \\ \hline
 \mbox{G2} & \multicolumn{2}{c|}{\ba[t]{c@{\,}l} 
               -\frac{\sqrt{2}}{3} x_3
                \{ & (x_1+y_1-2 z)(2 y_1+z)-2 (2 x_1+z)         \\
                   & +(-1)^J (x_1 -y_1) (1-y_1 -2 z) \} \ea }   \\ \hline
 \mbox{G1} & \multicolumn{2}{c|}{\ba[t]{c@{\,}l} 
                \frac{2 \sqrt{2}}{3} x_3
                \{ & (x_1-y_1)(2 y_1-z)                         \\ 
                   & -(-1)^J ((x_1+z)(2 y_1-z)-(x_1+y_1+2 z)) \} \ea }
                                                                \\ \hline
 \mbox{Q}  & \multicolumn{2}{c|}{\ba[t]{c@{\,}l} 
               -\frac{\sqrt{2}}{27} x_3 
                \{ & \half (2y_1+z)(2y_1-z) - (x_1+y_1+z)       \\  
          & +(-1)^J (\half (2x_1+z)(2y_1-z) - (x_1+y_1+z)  ) \} \ea }
                                                                \\ 
 \hline\hline
\ea
}
\caption{(b) Numerators of the graphs of group 3 with helicity (+,--,+).}
\label{tab:grp3b}
\end{table}

\addtocounter{table}{-1}

\null 
\begin{table}
\centerline{
\ba[t]{|c|c|c|c|} \hline\hline
 \mbox{I.P.} & \multicolumn{2}{c|}{\mbox{Num. for graphs of Group 3 with}} 
             & \mbox{X-Prop.}                                   \\
             & \multicolumn{2}{|c|}{(\l_1,\l_2,\l_3) = (-,+,+)} 
             &                                                  \\ \cline{2-3}
             & \mbox{\hspace{1.2cm} J=1 \hspace{1.2cm}} 
             & \mbox{J=2} &                                     \\ \hline
 \mbox{U3} & \emptyset & \emptyset
           & \frac{1}{-z(x_3-z) \mc^2}                          \\ \hline
 \mbox{L3} & \multicolumn{2}{c|}{
            -(-1)^J \frac{16 \sqrt{2}}{27} z(x_1-y_1)(1-y_1-z) }
           & \frac{1}{-z(y_3-z) \mc^2}                          \\ \hline
 \mbox{U2} & \emptyset & \emptyset
           & \frac{1}{-z(x_2-z) \mc^2}                          \\ \hline
 \mbox{M2} & \emptyset & \emptyset
           & \frac{1}{x_3 (1-y_1) \mc^2}                        \\ \hline
 \mbox{L2} & \multicolumn{2}{c|}{
            -(-1)^J \frac{8 \sqrt{2}}{27} z (x_1-y_1)(1-y_1-z) }   
           & \frac{1}{-z(y_2-z) \mc^2}                          \\ \hline
 \mbox{U1} & \emptyset & \emptyset
           & \frac{1}{-z(x_1-z) \mc^2}                          \\ \hline
 \mbox{L1} & \emptyset & \emptyset
           & \frac{1}{-z(y_1-z) \mc^2}                          \\ \hline
 \mbox{G3} & \multicolumn{2}{c|}{
            -(-1)^J\frac{2 \sqrt{2}}{3} (x_3+z)(x_1-y_1)(1-y_1-z)  } 
           & \frac{1}{x_3 y_3 \mc^2}                            \\ \hline
 \mbox{G2} & \emptyset & \emptyset
           & \frac{1}{(1-x_1)(1-y_1) \mc^2}                     \\ \hline
 \mbox{G1} & \emptyset & \emptyset
           & \frac{1}{x_1 y_1 \mc^2}                            \\ \hline
 \mbox{Q } & \emptyset & \emptyset
           & \frac{1}{[(x_1-z_1)(y_1-z_1)-1/4] \mc^2}           \\ 
 \hline\hline
\ea
}
\caption{(c) Numerators and the additional propagators of the graphs of 
group 3 with helicity (--,+,+).}
\label{tab:grp3c}
\end{table}

\vfill
\eject

\subsection{Group 4}
\label{a:grp4}

\bfi[t]
\setlength{\unitlength}{1.0mm}
\bpi{(100,40)}

\put( 0,40){\line(1,0){160}}
\put( 0, 0){\line(1,0){160}}
\put( 0, 0){\line(0,1){40}}
\put(50, 0){\line(0,1){40}}
\put(160, 0){\line(0,1){40}}

\put(5,5){
\psfig{figure=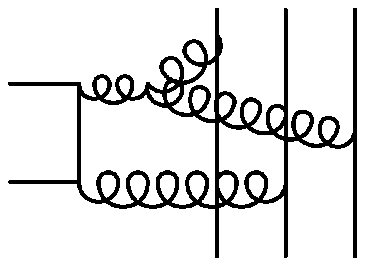,width=4.0cm}}

\put(60,30){Basic Propagators:}
\put(60,20){$  \frac{1}{\{ (z_2-x_2)(z_2-y_2) - 1/4 \} \mc^2} 
               \frac{1}{(1-x_2)(1-y_2) \mc^2 +\r^2 }
               \frac{1}{x_1 y_1 \mc^2}  $}
\put(60,12){$  \times \frac{1}{x_2 y_2 \mc^2}
                      \frac{1}{x_3 y_3 \mc^2}  $}
\epi
\caption{Basic graph of Group 4}
\label{f:grp4}
\efi

\null 
\begin{table}[h]
\centerline{
\ba[t]{|c|c|c|c|c|} \hline\hline
 \mbox{I.P.} & \multicolumn{4}{c|}
               {\mbox{Num. for graphs of Group 4 with}\; (\l_1,\l_2,\l_3)} 
                                                                \\ \cline{2-5}
             & \multicolumn{2}{c|}{(+,+,-)}
             & \multicolumn{2}{c|}{(+,-,+)}                     \\ \cline{2-5}
             & \mbox{\hspace{1.5cm} J=1 \hspace{1.5cm}} 
             & \mbox{J=2} 
             & \mbox{\hspace{1.0cm} J=1 \hspace{1.0cm}} 
             & \mbox{J=2}                                       \\ \hline
 \mbox{U3} & \multicolumn{2}{|c|}{\ba[t]{c@{\,}l}
              -\frac{\sqrt{2}}{3} (x_3-z) 
           \{ & (x_1-x_3+z)+(y_1-y_3+z)                         \\
              & -2 (x_1-x_3+z)(z_2-y_2)                         \\
              & -2 (y_1-y_3+z)(z_2-x_2)  \} \ea }
           & \multicolumn{2}{c|}
             {(-1)^J \frac{2 \sqrt{2}}{3} z(x_2-y_2) (2 x_1+x_3-z) }
                                                                \\ \hline 
 \mbox{L3} & \multicolumn{2}{|c|}{\ba[t]{c@{\,}l}
               \frac{\sqrt{2}}{12} (y_3-z) 
           \{ & (x_1-x_3+z)+(y_1-y_3+z)                         \\
              & -2 (x_1-x_3+z)(z_2-y_2)                         \\
              & -2 (y_1-y_3+z)(z_2-x_2)  \} \ea }
           & \multicolumn{2}{c|}
             {(-1)^J \frac{\sqrt{2}}{6} z(x_2-y_2) (2 y_1+y_3-z) }
                                                                \\ \hline 
 \mbox{U2} & \emptyset & \emptyset & \emptyset & \emptyset      \\ \hline 
 \mbox{L2} & \emptyset & \emptyset & \emptyset & \emptyset      \\ \hline 
 \mbox{U1} & \multicolumn{2}{c|}
              {(+, -, +) + (-, +, +)}
           & \multicolumn{2}{c|}
              {(-1)^J \frac{2 \sqrt{2}}{3} z(x_2-y_2) (x_1+2 x_3-z) }
                                                                \\ \hline 
 \mbox{L1} & \multicolumn{2}{c|}
              {(+, -, +) + (-, +, +)}
           & \multicolumn{2}{c|}
              {(-1)^J \frac{\sqrt{2}}{6} z(x_2-y_2) (y_1+2 y_3-z) }
                                                                \\ \hline 
 \mbox{G3} & \multicolumn{2}{|c|}{\ba[t]{c@{\,}l}
              -\frac{5 \sqrt{2}}{12} (x_3-y_3) 
           \{ & (x_1-x_3+z)+(y_1-y_3+z)                         \\
              & -2 (x_1-x_3+z)(z_2-y_2)                         \\
              & -2 (y_1-y_3+z)(z_2-x_2)  \} \ea }
           & \multicolumn{2}{|c|}{\ba[t]{c@{\,}l}             
             (-1)^J & \frac{5 \sqrt{2}}{12} (x_2-y_2)           \\
    \times & \{ (2 x_1+x_3-z)(y_3+z)                            \\
           & \  (2 y_1+y_3-z)(x_3+z)     \} \ea }               \\ \hline                
 \mbox{G2} & \emptyset & \emptyset & \emptyset & \emptyset      \\ \hline 
 \mbox{G1} & \multicolumn{2}{c|}
              {(+, -, +) + (-, +, +)}
           & \multicolumn{2}{|c|}{\ba[t]{c@{\,}l}             
             (-1)^J & \frac{5 \sqrt{2}}{12} (x_2-y_2)           \\
    \times & \{ (2 x_3+x_1-z)(y_1+z)                            \\
           & \  (2 y_3+y_1-z)(x_1+z)     \} \ea }               \\ \hline                
 \mbox{GR} & \emptyset & \emptyset & \emptyset & \emptyset      \\ \hline
 \mbox{4G} & -\frac{5}{ \sqrt{2}} (x_2-y_2)                     
           & -\frac{5}{3\sqrt{2}} (x_2-y_2)                     
           & \multicolumn{2}{c|}
             {-(-1)^J \frac{5 \sqrt{2}}{3} (x_2-y_2)}           \\ \hline
 \mbox{Q } & \emptyset & \emptyset & \emptyset & \emptyset      \\ 
 \hline\hline
\ea
}
\caption{(a) Numerators of the graphs of group 4 with helicity (+,+,-) 
and (+,--,+).}
\label{tab:grp4a}
\end{table}

\addtocounter{table}{-1}

\null 
\begin{table}[h]
\centerline{
\ba[t]{|c|c|c|c|} \hline\hline
 \mbox{I.P.} & \multicolumn{2}{c|}{\mbox{Num. for graphs of Group 4 with}} 
             & \mbox{X-Prop.}                                   \\
             & \multicolumn{2}{|c|}{(\l_1,\l_2,\l_3) = (-,+,+)} 
             &                                                  \\ \cline{2-3}
             & \mbox{\hspace{1.5cm} J=1 \hspace{1.5cm}} 
             & \mbox{J=2} &                                     \\ \hline
 \mbox{U3} & \multicolumn{2}{c|}
              {(+, +, -) + (+, -, +)}                             
           & \frac{1}{-z(x_3-z) \mc^2}                          \\ \hline
 \mbox{L3} & \multicolumn{2}{c|}
              {(+, +, -) + (+, -, +)}
           & \frac{1}{-z(y_3-z) \mc^2}                          \\ \hline
 \mbox{U2} & \emptyset & \emptyset  & \mbox{n. g.}              \\ \hline
 \mbox{L2} & \emptyset & \emptyset  & \mbox{n. g.}              \\ \hline
 \mbox{U1} & \multicolumn{2}{|c|}{\ba[t]{c@{\,}l}
              -\frac{\sqrt{2}}{3} (x_1-z) 
           \{ & (x_3-x_1+z)+(y_3-y_1+z)                         \\
              & -2 (x_3-x_1+z)(z_2-y_2)                         \\
              & -2 (y_3-y_1+z)(z_2-x_2)  \} \ea }
           & \frac{1}{-z(x_1-z) \mc^2}                          \\ \hline
 \mbox{L1} & \multicolumn{2}{|c|}{\ba[t]{c@{\,}l}
               \frac{\sqrt{2}}{12} (y_1-z) 
           \{ & (x_3-x_1+z)+(y_3-y_1+z)                         \\
              & -2 (x_3-x_1+z)(z_2-y_2)                         \\
              & -2 (y_3-y_1+z)(z_2-x_2)  \} \ea }
           & \frac{1}{-z(y_1-z) \mc^2}                          \\ \hline
 \mbox{G3} & \multicolumn{2}{c|}
              {(+, +, -) + (+, -, +)}                             
           & \frac{1}{x_3 y_3 \mc^2}                            \\ \hline
 \mbox{G2} & \emptyset & \emptyset & \mbox{n. g.}               \\ \hline
 \mbox{G1} & \multicolumn{2}{|c|}{\ba[t]{c@{\,}l}
              -\frac{5 \sqrt{2}}{12} (x_1-y_1) 
           \{ & (x_3-x_1+z)+(y_3-y_1+z)                         \\
              & -2 (x_3-x_1+z)(z_2-y_2)                         \\
              & -2 (y_3-y_1+z)(z_2-x_2)  \} \ea }
           & \frac{1}{x_1 y_1 \mc^2}                            \\ \hline
 \mbox{GR} & \emptyset & \emptyset & \mbox{n. g.}               \\ \hline
 \mbox{4G} & -\frac{5}{ \sqrt{2}} (x_2-y_2)                     
           & -\frac{5}{3\sqrt{2}} (x_2-y_2)                     
           & \frac{1}{\mc^2}                                    \\ \hline
 \mbox{Q } & \emptyset & \emptyset & \mbox{n. g.}               \\ 
 \hline\hline
\ea
}
\caption{(b) Numerators and the additional propagators of the graphs of 
group 4 with helicity (--,+,+).}
\label{tab:grp4b}
\end{table}

\vfill
\eject

\subsection{Group 5}
\label{a:grp5}

\bfi[t]
\setlength{\unitlength}{1.0mm}
\bpi{(100,40)}

\put( 0,40){\line(1,0){160}}
\put( 0, 0){\line(1,0){160}}
\put( 0, 0){\line(0,1){40}}
\put(50, 0){\line(0,1){40}}
\put(160, 0){\line(0,1){40}}

\put(5,5){
\psfig{figure=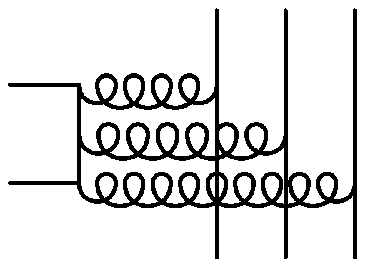,width=4.0cm}}

\put(60,30){Basic Propagators:}
\put(60,20){$  \frac{1}{\{ (z_1-x_1)(z_1-y_1) -1/4 \} \mc^2} 
               \frac{1}{\{ (z_2-x_3)(z_2-y_3) -1/4 \} \mc^2}
               \frac{1}{x_1 y_1 \mc^2}  $}
\put(60,12){$  \times \frac{1}{x_2 y_2 \mc^2}
                      \frac{1}{x_3 y_3 \mc^2}  $}
\epi
\caption{Basic graph of Group 5}
\label{f:grp5}
\efi

\null 
\begin{table}[h]
\centerline{
\ba[t]{|c|c|c|} \hline\hline
 \mbox{I.P.} & \multicolumn{2}{c|}{\mbox{Num. for graphs of Group 5 with}} \\
             & \multicolumn{2}{|c|}{(\l_1,\l_2,\l_3) = (+,+,-)} \\ \cline{2-3}
             & \mbox{\hspace{2.1cm} J=1 \hspace{2.1cm}} 
             & \mbox{J=2}                                       \\ \hline
 \mbox{U3} & \emptyset & \emptyset                              \\ \hline
 \mbox{L3} & \emptyset & \emptyset                              \\ \hline
 \mbox{U2} & \multicolumn{2}{c|}
            {-(-1)^J \frac{\sqrt{2}}{12} z (2x_1+z)(2x_3+z)}    \\ \hline
 \mbox{L2} & \multicolumn{2}{c|}
            {-(-1)^J \frac{\sqrt{2}}{12} z (2y_1+z)(2y_3+z)}    \\ \hline
 \mbox{U1} & \multicolumn{2}{|c|}{\ba[t]{c@{\,}l}
            & \;\; (-1)^J \frac{\sqrt{2}}{4} z                      
              \{ 1 -2 x_2                                       
                -4 (x_2+x_3-z_1)(y_3-z_1)    \} \ea }           \\ \hline     
 \mbox{L1} & \multicolumn{2}{|c|}{\ba[t]{c@{\,}l}
            & -(-1)^J \frac{\sqrt{2}}{6} z                      
              \{ 1 -2 y_2      				      
                -4 (y_2+y_3-z_1)(x_3-z_1)    \} \ea }           \\ \hline
 \mbox{G3} & \emptyset & \emptyset                              \\ \hline 
 \mbox{G2} & \emptyset & \emptyset                              \\ \hline
 \mbox{G1} & \multicolumn{2}{c|}{\ba[t]{c@{\,}l}
              -(-1)^J \frac{5}{12 \sqrt{2}} 
           \{ & (x_1 -y_1)(1 -4(x_3-z_1)(y_3-z_1))              \\
              & +2 x_2 (2 y_3+z)(y_1+z)                         \\
              & -2 y_2 (2 x_3+z)(x_1+z) \} \ea }                \\ \hline          
 \mbox{UQ} & \multicolumn{2}{|c|}{\ba[t]{c@{\,}l}
                -(-1)^J \frac{5 \sqrt{2}}{108}                  
           \{ &   (x_3-y_3) -2z (x_2-y_2) +4(x_1-y_1)(1-z_1)^2  \\
              & -4(x_1(1-x_3) -x_2 z_1)(1-y_1-z_1)              \\
              & +4(y_1(1-y_3) -y_2 z_1)(1-x_1-z_1) \} \ea }     \\ \hline
 \mbox{LQ} & \emptyset & \emptyset                              \\ 
 \hline\hline
\ea
}
\caption{(a) Numerators of the graphs of group 5 with helicity (+,+,--).}
\label{tab:grp5a}
\end{table}

\addtocounter{table}{-1}

\null 
\begin{table}
\centerline{
\ba[t]{|c|c|c|} \hline\hline
 \mbox{I.P.} & \multicolumn{2}{c|}{\mbox{Num. for graphs of Group 5 with}} \\
             & \multicolumn{2}{|c|}{(\l_1,\l_2,\l_3) = (+,-,+)} \\ \cline{2-3}
             & \mbox{\hspace{2.5cm} J=1 \hspace{2.5cm}} 
             & \mbox{J=2}                                       \\ \hline
 \mbox{U3} & \multicolumn{2}{|c|}{\ba[t]{c@{\,}l}
             -\frac{\sqrt{2}}{6} 
         \{ & 2 (x_3-z) [1 -x_2 -y_2 -2 (1-x_3-z_1) (y_1 -z_1)  \\	
            & \mbox{\hskip 1.80cm} -2 (1-y_3-z_1) (x_1 -z_1) ]  \\
            & + (-1)^J z [1 -2 x_2 -4 (1-x_3-z_1)(y_1-z_1) ]  \} \ea }
                                                                \\ \hline
 \mbox{L3} & \multicolumn{2}{|c|}{\ba[t]{c@{\,}l}
              \frac{\sqrt{2}}{4} 
         \{ & 2 (y_3-z) [1 -x_2 -y_2 -2 (1-x_3-z_1) (y_1 -z_1)  \\	
            & \mbox{\hskip 1.80cm} -2 (1-y_3-z_1) (x_1 -z_1) ]  \\
            & + (-1)^J z [1 -2 y_2 -4 (1-y_3-z_1)(x_1 -z_1) ]  \} \ea }  
                                                                \\ \hline
 \mbox{U2} & \multicolumn{2}{|c|}{\ba[t]{c@{\,}l}
             -\frac{\sqrt{2}}{6} (x_2-z) 
           \{ 1 - &\fx (x_2+y_2 -2z)                            \\
                +2 (x_3-z_1)(y_1-z_1)  				
              & +2 (y_3-z_1)(x_1-z_1) \} \ea  } 		\\ \hline
 \mbox{L2} & \multicolumn{2}{|c|}{\ba[t]{c@{\,}l}
             -\frac{\sqrt{2}}{6} (y_2-z) 
           \{ 1 - &\fx (x_2+y_2 -2z)                            \\
                +2 (x_3-z_1)(y_1-z_1)  				
              & +2 (y_3-z_1)(x_1-z_1) \} \ea  } 		\\ \hline
 \mbox{U1} & \multicolumn{2}{|c|}{\ba[t]{c@{\,}l}
              \frac{\sqrt{2}}{4} 
         \{ & 2 (x_1-z) [1 -x_2 -y_2 -2 (1-x_1-z_1) (y_3-z_1)  \\	
            & \mbox{\hskip 1.80cm} -2 (1-y_1-z_1) (x_3 -z_1) ] \\
            & + (-1)^J z [1 -2 x_2 -4 (1-x_1-z_1)(y_3-z_1) ]  \} \ea }    
                                                               \\ \hline
 \mbox{L1} & \multicolumn{2}{|c|}{\ba[t]{c@{\,}l}
             -\frac{\sqrt{2}}{6} 
         \{ & 2 (y_1-z) [1 -x_2 -y_2 -2 (1-x_1-z_1) (y_3-z_1)  \\	
            & \mbox{\hskip 1.80cm} -2 (1-y_1-z_1) (x_3 -z_1) ] \\
            & + (-1)^J z [1 -2 y_2 -4 (1-y_1-z_1)(x_3-z_1) ]  \} \ea }   
                                                               \\ \hline 
 \mbox{G3} & \multicolumn{2}{c|}{\ba[t]{c@{\,}l}
             -\frac{5}{12 \sqrt{2}} 
         \{ & 4 (x_3-y_3) [1-y_2-x_2 (2 y_1+z) -2 (x_1-z_1) (y_1-y_3+z) ] 
                                                                \\
            &-(-1)^J [2 z ( x_2(2 y_1+z) - y_2 (2 x_1+z) )      \\
            &+ x_3 (1 - 2 y_2 - 4 (x_1-z_1)(1-y_3-z_1) )        \\
            &- y_3 (1 - 2 x_2 - 4 (y_1-z_1)(1-x_3-z_1) ) ]  \} \ea } 
                                                                \\ \hline
 \mbox{G2} & \emptyset & \emptyset                              \\ \hline  
 \mbox{G1} & \multicolumn{2}{c|}{\ba[t]{c@{\,}l}
              \frac{5}{12 \sqrt{2}} 
         \{ & 4 (x_1-y_1) [1-y_2-x_2 (2 y_3+z) +2 (x_3-z_1) (y_1-y_3-z) ] 
                                                                \\
            &-(-1)^J [2 z ( x_2(2 y_3+z) - y_2(2 x_3+z) )       \\
            &+ x_1 (1 - 2 y_2 - 4 (x_3-z_1)(1-y_1-z_1) )        \\
            &- y_1 (1 - 2 x_2 - 4 (y_3-z_1)(1-x_1-z_1) ) ]  \} \ea } 
                                                                \\ \hline
 \mbox{UQ} & \multicolumn{2}{c|}{\ba[t]{c@{\,}l} 
                  -\frac{5 \sqrt{2}}{108}                       
             \{ &  z (2+z)(x_1-y_1)
                  +(z(2 -z)+2(x_1+y_1))(x_2-y_2)                \\
                & +4(z_1-y_3) x_1^2 -4(z_1-x_3) y_1^2 \} \ea }  \\ \hline
 \mbox{LQ} & \multicolumn{2}{c|}{\ba[t]{c@{\,}l} 
            \;\;\;  \frac{5 \sqrt{2}}{108}                       
             \{ &  z (2+z)(x_3-y_3)                             
                  +(z(2 -z)+2(x_3+y_3))(x_2-y_2)                \\
                & +4(z_1-y_1) x_3^2 -4(z_1-x_1) y_3^2 \} \ea }  \\ 
 \hline\hline
\ea
}
\caption{(b) Numerators of the graphs of group 5 with helicity (+,--,+).}
\label{tab:grp5b}
\end{table}

\addtocounter{table}{-1}

\null 
\begin{table}
\centerline{
\ba[t]{|c|c|c|c|} \hline\hline
 \mbox{I.P.} & \multicolumn{2}{c|}{\mbox{Num. for graphs of Group 5 with}} 
             & \mbox{X-Prop.}                                   \\
             & \multicolumn{2}{|c|}{(\l_1,\l_2,\l_3) = (-,+,+)} 
             &                                                  \\ \cline{2-3}
             & \mbox{\hspace{2.0cm} J=1 \hspace{2.0cm}} 
             & \mbox{J=2} &                                     \\ \hline
 \mbox{U3} & \multicolumn{2}{c|}
             {-(-1)^J \frac{\sqrt{2}}{6} z
             \{1 -2 x_2 -4 (x_1+x_2-z_1)(y_1-z_1) \} }
           & \frac{1}{-z(x_3-z) \mc^2}                          \\ \hline
 \mbox{L3} & \multicolumn{2}{c|}
             {\; (-1)^J \frac{\sqrt{2}}{4} z 
             \{1 -2 y_2 -4 (y_1+y_2-z_1)(x_1-z_1) \} }
           & \frac{1}{-z(y_3-z) \mc^2}                          \\ \hline
 \mbox{U2} & \multicolumn{2}{c|}
            {-(-1)^J \frac{\sqrt{2}}{12} z (2 x_1+z)(2 x_3+z) } 
           & \frac{1}{-z(x_2-z) \mc^2}                          \\ \hline
 \mbox{L2} & \multicolumn{2}{c|}
            {-(-1)^J \frac{\sqrt{2}}{12} z (2 y_1+z)(2 y_3+z) } 
           & \frac{1}{-z(y_2-z) \mc^2}                          \\ \hline
 \mbox{U1} & \emptyset & \emptyset 
           & \frac{1}{-z(x_1-z) \mc^2}                          \\ \hline
 \mbox{L1} & \emptyset & \emptyset 
           & \frac{1}{-z(y_1-z) \mc^2}                          \\ \hline
 \mbox{G3} & \multicolumn{2}{c|}{\ba[t]{c@{\,}l}
               (-1)^J \frac{5}{12 \sqrt{2}}                    
           \{ & (x_3 -y_3)(1 -4(x_1-z_1)(y_1-z_1))              \\
              & +2 x_2 (2 y_1+z)(y_3+z)                         \\
              & -2 y_2 (2 x_1+z)(x_3+z) \} \ea }                         
           & \frac{1}{x_3 y_3 \mc^2}                            \\ \hline
 \mbox{G2} & \emptyset & \emptyset & \mbox{n. g.}               \\ \hline
 \mbox{G1} & \emptyset & \emptyset 
           & \frac{1}{x_1 y_1 \mc^2}                            \\ \hline
 \mbox{UQ} & \emptyset & \emptyset 
           & \frac{1}{[(x_1-z_1)(y_1-z_1)-1/4] \mc^2}           \\ \hline
 \mbox{LQ} & \multicolumn{2}{|c|}{\ba[t]{c@{\,}l}
                 (-1)^J \frac{5 \sqrt{2}}{108}                  
           \{ &   (x_1-y_1) -2z (x_2-y_2)+4(x_3-y_3)(1-z_1)^2   \\
              & -4(x_3(1-x_1) -x_2 z_1)(1-y_3-z_1)              \\
              & +4(y_3(1-y_1) -y_2 z_1)(1-x_3-z_1) \} \ea }     
           & \frac{1}{[(x_3-z_1)(y_3-z_1)-1/4] \mc^2}           \\
 \hline\hline
\ea
}
\caption{(c) Numerators and the additional propagators of the graphs of 
group 5 with helicity (--,+,+).}
\label{tab:grp5c}
\end{table}

\vfill
\eject

\subsection{Group 6}
\label{a:grp6}

\bfi[t]
\setlength{\unitlength}{1.0mm}
\bpi{(100,40)}

\put( 0,40){\line(1,0){160}}
\put( 0, 0){\line(1,0){160}}
\put( 0, 0){\line(0,1){40}}
\put(50, 0){\line(0,1){40}}
\put(160, 0){\line(0,1){40}}

\put(5,5){
\psfig{figure=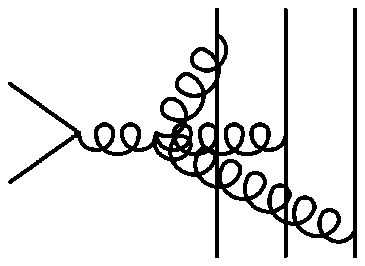,width=4.0cm}}

\put(60,30){Basic Propagators:}
\put(60,20){$  \frac{1}{\mc^4} \frac{1}{x_1 y_1 \mc^2} 
               \frac{1}{x_2 y_2 \mc^2} \frac{1}{x_3 y_3 \mc^2}   $}

\epi
\caption{Basic graph of Group 6}
\label{f:grp6}
\efi

\null 
\begin{table}[h]
\centerline{
\ba[t]{|c|c|c|c|c|} \hline\hline
 \mbox{I.P.} & \multicolumn{4}{c|}
               {\mbox{Num. for graphs of Group 6 with}\; (\l_1,\l_2,\l_3)} 
                                                                \\ \cline{2-5}
             & \multicolumn{2}{c|}{(+,+,-)}
             & \multicolumn{2}{c|}{(+,-,+)}                     \\ \cline{2-5}
             & \mbox{\hspace{0.6cm} J=1 \hspace{0.6cm}} 
             & \mbox{J=2} 
             & \mbox{\hspace{0.6cm} J=1 \hspace{0.6cm}} 
             & \mbox{J=2}                                       \\ \hline
 \mbox{U3} & \multicolumn{2}{c|}{ 2\sqrt{2}(x_3-z) }
           & \multicolumn{2}{c|}{ 
               -\sqrt{2} \{ x_3-z+(-1)^J 2 z \} }               \\ \hline
 \mbox{L3} & \multicolumn{2}{c|}{ 2\sqrt{2}(y_3-z) }
           & \multicolumn{2}{c|}{ 
               -\sqrt{2} \{ y_3-z+(-1)^J 2 z \} }               \\ \hline
 \mbox{U2} & \multicolumn{2}{c|}{ 
               -\sqrt{2} \{ x_2-z+(-1)^J 2 z \} }             
           & \multicolumn{2}{c|}{ 2\sqrt{2}(x_2-z) }            \\ \hline
 \mbox{L2} & \multicolumn{2}{c|}{ 
               -\sqrt{2} \{ y_2-z+(-1)^J 2 z \} }             
           & \multicolumn{2}{c|}{ 2\sqrt{2}(y_2-z) }            \\ \hline
 \mbox{U1} & \multicolumn{4}{c|}{ 
               -\sqrt{2} \{ x_1-z+(-1)^J 2 z \} }               \\ \hline
 \mbox{L1} & \multicolumn{4}{c|}{ 
               -\sqrt{2} \{ y_1-z+(-1)^J 2 z \} }               \\ \hline
 \mbox{G3} & \emptyset & \emptyset & \emptyset & \emptyset      \\ \hline 
 \mbox{G2} & \emptyset & \emptyset & \emptyset & \emptyset      \\ \hline
 \mbox{G1} & \emptyset & \emptyset & \emptyset & \emptyset      \\ 
 \hline\hline
\ea
}
\caption{(a) Numerators of the graphs of group 6 with helicity (+,+,--) 
and (+,--,+).}
\label{tab:grp6a}
\end{table}

\addtocounter{table}{-1}

\null 
\begin{table}
\centerline{
\ba[t]{|c|c|c|c|} \hline\hline
 \mbox{I.P.} & \multicolumn{2}{c|}{\mbox{Num. for graphs of Group 6 with}} 
             & \mbox{X-Prop.}                                   \\
             & \multicolumn{2}{|c|}{(\l_1,\l_2,\l_3) = (-,+,+)} 
             &                                                  \\ \cline{2-3}
             & \mbox{\hspace{0.6cm} \mbox{J=1} \hspace{0.6cm}}
             & \mbox{J=2} &                                     \\ \hline
 \mbox{U3} & \multicolumn{2}{c|}{ 
               -\sqrt{2} \{ x_3-z+(-1)^J 2 z \} }
           & \frac{1}{-z(x_3-z) \mc^2}                          \\ \hline
 \mbox{L3} & \multicolumn{2}{c|}{ 
               -\sqrt{2} \{ y_3-z+(-1)^J 2 z \} }
           & \frac{1}{-z(y_3-z) \mc^2}                          \\ \hline
 \mbox{U2} & \multicolumn{2}{c|}{ 
               -\sqrt{2} \{ x_2-z+(-1)^J 2 z \} }             
           & \frac{1}{-z(x_2-z) \mc^2}                          \\ \hline
 \mbox{L2} & \multicolumn{2}{c|}{ 
               -\sqrt{2} \{ y_2-z+(-1)^J 2 z \} }             
           & \frac{1}{-z(y_2-z) \mc^2}                          \\ \hline
 \mbox{U1} & \multicolumn{2}{c|}{ 2\sqrt{2} (x_1-z) } 
           & \frac{1}{-z(x_1-z) \mc^2}                          \\ \hline
 \mbox{L1} & \multicolumn{2}{c|}{ 2\sqrt{2} (y_1-z) } 
           & \frac{1}{-z(y_1-z) \mc^2}                          \\ \hline
 \mbox{G3} & \emptyset & \emptyset & \mbox{n. g.}               \\ \hline
 \mbox{G2} & \emptyset & \emptyset & \mbox{n. g.}               \\ \hline
 \mbox{G1} & \emptyset & \emptyset & \mbox{n. g.}               \\ 
 \hline\hline
\ea
}
\caption{(b) Numerators and the additional propagators of the graphs of 
group 6 with helicity (--,+,+).}
\label{tab:grp6b}
\end{table}

\clearpage
\vfill
\eject

\subsection{Group 7}
\label{a:grp7}

\bfi[t]
\setlength{\unitlength}{1.0mm}
\bpi{(100,40)}

\put( 0,40){\line(1,0){160}}
\put( 0, 0){\line(1,0){160}}
\put( 0, 0){\line(0,1){40}}
\put(50, 0){\line(0,1){40}}
\put(160, 0){\line(0,1){40}}

\put(5,5){
\psfig{figure=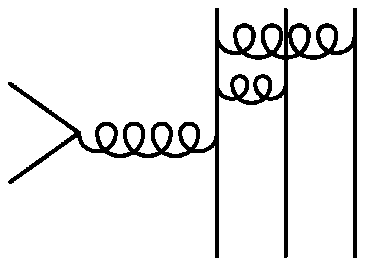,width=4.0cm}}

\put(60,30){Basic Propagators:}
\put(60,20){$  \frac{1}{\mc^2} \frac{1}{(1-y_1) \mc^2} 
               \frac{1}{(1-x_2) y_3 \mc^2} \frac{1}{x_2 y_2 \mc^2} 
               \frac{1}{x_3 y_3 \mc^2}  $}
\epi
\caption{Basic graph of Group 7}
\label{f:grp7}
\efi

\null 
\begin{table}[h]
\centerline{
\ba[t]{|c|c|c|c|c|} \hline\hline
 \mbox{I.P.} & \multicolumn{4}{c|}
               {\mbox{Num. for graphs of Group 7 with}\; (\l_1,\l_2,\l_3)} 
                                                                \\ \cline{2-5}
             & \multicolumn{2}{c|}{(+,+,-)}
             & \multicolumn{2}{c|}{(+,-,+)}                     \\ \cline{2-5}
             & \mbox{\hspace{1.2cm} J=1 \hspace{1.2cm}} 
             & \mbox{J=2} 
             & \mbox{\hspace{0.7cm} J=1 \hspace{0.7cm}} 
             & \mbox{J=2}                                       \\ \hline
 \mbox{U3} & \multicolumn{2}{c|}{
                    \frac{14\sqrt{2}}{27} (1-z)(x_3-z)(y_3-z) }        
           & \multicolumn{2}{|c|}{\ba[t]{c@{\,}l} 
            -(-1)^J \frac{14\sqrt{2}}{27} & z(1-x_2-z)          \\      
                                          & \times (1-y_1-z) \ea }
                                                                \\ \hline
 \mbox{L3} & \multicolumn{2}{c|}{
               \frac{32 \sqrt{2}}{27} (1-z) (y_3-z)^2 }
           & \emptyset & \emptyset                              \\ \hline
 \mbox{U2} & \multicolumn{2}{c|}{
              -\frac{40 \sqrt{2}}{27} (1-z) (x_2-z) y_3 }
           & \emptyset & \emptyset                              \\ \hline
 \mbox{L2} & \multicolumn{2}{c|}{\ba[t]{c@{\,}l} 
               \frac{32 \sqrt{2}}{27} y_3 
               \{ & (1-z)(y_2-z)                                \\
                  & +(-1)^J z (1-y_1-z) \} \ea }  
           & \emptyset & \emptyset                              \\ \hline
 \mbox{U1} & \multicolumn{2}{c|}{
               \frac{26 \sqrt{2}}{27} (1-z)(x_1-z)(y_3-z) } 
           & \multicolumn{2}{|c|}{\ba[t]{c@{\,}l} 
               (-1)^J \frac{26 \sqrt{2}}{27} & z (1-x_2-z)      \\
                                             & \times (1-y_1-z) \ea }
                                                                \\ \hline
 \mbox{UM1}& \multicolumn{2}{c|}{
              -\frac{8 \sqrt{2}}{27} y_3 (1-z)(y_3-z) } 
           & \emptyset & \emptyset                              \\ \hline
 \mbox{LM1}& \multicolumn{2}{c|}{
              -\frac{64 \sqrt{2}}{27} y_3 \{ (1-z)-(-1)^J (1-y_1-z) \} } 
           & \emptyset & \emptyset                              \\ \hline
 \mbox{L1} & \multicolumn{2}{c|}{
              -\frac{64 \sqrt{2}}{27} y_3 \{ (y_1-z) +(-1)^J z \} }
           & \emptyset & \emptyset                              \\ \hline
 \mbox{G3} & \multicolumn{2}{c|}{
              -\frac{2 \sqrt{2}}{3} (1-z)(x_3-y_3) (y_3-z) }
           & \multicolumn{2}{|c|}{\ba[t]{c@{\,}l} 
        (-1)^J \frac{\sqrt{2}}{3} & (1-x_2-z)                   \\
                                  & \times (1-y_1-z)            \\
                                  & \times (y_3+z)  \ea }       \\ \hline
 \mbox{G2} & \multicolumn{2}{c|}{\ba[t]{c@{\,}l} 
              -\frac{4 \sqrt{2}}{3} y_3 
               \{ & 2 (1-z)(x_2-y_2)                            \\
                  & -(-1)^J (x_2+z)(1-y_1-z) \} \ea }
           & \emptyset & \emptyset                              \\ 
 \hline\hline
\ea
}
\caption{(a) Numerators of the graphs of group 7 with helicity (+,+,--) and (+,--,+).}
\label{tab:grp7a}
\end{table}

\addtocounter{table}{-1}

\null 
\begin{table}
\centerline{
\ba[t]{|c|c|c|c|} \hline\hline
 \mbox{I.P.} & \multicolumn{2}{c|}{\mbox{Num. for graphs of Group 7 with}} 
             & \mbox{X-Prop.}                                   \\
             & \multicolumn{2}{|c|}{(\l_1,\l_2,\l_3) = (-,+,+)} 
             &                                                  \\ \cline{2-3}
             & \mbox{\hspace{1.3cm} J=1 \hspace{1.3cm}} 
             & \mbox{J=2} &                                     \\ \hline
 \mbox{U3} & \multicolumn{2}{c|}{
            -(-1)^J \frac{14\sqrt{2}}{27} z(1-x_2-z)(1-y_1-z) }
           & \frac{1}{-z (x_3-z) \mc^2}                         \\ \hline
 \mbox{L3} & \emptyset & \emptyset 
           & \frac{1}{-z (y_3-z) \mc^2}                         \\ \hline
 \mbox{U2} & \emptyset & \emptyset 
           & \frac{1}{-z (x_2-z) \mc^2}                         \\ \hline
 \mbox{L2} & \multicolumn{2}{c|}{
             (-1)^J \frac{32 \sqrt{2}}{27} z y_3 (1-y_1-z) }
           & \frac{1}{-z (y_2-z) \mc^2}                         \\ \hline
 \mbox{U1} & \emptyset & \emptyset 
           & \frac{1}{-z (x_1-z) \mc^2}                         \\ \hline
 \mbox{UM1}& \multicolumn{2}{c|}{
            -(-1)^J \frac{8 \sqrt{2}}{27} y_3 (1-x_2-z) (1-y_1-z) }
           & \frac{1}{(1-x_2) y_3 \mc^2}                        \\ \hline
 \mbox{LM1}& \emptyset & \emptyset 
           & \frac{1}{(1-y_1) \mc^2 }                           \\ \hline
 \mbox{L1} & \emptyset & \emptyset 
           & \frac{1}{-z (y_1-z) \mc^2}                         \\ \hline
 \mbox{G3} & \multicolumn{2}{c|}{
        (-1)^J \frac{\sqrt{2}}{3}  (1-x_2-z)(1-y_1-z) (y_3+z) }
           & \frac{1}{x_3 y_3 \mc^2}                            \\ \hline
 \mbox{G2} & \multicolumn{2}{c|}{
        (-1)^J \frac{4 \sqrt{2}}{3} y_3 (1-y_1-z)(x_2+z) }   
           & \frac{1}{x_2 y_2 \mc^2}                            \\ 
 \hline\hline
\ea
}
\caption{(b) Numerators and the additional propagators of the graphs of 
group 7 with helicity (--,+,+).}
\label{tab:grp7b}
\end{table}

\clearpage
\vfill
\eject

\subsection{Group 8}
\label{a:grp8}

\bfi[t]
\setlength{\unitlength}{1.0mm}
\bpi{(100,40)}

\put( 0,40){\line(1,0){160}}
\put( 0, 0){\line(1,0){160}}
\put( 0, 0){\line(0,1){40}}
\put(50, 0){\line(0,1){40}}
\put(160, 0){\line(0,1){40}}

\put(5,5){
\psfig{figure=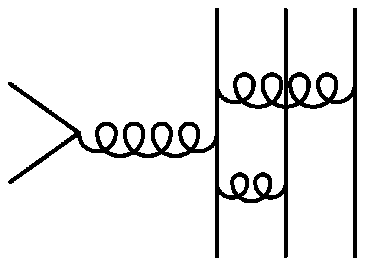,width=4.0cm}}

\put(60,30){Basic Propagators:}
\put(60,20){$  \frac{1}{\mc^2} \frac{1}{x_2 (1-y_3) \mc^2} 
               \frac{1}{(1-x_2) y_3 \mc^2} \frac{1}{x_2 y_2 \mc^2} 
               \frac{1}{x_3 y_3 \mc^2}  $}
\epi
\caption{Basic graph of Group 8}
\label{f:grp8}
\efi

\null 
\begin{table}[h]
\centerline{
\ba[t]{|c|c|c|c|c|} \hline\hline
 \mbox{I.P.} & \multicolumn{4}{c|}
               {\mbox{Num. for graphs of Group 8 with}\; (\l_1,\l_2,\l_3)} 
                                                                \\ \cline{2-5}
             & \multicolumn{2}{c|}{(+,+,-)}
             & \multicolumn{2}{c|}{(+,-,+)}                     \\ \cline{2-5}
             & \mbox{\hspace{0.8cm} J=1 \hspace{0.8cm}} 
             & \mbox{J=2} 
             & \mbox{\hspace{0.8cm} J=1 \hspace{0.8cm}} 
             & \mbox{J=2}                                       \\ \hline
 \mbox{U3} & \emptyset & \emptyset
           & \multicolumn{2}{c|}{
              (-1)^J \frac{40 \sqrt{2}}{27} x_2 z (1-x_2-z) }   \\ \hline
 \mbox{L3} & \emptyset & \emptyset & \emptyset & \emptyset      \\ \hline
 \mbox{U2} & \emptyset & \emptyset & \emptyset & \emptyset      \\ \hline
 \mbox{L2} & \multicolumn{2}{c|}{
              (-1)^J \frac{40 \sqrt{2}}{27} y_3 z (1-y_3-z) }
           & \emptyset & \emptyset                              \\ \hline
 \mbox{U1} & \emptyset & \emptyset
           & \multicolumn{2}{c|}{
              (-1)^J \frac{8  \sqrt{2}}{27} x_2 z (1-x_2-z) }   \\ \hline
 \mbox{UM1}& \emptyset & \emptyset & \emptyset & \emptyset      \\ \hline
 \mbox{LM1}& \emptyset & \emptyset & \emptyset & \emptyset      \\ \hline
 \mbox{L1} & \multicolumn{2}{c|}{
              (-1)^J \frac{8  \sqrt{2}}{27} y_3 z (1-y_3-z) }
           & \emptyset & \emptyset                              \\ \hline
 \mbox{G3} & \emptyset & \emptyset
           & \multicolumn{2}{c|}{
              (-1)^J \frac{4 \sqrt{2}}{3} x_2(1-x_2-z)(y_3+z) } \\ \hline
 \mbox{G2} & \multicolumn{2}{c|}{
              (-1)^J \frac{4 \sqrt{2}}{3} y_3(1-y_3-z)(x_2+z) } 
           & \emptyset & \emptyset                              \\  
 \hline\hline
\ea
}
\caption{(a) Numerators of the graphs of group 8 with helicity (+,+,--) and (+,--,+).}
\label{tab:grp8a}
\end{table}

\addtocounter{table}{-1}

\null 
\begin{table}
\centerline{
\ba[t]{|c|c|c|c|} \hline\hline
 \mbox{I.P.} & \multicolumn{2}{c|}{\mbox{Num. for graphs of Group 8 with}} 
             & \mbox{X-Prop.}                                   \\
             & \multicolumn{2}{|c|}{(\l_1,\l_2,\l_3) = (-,+,+)} 
             &                                                  \\ \cline{2-3}
             & \mbox{\hspace{1.6cm} J=1 \hspace{1.6cm}} 
             & \mbox{J=2} &                                     \\ \hline
 \mbox{U3} & \multicolumn{2}{c|}{
               \frac{40 \sqrt{2}}{27} x_2 
               \{ (x_3-z)(y_3-z) +(-1)^J z(1-x_2-z) \} }    
           & \frac{1}{-z (x_3-z) \mc^2 }                        \\ \hline
 \mbox{L3} & \multicolumn{2}{c|}{
              -\frac{32 \sqrt{2}}{27} x_2 (y_3-z)^2 }
           & \frac{1}{-z (y_3-z) \mc^2 }                        \\ \hline
 \mbox{U2} & \multicolumn{2}{c|}{
              -\frac{32 \sqrt{2}}{27} y_3 (x_2-z)^2 }
           & \frac{1}{-z (x_2-z) \mc^2 }                        \\ \hline
 \mbox{L2} & \multicolumn{2}{c|}{
               \frac{40 \sqrt{2}}{27} y_3 
               \{ (x_2-z)(y_2-z) +(-1)^J z(1-y_3-z) \} }
           & \frac{1}{-z (y_2-z) \mc^2 }                        \\ \hline
 \mbox{U1} & \multicolumn{2}{c|}{
              -\frac{8  \sqrt{2}}{27} x_2 (x_1-z)(y_3-z) }
           & \frac{1}{-z (x_1-z) \mc^2 }                        \\ \hline
 \mbox{UM1}& \multicolumn{2}{c|}{
              -\frac{64 \sqrt{2}}{27} x_2 y_3 
               \{ (y_3-z) -(-1)^J (1-x_2-z) \} }
           & \frac{1}{(1-x_2) y_3 \mc^2 }                       \\ \hline
 \mbox{LM1}& \multicolumn{2}{c|}{
              -\frac{64 \sqrt{2}}{27} x_2 y_3 
               \{ (x_2-z) -(-1)^J (1-y_3-z) \} }
           & \frac{1}{x_2 (1 -y_3) \mc^2}                       \\ \hline
 \mbox{L1} & \multicolumn{2}{c|}{
              -\frac{8 \sqrt{2}}{27} y_3 (x_2-z)(y_1-z) }      
           & \frac{1}{-z(y_1-z) \mc^2}                          \\ \hline 
 \mbox{G3} & \multicolumn{2}{c|}{\ba[t]{c@{\,}l} 
               \frac{4 \sqrt{2}}{3} x_2 
               \{ & 2(x_3-y_3)(y_3 -z)                          \\
                  & +(-1)^J (1-x_2-z) (y_3+z) \} \ea }          
           & \frac{1}{x_3 y_3 \mc^2 }                           \\ \hline 
 \mbox{G2} & \multicolumn{2}{c|}{\ba[t]{c@{\,}l} 
              -\frac{4 \sqrt{2}}{3} y_3 
               \{ & 2(x_2-y_2)(x_2 -z)                          \\
                  & -(-1)^J (1-y_3-z) (x_2+z) \} \ea }
           & \frac{1}{x_2 y_2 \mc^2 }                           \\ 
 \hline\hline
\ea
}
\caption{(b) Numerators and the additional propagators of the graphs of 
group 8 with helicity (--,+,+).}
\label{tab:grp8b}
\end{table}

\clearpage
\vfill
\eject

\subsection{Group 9}
\label{a:grp9}

\bfi[t]
\setlength{\unitlength}{1.0mm}
\bpi{(100,40)}

\put( 0,40){\line(1,0){160}}
\put( 0, 0){\line(1,0){160}}
\put( 0, 0){\line(0,1){40}}
\put(50, 0){\line(0,1){40}}
\put(160, 0){\line(0,1){40}}

\put(5,5){
\psfig{figure=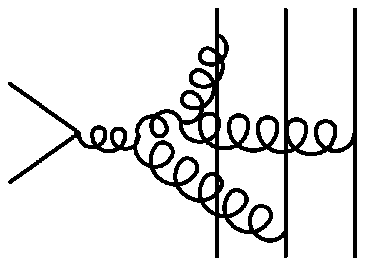,width=4.0cm}}

\put(60,30){Basic Propagators:}
\put(60,20){$  \frac{1}{\mc^2} \frac{1}{(1-x_2)(1-y_2) \mc^2 +\r^2}
               \frac{1}{x_1 y_1 \mc^2} \frac{1}{x_2 y_2 \mc^2}      $} 
\put(60,12){$  \times \frac{1}{x_3 y_3 \mc^2}  $}
\epi
\caption{Basic graph of Group 9}
\label{f:grp9}
\efi

\null 
\begin{table}[h]
\centerline{
\ba[t]{|c|c|c|c|c|} \hline\hline
 \mbox{I.P.} & \multicolumn{4}{c|}
               {\mbox{Num. for graphs of Group 9 with}\; (\l_1,\l_2,\l_3)} 
                                                                \\ \cline{2-5}
             & \multicolumn{2}{c|}{(+,+,-)}
             & \multicolumn{2}{c|}{(+,-,+)}                     \\ \cline{2-5}
             & \mbox{\hspace{1.5cm} J=1 \hspace{1.5cm}} 
             & \mbox{J=2} 
             & \mbox{\hspace{1.0cm} J=1 \hspace{1.0cm}} 
             & \mbox{J=2}                                       \\ \hline
 \mbox{U3} & \multicolumn{2}{|c|}{\ba[t]{c@{\,}l}
              -\frac{1}{\sqrt{2}} (x_3-z) 
               \{ & (1+x_2-z)(y_1-y_3+z)                        \\
                  & +(1+y_2-z)(x_1-x_3+z) \} \ea }
           & \multicolumn{2}{c|}
             {(-1)^J \sqrt{2} z(x_2-y_2) (2 x_1+x_3-z) }        \\ \hline 
 \mbox{L3} & \multicolumn{2}{|c|}{\ba[t]{c@{\,}l}
              -\frac{1}{\sqrt{2}} (y_3-z) 
               \{ & (1+x_2-z)(y_1-y_3+z)                        \\
                  & +(1+y_2-z)(x_1-x_3+z) \} \ea }
           & \multicolumn{2}{c|}
             {(-1)^J \sqrt{2} z(y_2-x_2) (2 y_1+y_3-z) }        \\ \hline
 \mbox{U2} & \emptyset & \emptyset & \emptyset & \emptyset      \\ \hline 
 \mbox{L2} & \emptyset & \emptyset & \emptyset & \emptyset      \\ \hline 
 \mbox{U1} & \multicolumn{2}{c|}{\ba[t]{c@{\,}l}
              -\sqrt{2} \{ 
             & \half (x_1-z) [(1+x_2-z)(y_3-y_1+z)              \\
             & +(1+y_2-z)(x_3-x_1+z)]                           \\
             & -(-1)^J z(x_2-y_2) (2 x_3+x_1-z) \} \ea }  
           & \multicolumn{2}{c|}
             {(-1)^J \sqrt{2} z(x_2-y_2) (2 x_3+x_1-z) }        \\ \hline
 \mbox{L1} & \multicolumn{2}{c|}{\ba[t]{c@{\,}l}
              -\sqrt{2} \{ 
             & \half (y_1-z) [(1+x_2-z)(y_3-y_1+z)              \\
             & +(1+y_2-z)(x_3-x_1+z)]                           \\
             & -(-1)^J z(y_2-x_2) (2 y_3+y_1-z) \} \ea }        
           & \multicolumn{2}{c|}
             {(-1)^J \sqrt{2} z(y_2-x_2) (2 y_3+y_1-z) }       \\ \hline
 \mbox{G3} & \emptyset & \emptyset & \emptyset & \emptyset      \\ \hline 
 \mbox{G2} & \emptyset & \emptyset & \emptyset & \emptyset      \\ \hline 
 \mbox{G1} & \emptyset & \emptyset & \emptyset & \emptyset      \\ \hline 
 \mbox{GR} & \emptyset & \emptyset & \emptyset & \emptyset      \\ \hline
 \mbox{4G1}& \emptyset & \emptyset & \emptyset & \emptyset      \\ \hline
 \mbox{4G2}& \emptyset & \emptyset & \emptyset & \emptyset      \\ 
 \hline\hline
\ea
}
\caption{(a) Numerators of the graphs of group 9 with helicity 
(+,+,--) and (+,--,+).}
\label{tab:grp9a}
\end{table}

\addtocounter{table}{-1}

\null 
\begin{table}
\centerline{
\ba[t]{|c|c|c|c|} \hline\hline
 \mbox{I.P.} & \multicolumn{2}{c|}{\mbox{Num. for graphs of Group 9 with}} 
             & \mbox{X-Prop.}                                   \\
             & \multicolumn{2}{|c|}{(\l_1,\l_2,\l_3) = (-,+,+)} 
             &                                                  \\ \cline{2-3}
             & \mbox{\hspace{1.4cm} J=1 \hspace{1.4cm}} 
             & \mbox{J=2} &                                     \\ \hline
 \mbox{U3} & \multicolumn{2}{c|}{\ba[t]{c@{\,}l}
             -\sqrt{2} \{ 
             & \half (x_3-z) [(1+x_2-z)(y_1-y_3+z)              \\
             & +(1+y_2-z)(x_1-x_3+z)]                           \\
             & -(-1)^J z(x_2-y_2) (2 x_1+x_3-z) \} \ea }  
           & \frac{1}{-z(x_3-z) \mc^2}                          \\ \hline
 \mbox{L3} & \multicolumn{2}{c|}{\ba[t]{c@{\,}l}
             -\sqrt{2} \{ 
             & \half (y_3-z) [(1+x_2-z)(y_1-y_3+z)              \\
             & +(1+y_2-z)(x_1-x_3+z)]                           \\
             & -(-1)^J z(y_2-x_2) (2 y_1+y_3-z) \} \ea }  
           & \frac{1}{-z(y_3-z) \mc^2}                          \\ \hline
 \mbox{U2} & \emptyset & \emptyset  & \mbox{n. g.}              \\ \hline
 \mbox{L2} & \emptyset & \emptyset  & \mbox{n. g.}              \\ \hline
 \mbox{U1} & \multicolumn{2}{|c|}{\ba[t]{c@{\,}l}
              -\frac{1}{\sqrt{2}} (x_1-z) 
               \{ & (1+x_2-z)(y_3-y_1+z)                        \\
                  & +(1+y_2-z)(x_3-x_1+z) \} \ea }
           & \frac{1}{-z(x_1-z) \mc^2}                          \\ \hline
 \mbox{L1} & \multicolumn{2}{|c|}{\ba[t]{c@{\,}l}
              -\frac{1}{\sqrt{2}} (y_1-z) 
               \{ & (1+x_2-z)(y_3-y_1+z)                        \\
                  & +(1+y_2-z)(x_3-x_1+z) \} \ea }
           & \frac{1}{-z(y_1-z) \mc^2}                          \\ \hline
 \mbox{G3} & \emptyset & \emptyset & \mbox{n. g.}               \\ \hline
 \mbox{G2} & \emptyset & \emptyset & \mbox{n. g.}               \\ \hline
 \mbox{G1} & \emptyset & \emptyset & \mbox{n. g.}               \\ \hline
 \mbox{GR} & \emptyset & \emptyset & \mbox{n. g.}               \\ \hline
 \mbox{4G1}& \emptyset & \emptyset & \mbox{none}                \\ \hline    
 \mbox{4G2}& \emptyset & \emptyset & \mbox{none}                \\ 
 \hline\hline
\ea
}
\caption{(b) Numerators and the additional propagators of the graphs of 
group 9 with helicity (--,+,+).}
\label{tab:grp9b}
\end{table}

\vfill
\eject

\subsection{Group 10}
\label{a:grp10}

\bfi[t]
\setlength{\unitlength}{1.0mm}
\bpi{(100,40)}

\put( 0,40){\line(1,0){160}}
\put( 0, 0){\line(1,0){160}}
\put( 0, 0){\line(0,1){40}}
\put(50, 0){\line(0,1){40}}
\put(160, 0){\line(0,1){40}}

\put(5,5){
\psfig{figure=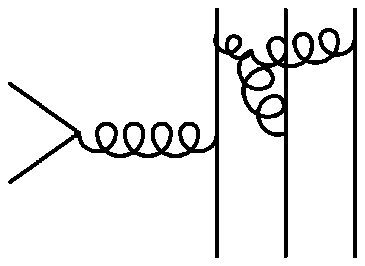,width=4.0cm}}

\put(60,30){Basic Propagators:}
\put(60,20){$  \frac{1}{\mc^2} \frac{1}{(1-y_1) \mc^2}        
               \frac{1}{(1-x_1)(1-y_1) \mc^2+\r^2} \frac{1}{x_2 y_2 \mc^2} $}
\put(60,12){$  \times \frac{1}{x_3 y_3 \mc^2}  $}
\epi
\caption{Basic graph of Group 10}
\label{f:grp10}
\efi

\null 
\begin{table}[h]
\centerline{
\ba[t]{|c|c|c|c|c|} \hline\hline
 \mbox{I.P.} & \multicolumn{4}{c|}
               {\mbox{Num. for graphs of Group 10 with}\; (\l_1,\l_2,\l_3)} 
                                                                \\ \cline{2-5}
             & \multicolumn{2}{c|}{(+,+,-)}
             & \multicolumn{2}{c|}{(+,-,+)}                     \\ \cline{2-5}
             & \mbox{\hspace{1.2cm} J=1 \hspace{1.2cm}} 
             & \mbox{J=2} 
             & \mbox{\hspace{1.2cm} J=1 \hspace{1.2cm}} 
             & \mbox{J=2}                                       \\ \hline
 \mbox{U3} & \multicolumn{2}{|c|}
              { \sqrt{2} (1-z) (x_3-z) (y_3-y_2-z) }
           & \multicolumn{2}{c|}{\ba[t]{c@{\,}l}
                & \sqrt{2} \{ (1-z) (x_3-z) (y_3-y_2-z)         \\
                & +(-1)^J z (y_2+y_3-z) (2 x_2+x_3-z) 
                      \} \ea }                                  \\ \hline 
 \mbox{L3} & \emptyset & \emptyset & \emptyset & \emptyset      \\ \hline 
 \mbox{U2} & \multicolumn{2}{|c|}{\ba[t]{c@{\,}l}
                & \sqrt{2} \{ (1-z) (x_2-z) (y_2-y_3-z)         \\
                & +(-1)^J z (y_2+y_3-z) (2 x_3+x_2-z) 
                      \} \ea }                                 
           & \multicolumn{2}{c|}
              { \sqrt{2} (1-z) (x_2-z) (y_2-y_3-z) }            \\ \hline
 \mbox{L2} & \emptyset & \emptyset & \emptyset & \emptyset      \\ \hline 
 \mbox{U1} & \emptyset & \emptyset & \emptyset & \emptyset      \\ \hline 
 \mbox{M1} & \emptyset & \emptyset & \emptyset & \emptyset      \\ \hline 
 \mbox{L1} & \emptyset & \emptyset & \emptyset & \emptyset      \\ \hline 
 \mbox{G3} & \multicolumn{2}{c|}
              { \sqrt{2} (1-z) (x_3-y_3) (y_3-y_2-z) }
           & \multicolumn{2}{c|}{\ba[t]{c@{\,}l}
                & \sqrt{2} \{ (1-z) (x_3-y_3) (y_3-y_2-z)       \\
                & \mbox{\hspace{0.8cm}} +(-1)^J (y_2+y_3-z)     \\
                & \mbox{\hspace{1.0cm}} \times 
                  (x_3 y_2+x_2 y_3+x_3 y_3                      \\
                & \mbox{\hspace{1.4cm}} +z (x_2+y_2-z)) 
                             \} \ea }                           \\ \hline
 \mbox{G2} & \multicolumn{2}{c|}{\ba[t]{c@{\,}l}
                & \sqrt{2} \{ (1-z) (x_2-y_2) (y_2-y_3-z)       \\
                & \mbox{\hspace{0.8cm}} +(-1)^J (y_2+y_3-z)     \\
                & \mbox{\hspace{1.0cm}} \times 
                  (x_3 y_2+x_2 y_3+x_2 y_2                      \\
                & \mbox{\hspace{1.4cm}} +z (x_3+y_3-z))
                             \} \ea }                             
           & \multicolumn{2}{c|}
             { \sqrt{2} (1-z) (x_2-y_2) (y_2-y_3-z) }           \\ \hline
 \mbox{GR} & \emptyset & \emptyset & \emptyset & \emptyset      \\ \hline
 \mbox{4G} & \multicolumn{2}{c|}{
               \sqrt{2} \{ 2 (1-z) -(-1)^J (y_2+y_3-z) \} }
           & \multicolumn{2}{c|}{
               \sqrt{2} \{ 2 (1-z) -(-1)^J (y_2+y_3-z) \} }     \\
 \hline\hline
\ea
}
\caption{(a) Numerators of the graphs of group 10 with helicity (+,+,-) 
and (+,--,+).}
\label{tab:grp10a}
\end{table}

\addtocounter{table}{-1}

\null 
\begin{table}
\centerline{
\ba[t]{|c|c|c|c|} \hline\hline
 \mbox{I.P.} & \multicolumn{2}{c|}{\mbox{Num. for graphs of Group 10 with}} 
             & \mbox{X-Prop.}                                   \\
             & \multicolumn{2}{|c|}{(\l_1,\l_2,\l_3) = (-,+,+)} 
             &                                                  \\ \cline{2-3}
             & \mbox{\hspace{1.3cm} J=1 \hspace{1.3cm}} 
             & \mbox{J=2} &                                     \\ \hline
 \mbox{U3} & \multicolumn{2}{c|}
               { (-1)^J \sqrt{2} z (y_2+y_3-z) (2 x_2+x_3-z) }
           & \frac{1}{-z(x_3-z) \mc^2}                          \\ \hline
 \mbox{L3} & \emptyset & \emptyset & \mbox{n. g.}               \\ \hline
 \mbox{U2} & \multicolumn{2}{c|}
               { (-1)^J \sqrt{2} z (y_2+y_3-z) (2 x_3+x_2-z) } 
           & \frac{1}{-z(y_3-z) \mc^2}                          \\ \hline
 \mbox{L2} & \emptyset & \emptyset  & \mbox{n. g.}              \\ \hline
 \mbox{U1} & \emptyset & \emptyset  & \mbox{n. g.}              \\ \hline
 \mbox{M1} & \emptyset & \emptyset  & \mbox{n. g.}              \\ \hline
 \mbox{L1} & \emptyset & \emptyset  & \mbox{n. g.}              \\ \hline
 \mbox{G3} & \multicolumn{2}{|c|}{\ba[t]{c@{\,}l}
                & (-1)^J \sqrt{2} (y_2+y_3-z)                   \\
                \times & (x_3 y_2+x_2 y_3+x_3 y_3+z (x_2+y_2-z))\\ 
                \ea }
           & \frac{1}{x_3 y_3 \mc^2}                            \\ \hline
 \mbox{G2} & \multicolumn{2}{|c|}{\ba[t]{c@{\,}l}
                & (-1)^J \sqrt{2} (y_2+y_3-z)                   \\
                \times & (x_3 y_2+x_2 y_3+x_2 y_2+z (x_3+y_3-z))\\
                \ea }
           & \frac{1}{x_2 y_2 \mc^2}                            \\ \hline
 \mbox{GR} & \emptyset & \emptyset & \mbox{n. g.}               \\ \hline
 \mbox{4G} & \multicolumn{2}{c|}
             { (-1)^J 2 \sqrt{2} (y_2+y_3-z) }
           & \frac{1}{\mc^2}                                    \\ 
 \hline\hline
\ea
}
\caption{(b) Numerators and the additional propagators of the graphs of 
group 10 with helicity (--,+,+).}
\label{tab:grp10b}
\end{table}

\null
\vfill\eject

\null
\section{Numerical Parameters}
\label{a:np}

Here we gather together our input as well as other parameters used in the
calculations in table form.

\subsection{Basic Parameters}
\label{a:base}
\begin{table}[h]
\centerline{
\ba[t]{|c|c|} \hline\hline
 \mbox{Symbol} & \mbox{Value}                     \\ \hline\hline
 \lqcd & 0.22 \; \mbox{GeV}                       \\ \hline
 m_c   & 1.50 \; \mbox{GeV}                       \\ \hline
 \m_0  & 1.00 \; \mbox{GeV}                       \\ \hline
 \m_R \;\; \mbox{(SHSA)} & m_c                    \\ \hline
 \m_R \; \mbox{(MHSA)} 
       & \mbox{square root of the largest}        \\ 
       & \mbox{virtuality, see section \ref{sec:cs}}
                                                  \\ \hline
 n_f   & 4                                        \\ 
 \hline\hline
\ea
}
\caption{Basic input parameters.}
\label{tab:b_param}
\end{table}

\subsection{Charmonium Parameters}
\label{a:charm}
\begin{table}[h]
\centerline{
\ba[t]{|c|l|l|} \hline\hline
 \mbox{Symbol} & \mbox{Value}                 & \mbox{Origin}     \\ \hline\hline
  |R'_P(0)|        & 0.220 \; \mbox{GeV}^{5/2}  
                   & \mbox{see ref. \cite{bks,mag&pet}}             \\ \hline
  f^{(8)}_{\c_1}   & 0.225 \times 10^{-3} \; \mbox{GeV}^2  
                   & \mbox{\em obtained from fit here}              \\ \hline
  f^{(8)}_{\c_2}   & 0.900 \times 10^{-3} \; \mbox{GeV}^2  
                   & \mbox{from ref. \cite{bks}}                    \\ \hline
  z = z_3          & 0.150                       
                   & \mbox{from ref. \cite{bks,bks2}}               \\ \hline
  z_1 = z_2 = (1-z)/2  & 0.425                
                   & \mbox{from ref. \cite{bks,bks2}}               \\ 
 \hline\hline
\ea
}
\caption{Charmonium parameters.}
\label{tab:c_param}
\end{table}

\null
\vfill\eject

\subsection{Baryon Wavefunction Parameters}
\label{a:baryon}
\begin{table}[h]
\centerline{
\ba[t]{|c|c|c|c|c|c|} \hline\hline
 \mbox{Baryon} & B_1 & B_2 & B_3 & B_4 & B_5                     \\ \hline\hline
  N       & \;\; 0.750  & \;\; 0.250  & \;\; 0.000 & \;\; 0.000 
          & \;\; 0.000                                           \\ \hline
  \Sigma  & \;\; 0.216  & \;\; 0.394  &     -0.293 &     -0.914 
          & \;\; 0.241                                           \\ \hline
  \Xi     & \;\; 1.106  & \;\; 0.050  &     -0.282 & \;\; 1.717 
          & -0.498                                               \\ \hline
  \Lambda &     -0.721  & \;\; 0.389  &     -0.150 &     -0.574 
          & \;\; 0.093                                           \\ \hline\hline
 \Delta   & \;\; 0.000 & \;\; 0.000 & \;\; 0.000 & \;\; 0.000 
          & \;\; 0.000                                           \\ \hline
 \Sigma^* &     -0.547 & \;\; 0.182 &     -0.216 &     -1.081 
          & \;\; 0.062                                           \\ \hline
 \Xi^*    & \;\; 0.540 &     -0.180 &     -0.382 & \;\; 1.742 
          &     -0.413                                           \\
 \hline\hline
\ea
}
\caption{Baryon wavefunction parameters derived in ref. \cite{bolz&kroll2} 
with the constituent strange quark mass $m_s = 350$ MeV, the octet baryon 
decay constant $f_{B_8} = 6.64 \times 10^{-3}$ GeV$^2$ and the transverse 
size parameter $a_{B_8} = 0.75$ GeV$^{-1}$ at the reference scale $\m_0$. 
The same for the decuplet baryons are $f_{B_{10}} = 0.0143$ GeV$^2$ and 
$a_{B_{10}} = 0.80$ GeV$^{-1}$. Note that in ref. \cite{bolz&kroll2} of
all the decuplet baryons, only the parameters of $\Delta$ were given.} 
\label{tab:baryon_param}
\end{table}

From these parameters, the mean-squared internal transverse momentum of 
the baryons $\r^2 = \lan \Kp^2 \ran$ can be worked out. The average value is 
$\r_{(8)} = 415.0$ MeV for the octet and $\r_{(10)} = 389.0$ MeV for 
the decuplet baryons. Their use as infrared cutoff was discussed  
in section \ref{sec:gg}.

\null
\vfill\eject

\vfill
\eject

\section*{Figure Captions}

\begin{itemize}

\item[\ref{f:sing_fig})]{Basic graphs that could contribute to $\c_J$ 
colour singlet decay. But actually, only graphs of type 
(a) can contribute.}

\item[\ref{f:sing})]{Graphs of type \fref{f:sing_fig} (a) can be
divided further into four groups.}

\item[\ref{f:oct})]{In addition to the graphs of type \fref{f:sing_fig},
these form the bases of further contributions in the colour
octet decay channel.}

\item[\ref{f:label_ex})]{Our labelling scheme as applied to (a) Group 2 and 
(b) Group 4.}

\item[\ref{f:eg1})]{Examples of complete colour octet graphs. (a) graph U1 and (b)
graph L2 of Group 5, and (c) graph L3 of Group 1.}

\item[\ref{f:eg2})]{Examples of complete colour octet graphs with 4-gluon vertex.
(a) graph 4G of Group 4 and (b) graph 4G of Group 10.}

\item[\ref{f:grp1})]{Basic graph of Group 1}

\item[\ref{f:grp2})]{Basic graph of Group 2}

\item[\ref{f:grp2d})]{Basic graph of Group 2'}

\item[\ref{f:grp3})]{Basic graph of Group 3}

\item[\ref{f:grp4})]{Basic graph of Group 4}

\item[\ref{f:grp5})]{Basic graph of Group 5}

\item[\ref{f:grp6})]{Basic graph of Group 6}

\item[\ref{f:grp7})]{Basic graph of Group 7}

\item[\ref{f:grp8})]{Basic graph of Group 8}

\item[\ref{f:grp9})]{Basic graph of Group 9}

\item[\ref{f:grp10})]{Basic graph of Group 10}

\end{itemize}

\section*{Table Captions}

\begin{itemize}

\item[\ref{tab:decuplet_B})]{The expansion coefficients of the distribution 
amplitudes $\f^{B_{10}}_{123}$ of the octet baryons considered in the $\c_J$ 
decay. The parameters associated with this set of coefficents are 
$f_{B_{10}}(\m_0) = 0.0143$ GeV$^2$ and $a_{B_{10}} = 0.80$ GeV$^{-1}$.}

\item[\ref{tab:sing_nucl})]{Clearly, the colour singlet contributions are 
insufficient in explaining the experimental data of $\c_J$ decay into 
$p\bar p$.}

\item[\ref{tab:results})]{The partial decay widths for $\c_J$ decay into octet 
and decuplet baryon-antibaryon pairs. The width of $\c_1 \lra N\bar N$ 
in parenthesis is to indicate that this value is a fit unlike all partial
widths of $\c_2$ which are predictions. Based on this fit, the rest of 
$\c_1$ widths are also predictions.}

\item[\ref{tab:results_cf})]{Comparing our results with the measured widths from 
the PDG \cite{pdg} data and from the BES collaboration \cite{bes}.
This branching ratio of $\c_1$ is a fit.}

\item[\ref{tab:grp1a}a)]{Numerators of the graphs of group 1 with helicity (+,+,--).}

\item[\ref{tab:grp1b}b)]{Numerators of the graphs of group 1 with helicity (+,--,+).}

\item[\ref{tab:grp1c}c)]{Numerators and the additional propagators of the graphs of 
group 1 with helicity (--,+,+).}

\item[\ref{tab:grp2a}a)]{Numerators of the graphs of group 2 with helicity 
(+,+,--) and (+,--,+).}

\item[\ref{tab:grp2b}b)]{Numerators and the additional propagators of the graphs of 
group 2 with helicity (--,+,+).}

\item[\ref{tab:grp2da}a')]{ Numerators of the graphs of group 2' with helicity 
(+,+,--).}

\item[\ref{tab:grp2db}b')]{ Numerators of the graphs of group 2' with helicity 
(+,--,+).}

\item[\ref{tab:grp2dc}c')]{ Numerators and the additional propagators of the graphs 
of group 2' with helicity (--,+,+).}

\item[\ref{tab:grp3a}a)]{Numerators of the graphs of group 3 with helicity (+,+,--).}

\item[\ref{tab:grp3b}b)]{Numerators of the graphs of group 3 with helicity (+,--,+).}

\item[\ref{tab:grp3c}c)]{Numerators and the additional propagators of the graphs of 
group 3 with helicity (--,+,+).}

\item[\ref{tab:grp4a}a)]{Numerators of the graphs of group 4 with helicity (+,+,--) 
and (+,--,+).}

\item[\ref{tab:grp4b}b)]{Numerators and the additional propagators of the graphs of 
group 4 with helicity (--,+,+).}

\item[\ref{tab:grp5a}a)]{Numerators of the graphs of group 5 with helicity (+,+,--).}

\item[\ref{tab:grp5b}b)]{Numerators of the graphs of group 5 with helicity (+,--,+).}

\item[\ref{tab:grp5c}c)]{Numerators and the additional propagators of the graphs of 
group 5 with helicity (--,+,+).}

\item[\ref{tab:grp6a}a)]{Numerators of the graphs of group 6 with helicity (+,+,--) 
and (+,--,+).}

\item[\ref{tab:grp6b}b)]{Numerators and the additional propagators of the graphs of 
group 6 with helicity (--,+,+).}

\item[\ref{tab:grp7a}a)]{Numerators of the graphs of group 7 with helicity (+,+,--) 
and (+,--,+).}

\item[\ref{tab:grp7b}b)]{Numerators and the additional propagators of the graphs of 
group 7 with helicity (--,+,+).}

\item[\ref{tab:grp8a}a)]{Numerators of the graphs of group 8 with helicity 
(+,+,--) and (+,--,+).}

\item[\ref{tab:grp8b}b)]{Numerators and the additional propagators of the graphs of 
group 8 with helicity (--,+,+).}

\item[\ref{tab:grp9a}a)]{Numerators of the graphs of group 9 with helicity 
(+,+,--) and (+,--,+).}

\item[\ref{tab:grp9b}b)]{Numerators and the additional propagators of the graphs 
of group 9 with helicity (--,+,+).}

\item[\ref{tab:grp10a}a)]{Numerators of the graphs of group 10 with helicity 
(+,+,--) and (+,--,+).}

\item[\ref{tab:grp10b}b)]{Numerators and the additional propagators of the 
graphs of group 10 with helicity (--,+,+).}

\item[\ref{tab:b_param})]{Basic input parameters.}

\item[\ref{tab:c_param})]{Charmonium parameters.}

\item[\ref{tab:baryon_param})]{Baryon wavefunction parameters derived in 
ref. \cite{bolz&kroll2} with the constituent strange quark mass $m_s = 350$ MeV, 
the octet baryon decay constant $f_{B_8} = 6.64 \times 10^{-3}$ GeV$^2$ and the 
transverse size parameter $a_{B_8} = 0.75$ GeV$^{-1}$ at the reference scale 
$\m_0$. The same for the decuplet baryons are $f_{B_{10}} = 0.0143$ GeV$^2$ and 
$a_{B_{10}} = 0.80$ GeV$^{-1}$. Note that in ref. \cite{bolz&kroll2} of
all the decuplet baryons, only the parameters of $\Delta$ were given.}

\end{itemize}

\end{document}